\documentclass[showpacs,twocolumn,aps,prb,superscriptaddress,10pt]{revtex4-1}
\usepackage{amsmath,amssymb,amsfonts,bm,graphicx}
\usepackage{braket}

\usepackage{graphicx}
\usepackage{amsmath}
\usepackage{bm}
\usepackage{hyperref}
\hypersetup{%
	pdftitle={Nano Terahertz}, %
	pdfauthor={Zhiyuan Sun},%
	pdfpagemode={UseNone},%
	pdfstartview={FitH},%
	breaklinks=true,%
	citecolor=blue,%
	colorlinks=true,%
	linkcolor=blue,%
	urlcolor=blue
}

\usepackage[T1]{fontenc}
\usepackage{fourier}
\usepackage{baskervald}

\newcommand{\unit}[1]{\,\mathrm{#1}} 
\newcommand{\equa}[1]{Eq.~\eqref{#1}} %

\graphicspath{{./}}

\begin{document}

\title{Collective modes and terahertz near-field response of superconductors}

\author{Zhiyuan Sun}
\affiliation{Department of Physics, Columbia University,
	538 West 120th Street, New York, NY 10027}

\author{M. M. Fogler}
\affiliation{Department of Physics, University of California San Diego, 9500 Gilman Drive, La Jolla, CA 92093}

\author{D. N. Basov}
\affiliation{Department of Physics, Columbia University,
	538 West 120th Street, New York, NY 10027}
\date{\today}

\author{Andrew J. Millis}
\affiliation{Department of Physics, Columbia University,
	538 West 120th Street, New York, NY 10027}
\affiliation{Center for Computational Quantum Physics, The Flatiron Institute, 162 5th Avenue, New York, NY 10010}
\date{\today}

\begin{abstract}
We theoretically study the low energy electromagnetic response of BCS type superconductors focusing on propagating collective modes that are observable with THz near field optics.
The interesting frequency and momentum range is $\omega < 2\Delta$ and $q < 1/\xi$ where $\Delta$ is the gap and $\xi$ is the coherence length.
We show that it is possible to observe the superfluid plasmons, amplitude (Higgs) modes, Bardasis-Schrieffer modes  and Carlson-Goldman modes using THz near field technique, although none of these modes couple linearly to far field radiation. Coupling of THz near field radiation to the amplitude mode requires particle-hole symmetry breaking while coupling to the Bardasis-Schrieffer mode does not and is typically stronger. For parameters appropriate to layered superconductors of current interest, the Carlson-Goldman mode appears in the near field reflection coefficient as a weak feature in the sub-THz frequency range. In a system of two superconducting layers with nanometer scale separation, an acoustic phase mode appears as the antisymmetric density fluctuation mode of the system. This mode produces well defined resonance peaks in the near-field THz response and has strong anticrossings with the Bardasis-Schrieffer and amplitude modes, enhancing their response. In a slab consisting of many layers of quasi-two dimensional superconductors, realized for example in samples of   high T$_c$ cuprate compounds,  many branches of propagating Josephson plasmon modes are found to couple to the THz near field radiation.
\end{abstract}

\maketitle

\section{Introduction}
\begin{figure}[b]
	\includegraphics[width= 0.9\linewidth]{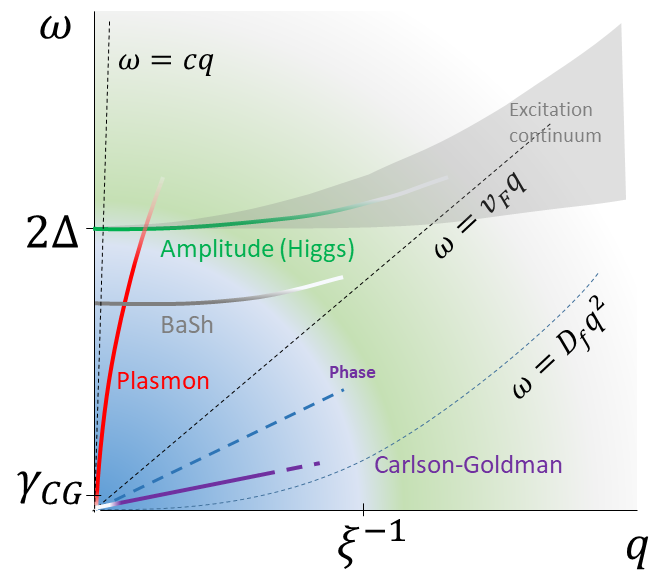}
	\caption{Schematic representation in the frequency-momentum plane of the collective modes that may appear in the electrodynamical response of a 2D superconductor. The blue area shows the low energy and long wave length region where weakly damped collective modes maybe observed.  Anti crossing between the plasmon and Higgs mode and the BaSh mode is not shown here. $c$ is the speed of light, $v_F$ is the fermi velocity and $D_f$ is the normal state diffusion coefficient.}
	\label{fig:parameter_regime}
\end{figure}

Electromagnetic (EM) response is a fundamental property of superconductors. The response to static electric and magnetic fields (infinite conductivity and the Meissner effect) are the defining properties of  the superconducting state. The response to the time dependent, very long wavelength transverse fields produced by far field radiation has been extensively studied \cite{Mattis1958, Anderson1958b,Zimmermann1991,Tinkham.1974, Basov2005a}. Superconductors are also characterized by a diversity of sub-gap collective modes \cite{Parks1969} including plasmons, acoustic phase modes, amplitude (Higgs) modes, the Carlson-Goldman modes, and the Bardasis-Schrieffer modes associated with fluctuations of subdominant order parameters, shown schematically in Fig.~\ref{fig:parameter_regime}. 
For superconductors  of current interest including cuprates \cite{Damascelli.2003,Basov2005a}, iron pnictides \cite{Stewart.2011}, NbSe$_2$ \cite{Sooryakumar1980} and MgB$_2$ \cite{Buzea_2001} the gap values
and the relevant collective modes are in the terahertz (THz) range. 
These modes couple weakly, if at all, to far field transverse photons.

Recent progress in cryogenic near field nano optics \cite{Ni2018} has enabled new generations of experiments probing the response of materials to short wavelength ($q\ll \omega/c$), primarily longitudinal, THz electric fields \cite{Basov2014a, Lundeberg2017, Dias2018,Wang.2020} radiated by a sharp metallic tip very close to the sample.
The essential new feature of the nano optics experiments 
as compared to traditional far field optics is the excitation of charge fluctuations in the material under study.
This information is encoded in the near field reflection coefficient $R_p(\omega,q)$ (see Appendix~\ref{appendix:rp}). 
In this paper we calculate the nano optics response $R_p(\omega,q)$ of a standard s-wave BCS superconductor, assuming a circular Fermi surface and focusing on the contribution of collective modes. Each of the modes we consider couples to charge fluctuations and is therefore in principle observable in nano optics experiments. We calculate in detail the matrix elements coupling each mode to charge excitations at nonzero momentum and from this the signal of the nano optical response.

Charge fluctuations are constrained by the continuity equation, the proper treatment of which requires a fluctuation calculation consistent with the relevant Ward Identities \cite{Anderson1958a,Schrieffer1999,Arseev2006, Yang2018}. We employ a one loop effective action method based on a Hubbard-Stratonovich transformation of the fundamental interacting electron system. This methodology, which is equivalent to a diagrammatic calculation including vertex corrections \cite{Schrieffer1999},  is an efficient way to take order parameter and charge fluctuations into account while respecting conservation laws.   

Fig.~\ref{fig:parameter_regime} shows many of the collective modes of interest in this paper.  We now discuss their physics and coupling to light.

The phase (Anderson-Bogoliubov-Goldstone)  mode (heavy blue dashed line in Fig.~\ref{fig:parameter_regime}) is the order parameter phase fluctuation. It is accompanied by a superfluid density oscillation \cite{Bogoljubov1958} which in the presence of long range Coulomb interaction converts this mode into a plasmon \cite{Anderson1963}.
In two dimensions (2D), the plasmon has a $\omega\sim \sqrt{q}$ dispersion, shown as the red solid line in Fig.~\ref{fig:parameter_regime} and discussed in section~\ref{sec:plasmons}. The plasmons directly couple to near field radiation. In multi layer systems, mutual screening leads to branches of acoustic (linearly dispersing) plasmons (not shown in Fig.~\ref{fig:parameter_regime}). These acoustic plasmons also couple to THz near field probes, see section~\ref{sec:double_layer}.

In the presence of abundant normal carriers as happens for example close to $T_c$, the Coulomb potential of the superfluid density fluctuation can be screened and as a result,
one finds two modes: a charge-neutral gapless mode (the so-called Carlson-Goldman (CG) mode \cite{Carlson1975}) in which the normal and superfluid densities oscillate out of phase, and the usual plasmon (phase mode) in which the
two densities oscillate in phase and the dispersion is controlled by the Coulomb interaction. 
The CG mode is discussed in section~\ref{sec:CG}.

The amplitude (Higgs) mode (green line in Fig.~\ref{fig:parameter_regime}) is the gap amplitude fluctuation \cite{Volkov.1974, Littlewood1982, Cea2014, Cea2015} which couples to EM linear response only when particle hole symmetry is broken and only at non-zero $q$ because electron density fluctuation is needed to locally perturb the density of states and then the gap. The coupling is suppressed by the small parameter $\Delta/E_F$, as discussed in section~\ref{sec:higgs}.
Therefore, the large momentum electric field from the near field tip \cite{Basov2014a, Lundeberg2017, Dias2018} would be the suitable probe of this mode, and the ideal samples are bilayer and multilayer superconductors, as we will show in the following.
The Higgs mode does couple to far field radiation in non equilibrium \cite{Krull.2014} or through third order nonlinear response as has been reported experimentally \cite{Matsunaga.2013,Katsumi2018,Nakamura2019,Cui.2019,Shimano2020}; the nonlinear coupling is however rather small and may not be sufficient to account for the measured signal \cite{Cea.2016,Udina.2019}.

The Bardasis-Schrieffer (BaSh)  mode \cite{Bardasis1961, Maiti2015, Maiti2016,Allocca2019,Muller.2019} is a fluctuation of a subdominant pairing order parameter, e.g., a $d$-wave fluctuation in an $s$-wave superconductor. It was proposed half a century ago \cite{Bardasis1961}, but has been very difficult to observe, although recent reports of its signature in  iron based superconductors \cite{Kretzschmar.2013,Bohm2014,Jost.2018} are very encouraging. The BaSh mode frequency is slightly below the gap for weak subdominant pairing and approaches zero as the subdominant pairing strength approaches the dominant one. 

The rest of the paper is organized as follows. In section~\ref{sec:Ginzburg_Landau} we present the Hubbard-Stratonovich transformation of the BCS Hamiltonian to obtain the Ginzburg-Landau effective action which includes the collective modes. Section~\ref{sec:EM_response} derives the linear EM response functions and their simple forms in the low energy limit. With the longitudinal optical conductivity, we analyse the properties of the collective modes in sections~\ref{sec:plasmons}, \ref{sec:higgs}, \ref{sec:BS} and \ref{sec:CG} with a focus on 2D. Section~\ref{sec:double_layer} discusses the acoustic plasmon mode in superconducting double layer which is promising to be observed in THz near field optics. We then discuss the cluster of hyperbolic Josephson plasmons in naturally layered superconductors in section~\ref{sec:bulk_layered} and show that they are greatly affected by the nonlocal correction to the optical conductivity and the discrete nature across the layers. 
Section~\ref{sec:discussion} is a summary and conclusion, with pointers to the relevant equations and figures, for readers uninterested in the details of the derivations.
Appendix~\ref{app:correlation_function} contains the definition and explicit forms of the correlation functions. Appendix~\ref{appendix:rp} has the derivation of the reflection coefficients. Appendix~\ref{appendix:two_fluid} has the derivation of the two fluid model.

\section{The effective Ginzburg-Landau action}
\label{sec:Ginzburg_Landau}
\subsection{The action of fermion, gap and electromagnetic fields}
The starting point is the BCS Lagrangian of attracting electrons coupled to electromagnetic (EM) field:
\begin{align}
L =& \int dr 
\left\{
\psi^\dagger 
\left[ \partial_\tau +
\xi(p-eA) + e\phi
\right]
\psi 
\right\}
\notag\\
&- \int dr dr^\prime g(r,r^\prime) \psi^\dagger(r)\psi^\dagger(r^\prime) \psi(r^\prime)\psi(r)
\notag\\
& - \int dr \frac{1}{16\pi} F^{\mu\nu} F_{\mu\nu}
\label{eqn:hamiltonian}
\end{align}
where $(\phi,\,\mathbf{A})=A^\mu$ is the EM field, $F_{\mu\nu} =\partial_\mu A_\nu-\partial_\nu A_\mu$, $\psi$ is the electron annihilation operator, $\xi(p)=\varepsilon(p) -\mu$, $p=-i\nabla$, $e$ is the electron charge and $g>0$ is the attractive interaction. We will be interested in relatively low frequency longitudinal EM fluctuations where the magnetic field can be neglected. Thus $\int dr \frac{1}{16\pi} F^{\mu\nu} F_{\mu\nu} \rightarrow \int dr \frac{1}{8\pi} E^2(r)$ for 3D and $\rightarrow \sum_q \frac{1}{4\pi |q|} E_{-q} E_q$ for a 2D plane embedded in 3D space. In the 2D formula $E_{q}$ is the Fourier component of the electric field on the 2D plane. Note that the EM field $A$ has an UV cutoff which is much smaller than the fermi momentum such that it mediates only the smooth part of the Coulomb interaction between the electrons. Here, the high energy part of the photon has been integrated out; together with the phonons or other pairing modes this  results in an effective interaction $g$. For simplicity, we don't explicitly notate the photons in what follows. 

Performing the Hubbard-Stratonovich transformation of the path integral $Z=\int D[\bar{\psi},\psi] e^{-S}$ in the pairing channel gives
\begin{align}
Z = & \int D[A] D[\bar{\psi},\psi] D[\bar{\Delta},\Delta]
e^{-S[\psi, A, \Delta]}
\end{align}
where the action
\begin{align}
S = \int d\tau dr 
\Bigg\{
&
\psi^\dagger 
G_{A\Delta}^{-1}
\psi
+ 
\sum_l \frac{1}{2 g_l}|\Delta_l|^2 
\Bigg\}
\label{eqn:HS_action}
\end{align}
describes coupled dynamics of the fermion field $\psi$, the EM field $A$ and the gap $\Delta_l$ with $l$ denoting the pairing angular momentum. Note that we have neglected detailed structure in $g$ that is not important to our conclusions. 

The fermion propagator is 
\begin{align}
G_{A\Delta}^{-1}
=
\begin{pmatrix}
\partial_\tau  + e\phi + \xi(p-eA) & \sum_l \Delta_l f_l(p) \\
\sum_l \bar{\Delta}_l \bar{f}_l(p)   &  \partial_\tau - e\phi - \xi(-p-eA)
\end{pmatrix}
\,
\end{align}
where the index $l$ labels different pairing channels and $f_l(p)$ describes the momentum dependence of the pairing function in each channel. For simplicity, we consider only spin singlet pairing with the Nambu spinors being $\psi^\dagger = (\psi_{\uparrow}^\dagger,\psi_{\downarrow})$. In the BCS regime of two dimensional superconductors, we can choose $f_l=\cos (l\theta_k)$ or $\sin (l\theta_k)$ and the corresponding pairing interaction is $g_l= \frac{1}{2\pi} \int d\theta \cos(l\theta) g\left(2 k_F \sin(\theta/2) \right)$. Note that for $l=0$, the $1/2\pi$ factor should be changed to $1/4\pi$. We assume that the $l=0$ component of the interaction is the strongest and thus the ground state has $s$-wave pairing but with only minor variations our formalism can be rewritten to cover other cases. We successively integrate out the fermion field to obtain the Ginzburg Landau action for the gap and EM field, and then the gap to obtain the action for EM field which gives the information of the EM response functions.

\subsection{Integrating out the fermions}
Expanding in the EM field, the Lagrangian density is  
\begin{align}
\mathcal{L} = 
\psi^\dagger G_\Delta^{-1} \psi + J^P_\mu A^\mu + \frac{1}{2} D_{ij} A_i A_j + O(A^3) + \sum_l \frac{1}{2g_l}|\Delta_l|^2 
\label{eqn:lagrangian}
\end{align}
where 
the Gor'kov Green's function for the Bogoliubov quasi particle is
\begin{align}
G_\Delta^{-1}
&=
\begin{pmatrix}
\partial_\tau  + \xi(p) & \sum_l \Delta_l f_l(p) \\
\sum_l \bar{\Delta}_l \bar{f}_l(p)   &  \partial_\tau - \xi(-p) 
\end{pmatrix}
\,,
\end{align}
the paramagnetic contribution to the current is 
\begin{align}
J^P_\mu
&=
e \psi^\dagger (\sigma_3,\, -\mathbf{v} \sigma_0) \psi = (\rho, -\mathbf{j}^P)
\,
\end{align}
and the diamagnetic `Drude' kernel is 
\begin{align}
D_{ij}
&=
e^2 \psi^\dagger \sigma_3 (\partial_{p_i p_j} \varepsilon) \psi 
\,.
\end{align}
Note that we have assumed inversion symmetry of $\varepsilon(p)$.
After integrating out the fermions, the action becomes
\begin{align}
S(\Delta,A) &= \mathrm{Tr} \ln G_{A\Delta} +   \int d\tau dr  \sum_l \frac{1}{2g_l}|\Delta_l|^2
\label{eqn:GL_action_1}
\end{align}
where the trace is over the frequency, momentum and spinor degrees of freedom of $\psi$.

It is convenient to split the action into mean field and fluctuation parts:
\begin{align}
S(\Delta,A) &= S_{\mathrm{mean} \,\, \mathrm{field}} + S_{\mathrm{fluctuation}}
\,
\label{eqn:GL_action_1}
\end{align}
where in the mean field part the trace is evaluated with space and time independent order parameters and the fluctuation part is the difference between the mean field action and the full action, and will be expanded in powers of small fluctuations around a homogeneous solution.

\subsection{Mean field as the saddle point}
Assuming the ground state has $s$-wave pairing with $\Delta$ independent of momentum, the mean field free energy is 
\begin{align}
S_{\mathrm{mean} \,\, \mathrm{field}}/V 
&= \frac{1}{2g} \Delta \bar{\Delta} + 
\sum_{\omega_n, k} \ln \left((i\omega_n)^2 - E_k^2 \right) 
\notag\\
&= \frac{1}{2g} \Delta \bar{\Delta} -
\sum_{k} \left[\frac{2}{\beta}\ln \left(1 + e^{-\beta E_k} \right) + E_k\right]
\,
\label{eqn:mean_field_energy}
\end{align}
where $E_k = \sqrt{\xi_k^2 + |\Delta|^2}$ is the quasi particle energy with gap $\Delta$. Minimization of $S_{\mathrm{mean} \,\, \mathrm{field}}$ with respect to $\Delta$ yields the gap equation
\begin{align}
\frac{1}{g}\Delta
= \sum_k \frac{\Delta}{E_k} (1-2f(E_k))
\,
\end{align}
where $f$ is the fermi distribution function.
At zero temperature, the integral in \equa{eqn:mean_field_energy} can be done and we obtain the condensation energy relative to the normal state:
\begin{align}
F &= S_{\mathrm{mean} \,\, \mathrm{field}}/V + \sum_{k} \xi_k
= \frac{1}{2g} \Delta \bar{\Delta} + \sum_{k} (\xi_k - E_k)
\notag\\
&\approx \frac{1}{2g} |\Delta|^2 - \frac{1}{2}\nu |\Delta|^2 \ln \frac{2\omega_D}{|\Delta|}
\,
\end{align}
where $\nu$ is the density of state at the fermi level of the normal state. The gap at zero temperature is thus $\Delta_0=2\omega_D e^{-\frac{1}{g\nu}-\frac{1}{2}}$ for $g\nu \ll 1$. Without loss of generality, we pick the mean field gap $\Delta$ to be real. The coherence length is defined as $\xi=v_F/\Delta$. Note that the free energy is non analytic around $\Delta=0$, i.e., the expansion coefficients in small $\Delta$ all diverge. The Ginzburg Landau expansion in powers of $\Delta$ is only possible at finite temperature and accurate close to $T_c$.

The validity of this `mean field plus fluctuation' approach is based on validity of the mean field theory. In other words, the quantum/thermal fluctuations of the order parameters have to be small. 
The latter is suppressed by the Ginzburg parameter $G=\frac{\Delta_0}{E_0 \xi^d} \sim \left( \frac{\Delta_0}{E_F} \right)^{d-1} \ll 1$, the small parameter of the mean field theory \cite{Altland.2010} where $E_0=\nu\Delta_0^2/4$ is the condensation energy density and $d$ is the space dimension. For conventional superconductors and charge/spin density waves, the accuracy of mean field approach has been extensively verified. 
For larger G, fluctuation corrections will change quantitative relationships (for example the relationship
between the model velocity and the fermi velocity) but will not change qualitative features including the 
qualitative nature of the dispersion relation (which modes are linear, which gapped). These effects  can be
accounted for by a diagrammatic expansion in the nonlinear coupling terms of the fluctuations. Renormalization group based approach could improve the accuracy of the perturbation theory. The most important qualitative effect of fluctuations is that in small superfluid stiffness superconductors at and below 2D, long wavelength fluctuations of the phase modes can be important, leading to physics of topological vortices and
Berezinskii-Kosterlitz-Thouless transition \cite{Kosterlitz_1973}. This is beyond the scope of the current paper.

\subsection{Fluctuations}
The fluctuations include those of the EM field, the $s$-wave gap and the subdominant pairing order parameters.  The $s$-wave gap fluctuation can be separated into amplitude and phase: $\Delta = (\Delta_0 + \Delta(r,t))e^{i2\theta(r,t)}$. 
It is convenient to perform a local gauge transformation\cite{Altland.2010} 
\begin{align}
\psi \rightarrow
\begin{pmatrix}
e^{i\theta}& 0\\
0& e^{-i\theta}
\end{pmatrix}
\psi
\end{align}
after which  \equa{eqn:HS_action} retains its form but with $\Delta$ real and the EM field changed to the gauge invariant ones
\begin{align}
e A_\mu
\rightarrow \partial_\mu \theta + eA_\mu  =
(i\partial_\tau \theta + e\phi, \nabla \theta-e\mathbf{A} )
\,.
\end{align}
In the effection action, the EM field always comes together with the phase fluctuations in the above gauge invariant form and therefore couples directly to the phase mode. The appearance of the other modes in the EM response can be inferred simply from their coupling to the phase mode. 

The action of fluctuations $\Delta_q$ around the mean field gap can be decomposed as 
\begin{align}
S_{\mathrm{fluctuation}} = S_{\theta} + S_{\Delta} + S_{\text{BaSh}}+S_{c}
+ \mathrm{nonlinear \,\, terms}
\label{eqn:s_fluctuation}
\end{align}
where 
\begin{align}
S_{\theta} =& \frac{1}{2} \sum_q K^{\mu \nu}(q) (\partial_\mu \theta + eA_\mu)_{-q} (\partial_\nu \theta + eA_\nu)_q 
\end{align}
is the phase action,
\begin{align}
S_{\Delta} =& \frac{1}{2} \sum_{q} G_a(q)^{-1} 
\Delta_{-q} \Delta_q 
\end{align}
is the amplitude action and
\begin{align}
S_{\text{BaSh}} =& \frac{1}{2} \sum_q G_l(q)^{-1} 
\Delta_l(-q) \Delta_l(q)
\end{align}
is the action for the fluctuation of the sub dominant pairing channels, i.e., the Bardasis-Schrieffer modes. Note that the `$a$' in $G_a$ means `amplitude' and `$l$' in $G_l$ labels the angular momentum of the subdominant pairing channel. Finally, 
\begin{align}
S_{c} =& \sum_q  \left( C^\mu(q) \Delta_{-q}  + \sum_l B_l^\mu(q) \Delta_l(-q)   \right)
\left(\partial_\mu \theta + e A_\mu \right)_{q}
\label{eqn:amplitude_phase_coupling}
\end{align}
is the coupling between phase and amplitude/BaSh mode fluctuations. Global $U(1)$ symmetry under $\theta \rightarrow \theta + \delta$ is manifest here since the action depends only on derivatives of the phase. This ensures charge conservation. 
The kernels $K$, $G_{a,l}$, $C$ and $B$ will be defined and discussed in subsequent subsections. 

\subsection{Phase action}
The quadratic kernel for the phase action is
\begin{align}
K_{\mu\nu}(q) &=  \langle \hat{T} J^P_\mu(x) J^P_\nu (0)  \rangle \big|_q + 
\begin{pmatrix}
0 & 0 \\
0 & \langle D_{ij} \rangle
\end{pmatrix}
\notag\\
&=
\begin{pmatrix}
\chi^{(0)}_{\rho \rho} & \chi^{(0)}_{\rho \mathbf{j}} \\
\chi^{(0)}_{\mathbf{j} \rho} & \chi^{(0)}_{\mathbf{j} \mathbf{j}} + \langle D_{ij} \rangle
\end{pmatrix}
\,
\end{align}
where $\chi^{(0)}_{\rho \rho}$, $\chi^{(0)}_{\rho \mathbf{j}}$ and $\chi^{(0)}_{\mathbf{j} \mathbf{j}}$ are the density-density, density-current and current-current correlation functions evaluated at the mean field saddle point.
In the case of quadratic band, $\varepsilon = p^2/(2m)$, the system has Galilean invariance and $D_{ij} = n/m \delta_{ij}$ where $n$ is the total carrier density. 

\emph{Low energy limit---} At zero temperature, for $\omega \ll \Delta$ and $q \ll 1/\xi$, we have $\chi^{(0)}_{\rho \rho} \rightarrow \nu$, $\chi^{(0)}_{\rho \mathbf{j}} \sim \omega \mathbf{q}$ and $\chi^{(0)}_{\mathbf{j} \mathbf{j}} \sim q^2$, thus to leading order we have 
\begin{align}
K_{\mu\nu}(q) 
&=
\begin{pmatrix}
-\nu & 0 \\
0 &  D
\end{pmatrix}
\,
\end{align}
where $\nu$ is the normal state density of state at the fermi level and $D$ is assumed to be rotationally invariant. 
Therefore, the effective low energy Lagrangian of the phase fluctuation is \cite{Altland.2010}
\begin{align}
\mathcal{L} = -\frac{1}{2}\nu \left( \partial_t \theta + e\phi \right)^2 + \frac{1}{2}D \left(\nabla \theta - e\mathbf{A} \right)^2 \,
\label{eqn:action_superfluid}
\end{align}
which describes the Nambu-Goldstone mode with velocity $v_g=\sqrt{n/(m \nu)} = v_F/\sqrt{d}$ if EM field were not present, also known as the Anderson-Bogoliubov mode \cite{Anderson1958a,Anderson1958b}. Here $d$ is the space dimension. Due to the long range Coulomb interaction (coupling to EM field), the Goldstone mode does not actually exist but is shifted to the high frequency plasmons through the Anderson-Higgs mechanism\cite{Anderson1963}. 

\subsection{Amplitude action}
The inverse propagator for amplitude fluctuation is 
\begin{align}
G^{-1}_a(q) &= \frac{1}{g} + \chi_{\sigma_1 \sigma_1}(q) 
\,.
\end{align}
At zero momentum $q=0$ and rotated to real frequency, it has the analytical form
\begin{align}
G^{-1}_a(\omega) &= (4\Delta_{sc}^2 -\omega^2) F(\omega) 
\,
\end{align}
where $\Delta_{sc}=\Delta$ and
\begin{align}
F(\omega) = \sum_{k} \frac{1}{E_k(4E_k^2 - \omega^2)}
=
\frac{\nu}{4\Delta^2} \frac{2\Delta}{\omega} \frac{\mathrm{sin}^{-1}\left(\frac{\omega}{2\Delta}\right)}{\sqrt{1-\left(\frac{\omega}{2\Delta}\right)^2}}
\end{align}
describes the quasi-particle effects. Specifically, $F$ 
diverges as $\frac{1}{\sqrt{2\Delta - \omega}}$ as the frequency approaches $2\Delta$ and has an imaginary part above $2\Delta$ due to quasi-particle excitations, leading to power law decay in time of the amplitude mode \cite{Volkov.1974}. Thus $G^{-1}_a$ does not have a simple pole at $\omega=2\Delta$ and the Higgs amplitude mode is not well defined in the weak coupling BCS approximation \cite{Volkov.1974,Cea2015}, although it might lead to a clearly observable feature in nonlinear optics\cite{Matsunaga.2013,Katsumi2018,Nakamura2019,Cui.2019,Shimano2020,Krull.2014}. 

Nevertheless, beyond weak coupling or in systems with additional physics, the behavior may be different. For example, in systems with superconductivity coexisting with charge density wave (CDW), the quasi particle absorption gap $\Delta=\sqrt{\Delta_{sc}^2 + \Delta_{cdw}^2}$ is larger than the Higgs frequency $2\Delta_{sc}$ and the amplitude mode becomes a well defined collective mode \cite{Cea2014}. In this case, $F$ is well behaved around the Higgs pole and we can approximate the propagator by 
\begin{align}
G^{-1}_a(\omega) &= \frac{\nu}{4\Delta^2}(4\Delta_{sc}^2 + \frac{1}{d}v_F^2 q^2 -\omega^2)  
\,
\end{align}
where $d$ is space dimension and the $O(q^2)$ expansion can be found in Appendix~\ref{appendix:Higgs}. Thus the Higgs mode frequency disperses roughly as $\omega_{hq}^2 = 4\Delta_{sc}^2 + \frac{1}{d}v_F^2 q^2$ as in Ref.~\cite{Littlewood1982}. Moreover, coupling between the Higgs mode and a higher frequency CDW phonon may further lower the Higgs mode frequency\cite{Littlewood1982} and enhance its Raman matrix element, as has been proposed for $2H$-NbSe$_2$ \cite{Sooryakumar1980,Grasset.2018}. 

\subsection{Bardasis-Schrieffer mode action}
The inverse propagators for the fluctuations of the higher angular momentum pairing channels are 
\begin{align}
G^{-1}_l(q) &= \frac{1}{g_l} + \chi_{f_l(k)\sigma_2, f_l(k)\sigma_2}(q) 
\,
\end{align}
where the correlator is defined in Appendix~\ref{appendix:BS}.
Note that there are two directions for the fluctuations of the subdominant order parameters in the complex plane: perpendicular to the mean field gap (in the `imaginary' direction) and parallel to it (in the `real' direction). The BaSh modes \cite{Bardasis1961,Maiti2016,Allocca2019} are the `imaginary' fluctuations, i.e., in the $\sigma_2$ channel. This channel has  nonzero matrix elements to quasiparticles at the gap edge thus rendering $\chi_{\sigma_2,\sigma_2}(\omega)$ divergent as the frequency approaches the gap $2\Delta$ from below. As a result, the BaSh modes all have energies below $2\Delta$. The `real' modes, the fluctuations in the $\sigma_1$ channel, don't have poles and are not well-defined collective modes, as shown in Appendix~\ref{appendix:BS}.

In this paper we focus on the $d$-wave BaSh mode in an $s$-wave superconductor.  In two dimensions, there are two $d$-wave BaSh modes corresponding to $d_{x^2-y^2}$ and $d_{xy}$. We consider here the $d_{x^2-y^2}$ mode; considerations for the $d_{xy}$ mode are  similar. Taking the momentum to be along $x$, one finds that to leading order in momentum the inverse propagator of the $d_{x^2-y^2}$ mode is 
\begin{align}
G^{-1}_{\text{BaSh}}(\omega,q) &= \frac{1}{g_d} + \chi_{cos(2\theta_k)\sigma_2, cos(2\theta_k)\sigma_2}(\omega,q) 
\notag\\
&= \frac{1}{g_d}-\frac{1}{2g} -\frac{1}{2}\omega^2 F(\omega) + \frac{1}{16} \frac{\nu}{\Delta^2} v_F^2 q^2
\,.
\label{eqn:bs_propagator}
\end{align}
For $g_d \in (0,\, 2g)$, at zero momentum, it has a pole below $2\Delta$ which gives the mode frequency
\begin{align}
\omega_{\text{BaSh}}
=
2\Delta \left\{
\begin{array}{lc}
1- \frac{\pi^2}{32} (\nu g_d)^2 
&  
\,(g_d \ll 2g)
\\
\sqrt{ \frac{2}{ g_d \nu}- \frac{1}{g \nu } }
&  \,(g_d \rightarrow 2g)
\end{array}
\right. 
\label{eqn:sigma_n}
\end{align}
in the weak and strong BaSh fluctuation limits.
As $g_d$ changes from $0$ to $2g$, $\omega_{\text{BaSh}}$ goes from $2\Delta$ to $0$. For $g_d > 2g$, the ground state is no longer an $s$-wave one \cite{Maiti2015}. At finite momentum, the mode frequency shifts up as $q^2$ as shown in Appendix~\ref{appendix:BS}.

\subsection{Coupling terms}
In this section, we present the coupling matrix elements between light and the Higgs mode/BaSh mode, a main result of this paper. In the present formalism, the exact form of the coupling between phase and amplitude is
\begin{align}
C^\mu(q) = \left( \chi^{(0)}_{\rho \Delta},\,\, \chi^{(0)}_{\mathbf{j} \Delta} \right)
= \left( \chi^{(0)}_{\sigma_3 \sigma_1},\,\, \chi^{(0)}_{\mathbf{v} \sigma_0,\, \sigma_1} \right)
\,.
\label{eqn:phase_higgs_C}
\end{align}
The phase $\theta$ and BaSh mode $\Delta_l$ are odd under either time reversal or particle hole interchange because they are fluctuations in the `imaginary' direction on the complex plane; the amplitude fluctuation $\Delta$ is however even under these operations. Therefore,  linear coupling between phase and amplitude fluctuations breaks particle hole symmetry while linear coupling between phase and Bash modes does not. 

Taking the requirements of time reversal and inversion symmetry into account we find that the coupling coefficient can be expanded as  
\begin{align}
C^\mu(q) = \left( C_0 + O(q^2),\,\, C_i \omega \mathbf{q} + O(q^3) \right)
\,
\end{align}
where $C_0,C_i=0$ in a particle hole symmetric situation.

To study the linear EM response for $q \ll 1/\xi$, it is sufficient to keep the leading terms.
The simplest way to break the particle hole symmetry is to use an energy dependent electronic density of state (DOS), e.g., as in the parabolic band electron gas in three dimension. Assuming the DOS is $g(\xi) = \nu (1+\lambda \xi/E_F)$ (note that $\xi=\varepsilon-\mu$ should not be confused with the coherence length), we obtain from $\chi^{(0)}_{\sigma_3 \sigma_1}$ that 
\begin{align}
C_0 
\approx & - \lambda \nu \frac{\Delta}{2E_F}\sinh^{-1} \left( \frac{\omega_D}{\Delta} \right) 
\,
\label{eqn:higgs_coupling_C0}
\end{align}
and from $\chi^{(0)}_{\mathbf{j} \Delta}$ that 
\begin{align}
C_i =& \frac{1}{12d} \lambda \nu  \frac{\Delta}{E_F} \left(\frac{v_F}{\Delta}  \right)^2
\,,
\label{eqn:higgs_coupling_Ci}
\end{align}
see Appendix~\ref{appendix:coupling_constants} for detailed derivation.
The  factor $\lambda \Delta/E_F$ characterizes the strength of particle hole symmetry broken and is small in known superconductors.

In two dimensions, from inversion, time reversal symmetry, and that the BaSh fluctuation is $\pi/2$ out of phase relative to the static $s$-wave order parameter, the linear coupling coefficient between the phase and the $d_{x^2-y^2}$ BaSh mode can be written as 
\begin{align}
B^\mu =(B_0 \omega q^2,\, B_i q_x,\, -B_i q_y  )
= \left( \chi^{(0)}_{\sigma_3,\, \sigma_2 f_l(p)},\,\, \chi^{(0)}_{\mathbf{v} \sigma_0,\, \sigma_2 f_l(p)} \right)
\label{eqn:BS_phase_coupling}
\end{align}
where $B_i = i \pi \Delta v_F^2 F(\omega)$ describes coupling between electric field and the BaSh mode, see Appendix~\ref{appendix:coupling_constants}. Since the angular momentum change is $\delta l=2$ in exciting an s bound state to a d state, an inhomogeneous electric field is required to overcome the selection rule and thus the $B_i$ terms exist only at finite momentum. It will be shown that in the optical conductivity, at leading ($O(q^2)$) order the $B_0$ term does not contribute. Note that coupling to the BaSh mode does not require breaking particle hole symmetry and is not suppressed by the typically small parameter $\Delta/E_F$. Thus in general, the BaSh mode couples to EM more strongly than the Higgs mode.

\section{Electrodynamics of superconductors}
\label{sec:EM_response}
\subsection{Linear electromagnetic response}
\begin{figure}
	\includegraphics[width= \linewidth]{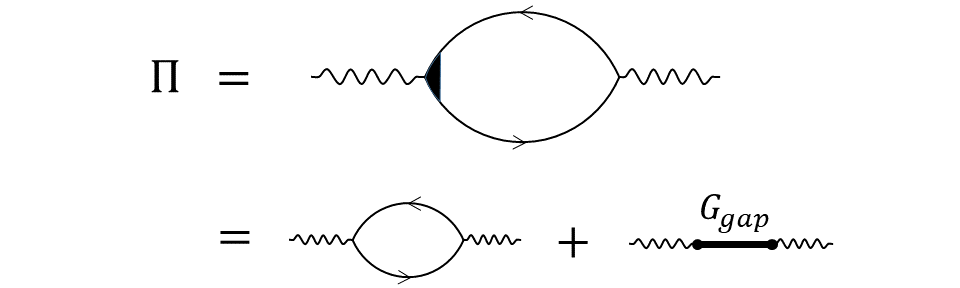}
	\caption{Diagrammatic representation of the EM linear response kernel $\Pi$, i.e., self energy of photon. First line is the photon self energy using the language of interacting electrons. The solid lines are electron Green's functions within the BCS approximation. The corresponding vertex correction should be included to restore the Ward identity \cite{Schrieffer1999}. Second line is the same thing but expressed in the language of coupling photons and order parameter fluctuations, as described by the Ginzburg-Landau action in \equa{eqn:GL_action_1}. The first term is the bare current correlation $K$ and the second term is the fluctuation contribution which corresponds to the vertex correction. }
	\label{fig:Feynman_Diagram}
\end{figure}
To obtain the linear EM response functions, one just needs to obtain the fluctuation action quadratic in the EM fields. Integrating out the amplitude and BaSh modes results in
\begin{align}
S =& \frac{1}{2} \sum_q K^{\mu \nu}(q) (\partial_\mu \theta + eA_\mu)_{-q} (\partial_\nu \theta + eA_\nu)_q 
\label{eqn:phase_action}
\end{align}
with the kernel modified to  
\begin{align}
K^{\mu \nu}(q) = 
&
\begin{pmatrix}
\chi^{(0)}_{\rho \rho} & \chi^{(0)}_{\rho \mathbf{j}} \\
\chi^{(0)}_{\mathbf{j} \rho} & \chi^{(0)}_{\mathbf{j} \mathbf{j}} + \langle D_{ij} \rangle
\end{pmatrix}
- G_a(q) C^\mu(q) C^\nu(q)
\notag\\
& - G_{\text{BaSh}}(q) B^\mu(q) B^\nu(q)
\label{eqn:full_K}
\,.
\end{align}
For longitudinal EM response, it is inappropriate to directly employ the `free' optical conductivity $\chi^{(0)}_{\mathbf{j} \mathbf{j}} + \langle D_{ij}\rangle$ or the density response $\chi^{(0)}_{\rho \rho}$ obtained from the BCS mean field Hamiltonian,  because the static mean field approximation   breaks global $U(1)$ gauge invariance and thus does not satisfy charge conservation: $K^{\mu \nu} q_\nu \neq 0$. The reason is that longitudinal EM fields excite order parameter phase fluctuations that are not captured by Bogoliubov quasi-particles.

The solution is to take into account the phase fluctuations\cite{Anderson1958a,Paramekanti.2000,Benfatto.2004}  which ensures charge conservation since the Euler-Lagrangian equation from \equa{eqn:phase_action} is just the continuity equation. By integrating out the phase (or equivalently by solving the Euler Lagrangian equation for the phase), one finally obtains the EM action 
\begin{align}
S =&  \frac{1}{2} \sum_q \Pi^{\mu \nu}(q) A_\mu(-q) A_\mu(q)
\end{align}
where 
\begin{align}
\Pi^{\mu \nu}(q) = K^{\mu \nu}(q) - \frac{q_\alpha q_\beta K^{\alpha \nu} K^{\mu \beta} }{q_a q_b K^{a b}}
\label{eqn:EM_response_tensor}
\end{align}
is the EM response tensor satisfying $J^\mu = \Pi^{\mu \nu} A_\nu$ and the continuity equation $\Pi^{\mu \nu} q_\nu = 0$. Specifically, $\Pi^{0 0} = \chi_{\rho\rho}$ is the  irreducible (with respect to the Coulomb interaction) density-density response (polarization function) and $\frac{i}{\omega}\Pi^{i j} = \sigma_{ij}$ is the optical conductivity. This formalism is equivalent to correcting the current vertex by electron electron interactions after which gauge invariance \cite{Anderson1958a} and thus Ward Identity \cite{Schrieffer1999} are recovered. The diagrammatic representation of \equa{eqn:EM_response_tensor} is shown in Fig.~\ref{fig:Feynman_Diagram}.   \equa{eqn:EM_response_tensor} contains all the information of linear coupling between EM field and the collective modes and will be frequently used in the following.

Note that in the clean limit, the optical conductivity from \equa{eqn:EM_response_tensor} has vanishing real parts above the gap due to momentum conservation, which means light does not break cooper pairs without the assistance of impurities or phonons. To account for optical absorption above the gap, it is therefore necessary to introduce the effect of disorder. The Mattis-Bardeen\cite{Mattis1958,Mahan1990} theory for optical absorption completely relaxes momentum conservation in the quasi particle excitation process, and has proven accurate in various BCS type superconductors. In this paper, we employ the Mattis-Bardeen formula to describe the optical conductivity above the gap:
\begin{align}
\sigma_1(\omega>2\Delta) = & \sigma_n(\omega) \Theta\left(\frac{\omega}{2\Delta}-1 \right) 
\notag\\
& \left[
\left(1+ \frac{2\Delta}{\omega}  \right)
E\left(\frac{\omega-2\Delta}{\omega+2\Delta} \right) -
\frac{4\Delta}{\omega}
K\left(\frac{\omega-2\Delta}{\omega+2\Delta} \right) 
\right]
\,
\label{eqn:Mattis-Bardeen}
\end{align}
where $\sigma_n$ is the normal state conductivity and $E(x),\, K(x)$ are the complete elliptic integrals. 

\subsection{The low energy limit}
At low temperature compared to $T_c$, in the low energy limit  $\omega \ll \Delta$ and $q \ll 1/\xi$
as shown in the blue region of Fig.~\ref{fig:parameter_regime}, the electrodynamics can be described by the Lagrangian \equa{eqn:action_superfluid} which leads to the longitudinal optical conductivity:  
\begin{align}
\sigma_s = i \frac{D_s/\pi}{\omega-v_g^2 q^2/\omega} 
\,,\quad D_s=\pi n_s e^2 / m
\,.
\label{eqn:superfluid_conductivity}
\end{align}
Here $n_s$ is the superfluid density and $n_s/m$ is the superfluid stiffness which in 3D is related to the magnetic penetration depth as $\lambda_B= \sqrt{\frac{c^2}{4\pi} \frac{m}{n_s e^2} }$.
This form closely resembles that of a hydrodynamic electron fluid \cite{Sun.2018,Torre2019,Sun2018b} except that damping is completely suppressed here by the gap. \equa{eqn:superfluid_conductivity} completely specifies the crossover between the Drude limit $\omega\gg v_F q$ and the Thomas-Fermi limit $\omega\ll v_F q$.
Note that the amplitude/BaSh mode and quasiparticle excitation don't enter here since they appear at higher energy.
For a clean and isotropic BCS superconductor, $v_g=v_F/\sqrt{d}$ at zero temperature and gradually decreases to zero as temperature is raised to $T_c$. 

At non-zero temperature, one should add the contribution of the normal carriers which makes the conductivity into the `two fluid' form derived in Appendix~\ref{appendix:two_fluid}:
\begin{align}
\sigma(\omega,q)  = \sigma_s + \sigma_n =
i \frac{D_s/\pi}{\omega-v_g^2 q^2/\omega}  + \sigma_n
\,.
\label{eqn:two_fluid}
\end{align}
For $\Delta \ll T$, an analytical formula for the normal fluid conductivity with non-zero scattering rate can be found from the Boltzmann equation\cite{Warren1960,Mahan1990,Abrikosov:1975}. In the simple limits, 
\begin{align}
\sigma_n 
=
\left\{
\begin{array}{lc}
\frac{ i D_n/\pi}{\omega + i\gamma}  &  \,(\omega \gg D_f q^2)
\\
-i \nu_n\frac{\omega}{q^2}
&  \,(\omega \ll D_f q^2, v_F q)
\end{array}
\right. 
\label{eqn:sigma_n}
\end{align}
where  $D_n = \pi n_n/m$, $n_n \approx n$ is the density of normal carriers, $\gamma$ is the scattering rate, $D_f=v_F^2/(d\gamma)$ is the normal state diffusion constant, $d$ is space dimension and $\nu_n$ is close to $\nu$ at temperate close to $T_c$.

\section{2D Plasmons}
\label{sec:plasmons}
For simplicity, we neglect the coupling to the amplitude mode in this section. The plasmons are the charge density fluctuations and can be found by the zeros of the dielectric function
\begin{align}
\epsilon = 1- V_q \chi_{\rho\rho} = 1+ V_q \frac{i q^2}{\omega} \sigma =0  \,
\label{eqn:dielectric}
\end{align}
where $V_q = 4\pi /q^2$ for three dimension and $V_q = 2\pi /|q|$ for 2D. Together with \equa{eqn:superfluid_conductivity}, we obtain the plasmon dispersion $\omega_p=\sqrt{2D_s q + v_g^2 q^2}$ for two dimension. For three dimension, \equa{eqn:dielectric} predicts $\omega_p=\sqrt{4\pi n_se^2/m + v_g^2 q^2} \gg \Delta$ which lies in the high energy regime beyond the limit of validity of our theory although the correct plasma frequency $\omega_p^2 = 4\pi n_se^2/m$ is obtained for a clean superconductor. 

In the low frequency limit $\omega \lesssim \omega_c$, the 2D plasmon
dispersion $\omega_p=\sqrt{2D_sq}$ approaches the edge of the continuum of vacuum
propagating photons $\omega=cq$ (recall that here $q$ is a two dimensional momentum and light modes disperse as $\omega=c\sqrt{q^2+k_z^2}$ for any $k_z$).   For lower frequencies the analysis given here requires modification, because the electric fields associated with the plasmons begin to extend far from the 2D sheet, so that the plasmon couples much less strongly to near field radiation. 
The critical frequency can be estimated as $\omega_c=2D_s/c$
corresponding to the energy $\hbar\omega_c=\frac{e^2}{\hbar c}\frac{\hbar^2n_s}{m}\sim \frac{1}{137}E_F^\ast$
where $E_F^\ast$ is the fermi energy equivalent to a two dimensional superfluid stiffness $n_s/m$. For a clean,
weakly correlated material $E_F^\ast$ is of eV-scale and the crossover frequency is of the order of $1$ THz. However, many superconductors of current interest \cite{Orenstein1990} have much lower $E_F^\ast$
so that the crossover frequency is well below the THz regime.

Assuming a doping level of $n=7\times 10^{13} \unit{cm}^{-2}$ (corresponding to the fermi momentum $k_F = 2\pi/(3 \unit{nm})$) and the fermi velocity of $v_F = 2.5\times 10^5 \unit{m/s}$, one obtains a wavelength of $\lambda \approx 180 \unit{\mu m}$ for the plasmon at $1 \unit{THz}$. This wavelength is close to that of the corresponding vacuum photon ($300 \unit{\mu m}$) although a substrate with large dielectric screening might make the plasmon wave length shorter.

\begin{figure}
	\includegraphics[width= 1.0 \linewidth]{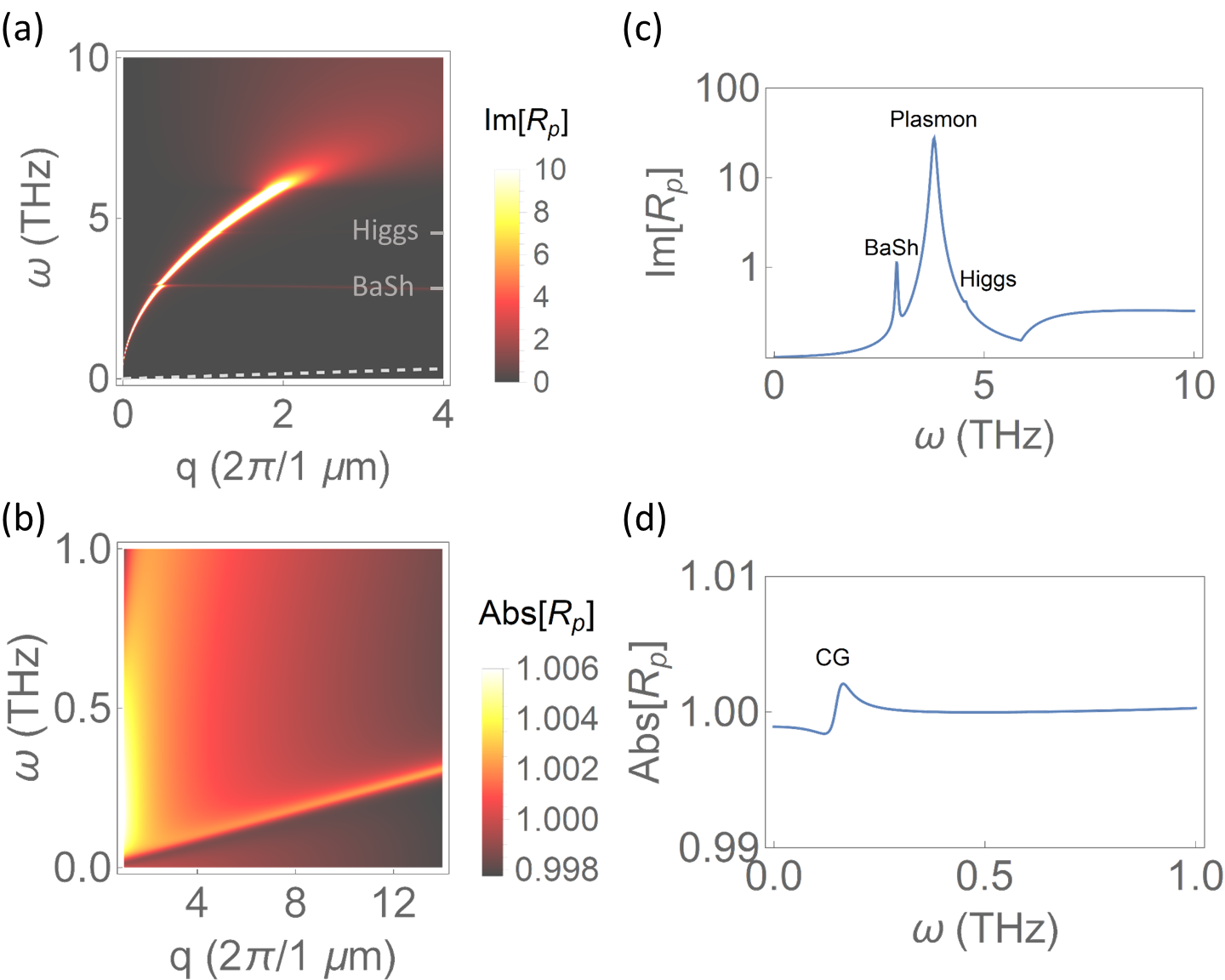}
	\caption{
		Near field reflection coefficients of a monolayer superconductor. (a) At $T=0 \unit{K}$, the dominant feature comes from the plasmon while there is very weak anti crossing with the BaSh mode at $3.0 \unit{THz}$. The coupling to Higgs mode is too weak to be seen. (b) At $T=79.6 \unit{K}$ close to $T_c =80 \unit{K}$, the plasmon is overdamped while the CG mode appears as a weak crossover of $R_p$. 
	Note the difference in color scales between (a) and (b).
	Right panel is the vertical line cuts of left panel at (c) $q=0.8 \times \unit{2\pi/(1 \mu m)}$ and (d) $q=7 \times \unit{2\pi/(1 \mu m)}$.
	The fermi momentum, velocity is $k_F = 2\pi/(3 \unit{nm})$, $v_F = 2.5\times 10^5 \unit{m/s} $, the normal state scattering rate is $\gamma= 30 \unit{THz}$, the gap at zero temperature is $\Delta= 3.0 \unit{THz}$ and $\kappa=0.4$, $\kappa_{\text{BaSh}}=1.5$.}
	\label{fig:Monolayer_plasmon_cg}
\end{figure}

Nevertheless, in a dirty superconductor with a large normal state scattering rate $\gamma \gg T_c$, the sub-gap plasmon frequency is mainly determined by the superfluid density which is only a part of the total density even at zero temperature: $n_s \sim n T_c/\gamma$.  At the same THz frequency far below $\gamma$, the plasmon wavelength is smaller by the factor $T_c/\gamma$ and they become more confined to the 2D plane.  Below the gap, weakly damped plasmons with dispersion $\omega = \sqrt{2D_s q}$ couple strongly to near field probe as shown by the near field reflection coefficient
\begin{align}
R_p = -\frac{1}{\epsilon} +1  = -\frac{1}{1+\frac{i 2\pi q}{\omega} \sigma} +1  \,
\label{eqn:rp}
\end{align}
in Fig.~\ref{fig:Monolayer_plasmon_cg} (a). For the ratio $T_c/\gamma =0.1$, the $1 \unit{THz}$ plasmon wave length is shrunk by the same factor to $18 \unit{\mu m}$.

\section{Higgs mode}
\label{sec:higgs}
The Higgs mode couples to phase fluctuation as shown in \equa{eqn:amplitude_phase_coupling},  manifests itself in the second term of \equa{eqn:full_K} and finally  enters the EM response through \equa{eqn:EM_response_tensor}.
The density response with the Higgs mode correction is thus 
\begin{align}
\chi_{\rho\rho}=\Pi^{00} = \frac{ \frac{q^2}{\omega^2} \frac{n}{m} \left(\nu - G_a C_0^2 \right) + \nu G_a C_i^2 q^4 }
{\nu -  G_a \left(C_0 +C_i q^2 \right)^2 - \frac{q^2}{\omega^2} \frac{n}{m}}
  \,.
\label{eqn:dielectric_higgs}
\end{align}
Since the $C_0$ and $C_i$ terms contribute terms at the same order, we take $C_i =0 $ to arrive at a simplified expression 
\begin{align}
\chi_{\rho\rho} = \frac{\frac{q^2}{\omega^2} \frac{n}{m} \left( \nu - G_a C_0^2 \right)  }
{\nu -  G_a C_0^2 - \frac{q^2}{\omega^2} \frac{n}{m}}
\,.
\label{eqn:dielectric_higgs2}
\end{align}
Thus the longitudinal optical conductivity is 
\begin{align}
\sigma(\omega,q) = i \frac{n_s e^2 / m}{\omega} \frac{1}{1- \frac{v_g^2 q^2}{\omega^2} \frac{1}{1- \kappa^2 \Delta^2/(\omega^2 - \omega_{hq}^2)}} 
\,
\label{eqn:sigma_higgs}
\end{align}
where
\begin{align}
\kappa = \lambda \frac{\Delta}{E_F} \sqrt{2\sinh^{-1} \left( \frac{\omega_D}{\Delta} \right) } 
\,
\label{eqn:kappa}
\end{align}
is the dimensionless coupling constant of the Higgs mode to EM and $\omega_{hq}= \sqrt{4\Delta_{sc}^2+ v_F^2q^2/d}$ is the Higgs mode frequency. Since $\lambda$ is order one and $\sinh^{-1} \left( \frac{\omega_D}{\Delta} \right)$ is not a large number, this coupling is simply suppressed by the small number $\frac{\Delta}{E_F}$. Note that at $q=0$, the optical conductivity reduces to the Drude form and there is no signature of Higgs mode showing why this mode cannot be observed in conventional far-field  THz linear response. In contrast, near field optical imaging technique has access to non-zero $q$ where the Higgs mode manifests itself through coupling to the plasmons. 

Specifically, for a monolayer superconductor, the coupled collective modes can be found as the poles of \equa{eqn:dielectric}. The weight of the Higgs pole in $R_p$ scales as $W_{higgs} \sim \kappa^2 v_g^2 q^2/\Delta$ for $\omega_h \gg \omega_p$, i.e., well before the Higgs mode crosses the plasmon.
Nevertheless, the most prominent signature of the Higgs mode is its anti crossing with the plasmon mode which happens roughly at $\omega_p(q)=\omega_{hq}$. A detailed solution of \equa{eqn:dielectric} gives the frequency splitting at the anti-crossing as 
\begin{align}
\delta \omega \approx \frac{\kappa \Delta}{2\omega_{hq}} \sqrt{\kappa^2 \Delta^2 + 4 v_g^2 q^2}
\label{eqn:higgs_splitting}
\end{align}
where $q$ is the momentum at the anti-crossing. Therefore, the splitting will be bigger if the anti-crossing happens at larger momentum.

\section{Bardasis-Schrieffer mode}
\label{sec:BS}
Optical excitation of the BaSh mode can be viewed as transition from an s bound state of the cooper pair to a d bound state. This is forbidden in far field optics for two reasons: first, unlike the Hydrogen atom case, uniform electric field exerts the same force on the two electrons and does not change the internal structure; second, both s-state and d-state have even parity which forbids the transition due to optical selection rule. Thus it is necessary to go to nonzero momentum for its nonzero coupling to EM field. Indeed, the coupling constant is proportional to $\xi q$ which is appreciable when the electric field becomes substantially nonuniform on the scale of a cooper pair size.

Plugging \equa{eqn:BS_phase_coupling} into \equa{eqn:full_K} and \equa{eqn:EM_response_tensor} gives the appearance of the BaSh mode in the longitudinal optical conductivity:
\begin{align}
\sigma(\omega,q) = i \frac{n_s e^2 / m}{\omega} \frac{1}{\frac{1}{1 + \kappa_{\text{BaSh}}^2 v_g^2q^2/(\omega^2 - \omega_{\text{BaSh}}^2)}- \frac{v_g^2 q^2}{\omega^2} } 
\,
\label{eqn:sigma_bs}
\end{align}
where $\kappa_{\text{BaSh}}= \frac{\pi}{\sqrt{2}}v_F^2/v_g^2 \sim 1$ is the dimensionless coupling constant between BaSh mode and EM.   The $B_0$ terms are higher order in $q$ and are neglected. Note that if the momentum $q$ is along x, the BaSh mode means the $d_{x^2-y^2}$ order parameter fluctuation. 

The BaSh mode couples to near field more strongly than the Higgs mode due to the absence of the $\Delta/E_F$ factor in the coupling constant $\kappa_{\text{BaSh}}$. Solving the pole equation, \equa{eqn:dielectric}, one obtains the frequency splitting at the anti-crossing between BaSh and plasmon
\begin{align}
\delta \omega \approx \kappa_{\text{BaSh}} v_g q
\label{eqn:bs_splitting}
\end{align}
which scales linearly with the momentum at the anti-crossing.

\section{Carlson-Goldman mode}
\label{sec:CG}
The CG mode is a superfluid density fluctuation accompanied by the counter flow of normal carriers such that the
Coulomb potential from the superfluid fluctuation is almost completely screened \cite{Carlson1975,schmid.1975,Artemenko.1976,Pethick1979,Artemenko1979,Kulik.1981,Goldman2007}. This screening requires a large density of normal carriers which is typically found near T$_c$. The velocity $v_g$ of the CG mode depends on the ratio between the superfluid density $n_s$ and superfluid susceptibility $\chi_s = \frac{\pi}{4} \frac{\Delta}{T_c} \nu$ and has different expressions in the clean\cite{Artemenko.1976} and dirty \cite{schmid.1975} limits:
\begin{align}
v_g = \sqrt{\frac{n_s}{m} / \chi_s}
=
\frac{v_F}{\sqrt{d}}
\left\{
\begin{array}{lc}
\sqrt{2\Delta/\gamma} & \gamma \gg T_c \,(\text{Dirty})
\\
\sqrt{\frac{7\zeta(3)}{\pi^3} \frac{\Delta}{T}}
& \gamma \ll T_c \,(\text{Clean})
\end{array}
\right. 
\label{eqn:vg}
\end{align}
where $\zeta(x)$ is the Riemann Zeta function and we have used the fact that  $n_s = n\frac{\pi \Delta^2}{2\gamma T_c}$ for dirty superconductors and $n_s = 2(1-T/T_c) n$ for clean superconductors close to $T_c$.

\begin{figure}
	\includegraphics[width=0.8 \linewidth]{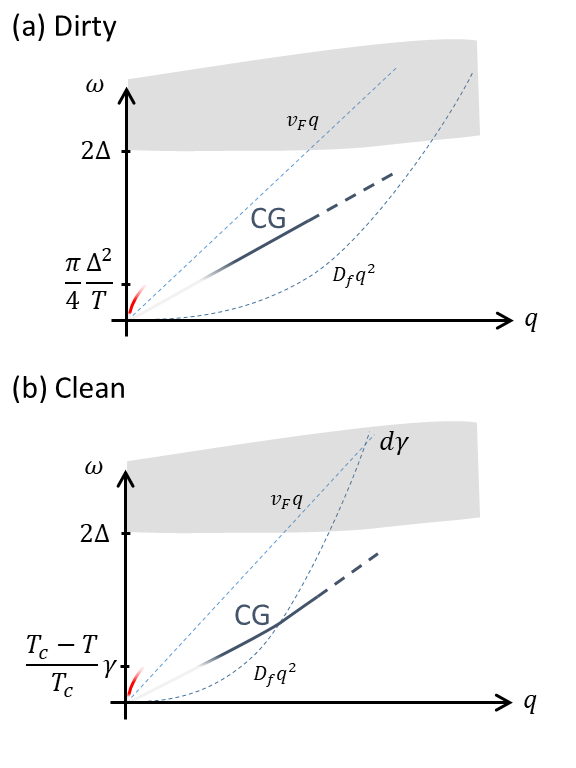}
	\caption{Schematics of the dispersion of the Carlson-Goldman mode on the frequency momentum plane. The CG mode speed is exaggerated. Since the theory \equa{eqn:vg} is accurate only for $q \ll \xi^{-1}$, the part of the dispersion beyond $\xi^{-1}$ is drawn as dashed. Gray region corresponds to the quasi particle pair excitation continuum.  Red solid line means under damped plasmon. In both the dirty and clean cases $\gamma, T_c \gg \Delta$ and $T_c-T \ll T_c$ are assumed.}
	\label{fig:CG}
\end{figure}

 Its dispersion can be derived from the two fluid conductivity \equa{eqn:two_fluid} by setting $\epsilon=0$ which yields
\begin{align}
\omega^3 + i \frac{\omega_n^2}{\gamma}\omega^2 - (\omega_s^2 + v_g q^2) \omega - i\frac{\omega_n^2}{\gamma} v_g^2 q^2 =0
\,
\label{eqn:CG_equation}
\end{align}
in the limit of $D_f q^2 \ll \omega \ll \gamma$. Note the plasma frequency $\omega_{s/n} = \sqrt{4D_{s/n}}$ in 2D and $\omega_{s/n} = \sqrt{2 D_{s/n}q}$ in 3D. Solving \equa{eqn:CG_equation} in the case of $ \omega_s \gg \omega \gg \frac{\omega_s^2}{\omega_n^2} \gamma$ renders the CG mode
\begin{align}
\omega = \sqrt{v_g^2 q^2 - \frac{1}{4}\frac{\omega_s^4}{\omega_n^4} \gamma^2} -i\frac{1}{2}\frac{\omega_s^2}{\omega_n^2} \gamma
\,.
\label{eqn:CG_solution}
\end{align}
At even lower frequency $\omega \ll \frac{\omega_s^2}{\omega_n^2} \gamma$ in 2D, the solution to \equa{eqn:CG_equation} gives the weakly damped plasmons 
\begin{align}
\omega = \sqrt{\omega_s^2-\frac{\omega_n^4}{4\gamma^2}} -i\frac{1}{2}\frac{\omega_n^2}{\gamma}
\,
\label{eqn:CG_plasmon}
\end{align}
where the $v_g q$ contribution has been neglected. Note that the effective Drude weight is $D_s$ in the low frequency regime of the two fluid model \equa{eqn:two_fluid}, similar to the collective mode called Demons in the hydrodynamic regime of the Dirac fluid\cite{Sun2016a,Sun.2018}. The schematic dispersion of the CG mode and plasmons in 2D are depicted in Fig.~\ref{fig:CG}. The damping rate $\frac{\omega_s^2}{2\omega_n^2} \gamma$  of the CG mode is equal to $\frac{\pi}{4}\frac{\Delta^2}{T}$ in the dirty case and $(1-T/T_c)\gamma$ in the clean case. 

\begin{figure}
	\includegraphics[width=0.6 \linewidth]{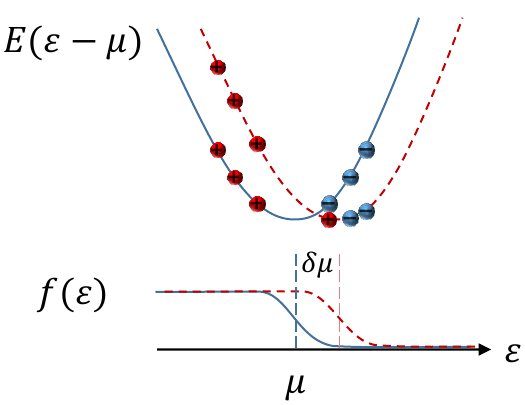}
	\caption{Physical picture of the local chemical potential shift and the quasi particle occupation in the CG mode. The plus and minus signs indicate the signs of the quasi particle charge. This quasi-particle distribution is referred to as `branch imbalance' or `charge mode' in the literature \cite{Tinkham,Pethick1979}.}
	\label{fig:CG_picture}
\end{figure}

Note that the CG mode can be understood as a sound with the standard sound velocity $\sqrt{\frac{n_s}{m}/\chi_s}$ and $\chi_s$ being the superfluid compressibility. The latter is smaller than $\nu$, the compressibility of the whole fluid in the low frequency thermal dynamic limit, because the super and normal fluids move out of phase in this relative high frequency regime. The local accumulation of superfluid causes the local chemical potential to shift up leading to a change of quasiparticle energy and charge. However, the quasiparticle occupation number relaxes too slowly and cannot adjust itself to this change\cite{Pethick1979}, resulting in `branch imbalance' \cite{Tinkham} as shown in Fig.~\ref{fig:CG_picture}. Normal impurities cannot relax this branch imbalance because in $s$-wave superconductors, the elastic scattering matrix element $u_k u_{k^\prime}-v_k v_{k^\prime}$ vanishes between hole like and electron like states at the same energy. Inelastic scattering due to, e.g., phonons, does relax branch imbalance and cause extra damping to the CG mode but we assume it to be small. In $d$-wave superconductors, the same matrix element is non-zero due to anisotropy of the gap which allows normal impurities to relax the branch imbalance and bring extra damping to the CG mode \cite{Artemenko1997}. 

In clean superconductors the CG mode can cross the diffusion line (Fig.~\ref{fig:CG}(b)) before reaching the gap, entering the regime $\omega \ll D_f q^2$ where the normal fluid part of the \equa{eqn:two_fluid} is in the Thomas-Fermi form in \equa{eqn:sigma_n}. The normal fluid still screens the CG mode but with a Thomas-Fermi screening character. The CG mode speed is slightly modified to $v_{CG} = \sqrt{v_g^2 + \frac{n_s}{\nu_n m}}$ in this regime but still remains close to $v_g$ since the second term is much smaller.

The original experiment using Josephson tunneling junctions by Carlson and Goldman\cite{Carlson1975} seems to be the only observation of this novel collective mode.  In the optical conductivity measured by far field optics, the CG mode might move part of the superfluid spectra weight to finite frequency due to smooth disorder\cite{Orenstein2003,Barabash2003}.
At non-zero momentum, being almost charge neutral, the CG mode appears as a very weak feature in the near field reflection coefficient: a one percent crossover of $\mathrm{Abs}[R_p]$ as shown by Fig.~\ref{fig:Monolayer_plasmon_cg}(b) plotted for a typical dirty superconductor close to its $T_c$.

\section{Double layer superconductor}
\label{sec:double_layer}
\begin{figure}
	\includegraphics[width=0.9 \linewidth]{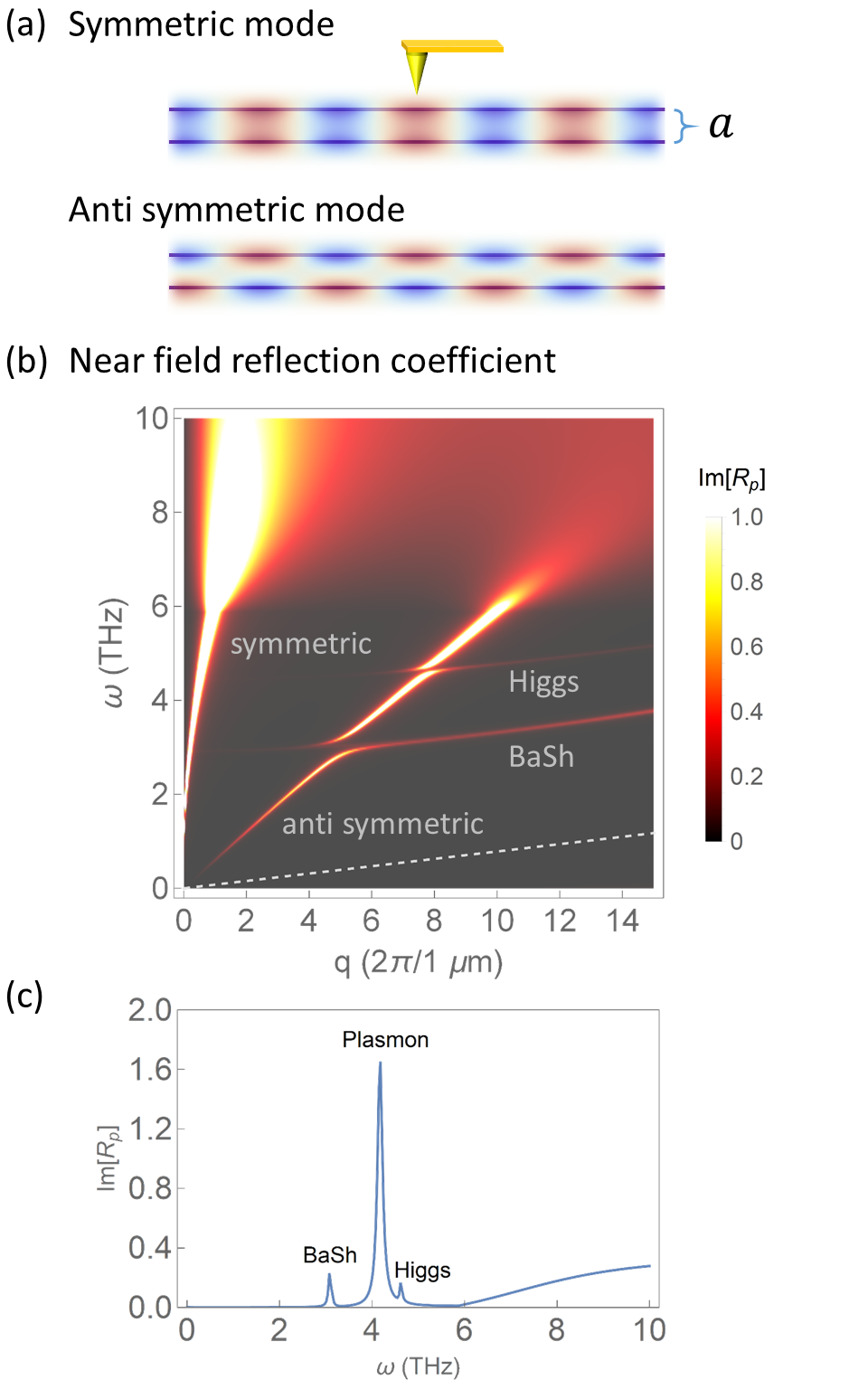}
	\caption{(a) The 2D system consisting of two superconducting layers. Colors represent the electric potential of the symmetric and anti symmetric modes. (b) Near field reflection coefficient of the double superconducting layer system. Josephson coupling is neglected. There is anti-crossing feature of the acoustic plasmon with the Higgs mode and the $d$-wave BaSh mode. Gray dashed line indicates the velocity of the Goldstone mode before coupling to EM. (c) Vertical line cut of (b) at momentum $q=7 \times \unit{2\pi/(1 \mu m)}$.
	The parameters are $k_F = 2\pi/(3 \unit{nm})$, $v_F = 2.5\times 10^5 \unit{m/s} $,  $\gamma= 30 \unit{THz}$, $n_s= 1.9 \times 10^{13} \unit{cm^{-2}}$, $a= 3 \unit{nm}$,  $\Delta= 3.0 \unit{THz}$, $\kappa=0.2$ and $\kappa_{\text{BaSh}}=0.4$. Higgs/BaSh mode frequencies are assumed to be $4.5 \unit{THz}$/$3.0 \unit{THz}$ at zero momentum.}
	\label{fig:double_layer_rp}
\end{figure}

In this section we consider the system made of two superconducting layers separated by a small distance $a$,  as shown in Fig.~\ref{fig:double_layer_rp}.  Each layer has an in plane conductivity described by \equa{eqn:superfluid_conductivity}  at low temperature. Density fluctuations in one plane may screen those in the other.  In the quasi static limit, the 2D plasmon dispersion can be obtained from the following eigenmode condition  
\begin{align}
\left( 1+\frac{2\pi i}{\omega} q \sigma \right)^2 + e^{-2aq} \left( \frac{2\pi q}{\omega} \sigma \right)^2 =0 \,
\label{eqn:double_layer_mode_equation}
\end{align}
which leads to two plasmon branches
\begin{align}
\omega_{\pm} = \sqrt{2Dq (1 \pm e^{-aq})  + v_g^2 q^2}  
\,.
\label{eqn:double_layer_plasmon_dispersion}
\end{align}
The upper branch is the symmetric mode whose dispersion follow the $\omega_{+} \sim \sqrt{q}$ law at small momentum. The lower (anti symmetric) branch is an acoustic mode which has the dispersion 
\begin{align}
\omega_{-} = \sqrt{2Da + v_g^2} \cdot q = v_{-} \cdot q \,
\label{eqn:acoustic_plasmon_dispersion}
\end{align}
for $q \ll 1/a$. This acoustic mode is charge fluctuations of the two layers which are out of phase such that the net charge fluctuation is near zero if looked at far away. In other words, the Coulomb interaction is mutually screened and is modified to the effective short range form $V(q)=2\pi(1-e^{-aq})/q$ that makes the mode acoustic. A nonzero Josephson coupling between the layers would give this mode a small gap equal to the `Josephson plasma frequency' $\omega_J=\sqrt{4\pi ea j_c/\hbar}$ where $j_c$ is the critical interlayer current density. An interlayer DC voltage that induces AC Josephson effect can parametrically generate these acoustic plasmons. This issue will be discussed   in a future publication.  In this section, we neglect the Josephson coupling between the layers which is weak for $a$ substantially larger than atomic scale.

Both modes correspond to non-zero momentum oscillations of the phase of the superconducting order parameter. This acoustic plasmon can be viewed as the Goldstone mode which recovers its acoustic nature because Coulomb interaction is greatly weakened. Its speed still has a large contribution $\sqrt{2Da} \sim \sqrt{\alpha k_F a} v_F$ from the residual Coulomb interaction where $\alpha= e^2/(\hbar v_F)$ is the `fine structure constant'. In BSCCO 2212 at typical doping\cite{Chiao2000}, $\alpha \approx 9$ since $v_F \approx 2.5 \times 10^5 \unit{m/s}$.   For $k_F=2\pi/(10 \unit{nm})$ and $a=3 \unit{nm}$, the ratio between the speeds of this acoustic mode and the original Goldstone mode is $v_{-}/v_g \approx 6$ which means they are at the same order of magnitude. Therefore, an accurate measurement of the acoustic plasmon dispersion would contain the information of the `Goldstone mode' speed.

In order for the acoustic mode to be observable to near field experiments, it should have substantial spectral weight in the the near field reflection coefficient
\begin{align}
R_p(\omega,q) = - \frac{2\pi i q \sigma}{\omega}  \frac{\epsilon + e^{-2aq} \left( 1-\frac{2\pi i}{\omega} q \sigma \right) }
{\epsilon^2  + e^{-2aq} \left( \frac{2\pi q}{\omega} \sigma \right)^2 }  \,
\label{eqn:r_p_double_layer}
\end{align}
derived in Appendix \ref{appendix:rp}.
Given the same amplitude of charge density oscillation in each layer, the electric field generated by the two layers tend to cancel each other since they are opposite in sign. The remaining field is weaker than the symmetric plasmon mode by a factor of $qa/2$ and the near field spectra weight is weaker by $(qa/2)^{3/2}$. Nevertheless, the acoustic mode is still visible as shown by the $R_p$ plotted in Fig.~\ref{fig:double_layer_rp} using the conductivity from \equa{eqn:EM_response_tensor}. 

Moreover, since the acoustic plasmon has higher momentum given the same frequency in the THz range, it has stronger coupling to the Higgs/BaSh modes. Thus there is more prominent anticrossing feature between the acoustic plasmon and Higgs/BaSh modes, as shown in Fig.~\ref{fig:double_layer_rp}. Note that \equa{eqn:higgs_splitting} and \equa{eqn:bs_splitting} apply to anti-crossings with both the symmetric and anti-symmetric modes. For example, the anti crossing of the BaSh mode with the acoustic plasmon happens at a momentum roughly $20$ times that with the symmetric plasmon, rendering the energy splitting $20$ times larger than the latter.

\section{Bulk layered superconductors}
\label{sec:bulk_layered}
In layered superconductors such as high T$_c$ cuprates, there is Josephson coupling between the layers and  the low temperature and subgap collective modes are the Josephson plasmons \cite{Basov2005a}. Considering only the phase degree of freedom, the Lagrangian for an evenly spaced layered superconductor is
\begin{align}
L = \int dr \sum_{n} 
\bigg[& -\frac{1}{2}\nu (\partial_t \theta_n + \phi_n)^2 + \frac{n_s}{2m} (\nabla \theta_n - \mathbf{A}_n)^2 
\notag\\
&-E_c \cos\left(\theta_{n+1} - \theta_{n} - \int_{n+1}^{n} A dz \right)
\bigg]
\,
\label{eqn:action_layered}
\end{align}
where $\theta_n(r)$, $\phi_n(r)$ and $A_n(r)$ are the phase, scalar and vector potentials on the nth layer and $E_c$ is the Josephson coupling energy per unit area and we have set $e=1$. For longitudinal fields we are interested in, we can choose the gauge where $A_n(r)= 0$. Due to continuous translational symmetry in plane and discrete one in z direction,  it is convenient to Fourier transform the fields into the `Bloch' form
\begin{align}
\theta_{n}(r) = \sum_{k_z, q} \theta_{k_z,q} e^{i(qr + k_z n a)}
\,
\label{eqn:theta_bloch}
\end{align}
where $a$ is the layer spacing, $q$ is the in plane momentum and $k_z \in (-\pi/a,\,\pi/a)$ is the lattice momentum in $z$ direction. The Lagrangian \equa{eqn:action_layered} is diagonalized as 
\begin{align}
L = \sum_{k_z, q}
\left[ -\frac{1}{2}\nu (\partial_t \theta_n + \phi_n)_q^2 + \left(\frac{n_s}{2m} q^2 
+E_c (1-\cos(ak_z)) \right) \theta_q^2 
\right]
\,. 
\label{eqn:action_layered_diag}
\end{align}
Solving the Euler-Lagrange equation of the phase and making use of the expression of the charge density $\rho = \nu (\partial_t \theta +\phi)$, we obtain the `nonlocal' polarization function
\begin{align}
\chi_{\rho \rho}(k_z, q) = \frac{\frac{n_s}{m} q^2 + 2 E_c \left(1- \cos(ak_z) \right)}
{\omega^2 - v_g^2 q^2 - \frac{1}{\nu} 2 E_c\left(1- \cos(ak_z) \right) }
\,. 
\label{eqn:polarization_layered_sc}
\end{align}
The Coulomb potential kernel is modified to
\begin{align}
V(k_z,q)= \frac{2\pi e^2}{q} \frac{\sinh(aq)}{\cosh(aq) -\cos(ak_z)}
\,. 
\label{eqn:coulomb_kernel}
\end{align}
The zeros of the dielectric function $\epsilon = 1- V(k_z, q)\chi_{\rho \rho}$ gives the dispersion of the collective modes 
\begin{align}
\omega^2 = \left(1/\nu + V(k_z,q) \right)
\left(\frac{n_s}{m}q^2 + 2 E_c\left(1- \cos(ak_z) \right) \right)
\,. 
\label{eqn:layered_modes_discrete}
\end{align}
In the long wave length limit $q,k_z \ll 1/a$, the Coulomb kernel reduces to that of the continuous limit and the mode dispersion simplifies to
\begin{align}
\omega = \sqrt{\omega_p^2 \frac{q^2}{q^2 + k_z^2} +  \omega_J^2 \frac{k_z^2}{q^2 + k_z^2} + v_g^2 q^2 +  v_z^2 k_z^2 
}
\,,
\label{eqn:layered_modes_continuous}
\end{align}
where $\omega_J = \sqrt{4\pi E_c a^2} $ is the Josephson plasma frequency, $\omega_p = \sqrt{4\pi n_s/m}$ is the in plane plasma frequency, $v_g=v_F/\sqrt{2}$ is the in plane Goldstone mode speed in the clean limit and $v_z=\frac{\omega_J}{\omega_p} v_g$ is the z axis Goldstone mode speed. These are the hyperbolic Josephson plasmons (HJP) extensively studied in the literature\cite{Pokrovsky1996,Stinson2014a,Zhou2014,Sun.2015,Basov2016} which can be viewed as mixtures of out-of-plane and in-plane plasmons.
Indeed, \equa{eqn:layered_modes_continuous} could be derived directly from the zeros of the continuous limit of the nonlocal dielectric function
\begin{align}
\epsilon(k_z,q) = 1- \frac{ \omega_p^2 \frac{q^2}{q^2 + k_z^2} +  \omega_J^2 \frac{k_z^2}{q^2 + k_z^2}}
{\omega^2 - v_g^2 q^2 - v_z^2 k_z^2 } 
\,
\label{eqn:sigma_layered_continuous}
\end{align}
which is defined as the external electrical potential divided by the total potential.

Alternatively, the long wavelength response can be described by the anisotropic dielectric function
\begin{align}
\epsilon_x(\omega,q,k_z) &= 1- \frac{ \omega_p^2 }
{\omega^2 - v_g^2 q^2 - v_z^2 k_z^2 } 
\,, \notag\\
\epsilon_z(\omega,q,k_z) &= 1- \frac{ \omega_j^2 }
{\omega^2 - v_g^2 q^2 - v_z^2 k_z^2 } 
\,
\label{eqn:dielectric_anisotropic}
\end{align}
and the collective mode dispersion is determined by $q^2 \epsilon_x + k_z^2 \epsilon_z =0$. This formalism is more convenient for calculating the reflection coefficient of a slab. To include the effect of Higgs and BaSh modes, one just needs to modify the in plane response $\epsilon_x$ in similar fashions as Eqs.~\eqref{eqn:sigma_higgs} and \eqref{eqn:sigma_bs}.

\begin{figure}
	\includegraphics[width=0.9 \linewidth]{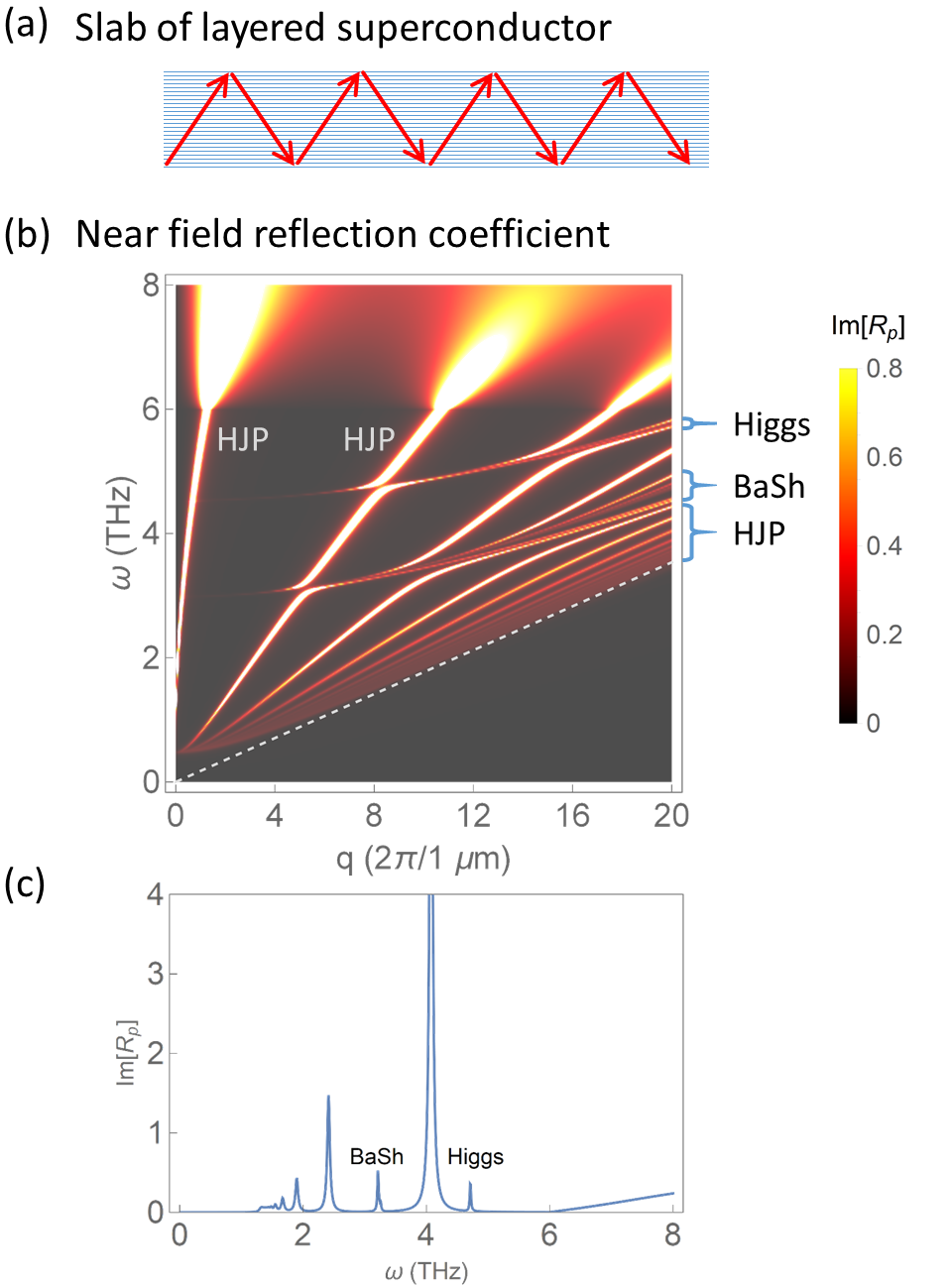}
	\caption{(a) Illustration of a slab made of layered superconductor and the propagating Josephson plasmons inside. (b) Near field reflection coefficient of a $10 \unit{nm}$ thick slab. Bright lines are due to the hyperbolic Josephson plasmons which anti cross with the Higgs modes and $d$-wave BaSh modes. Gray dashed line indicates the velocity $v_g$ of the in plane Goldstone mode. (c) Vertical line cut of the above at momentum $q=7 \times \unit{2\pi/(1 \mu m)}$. The effective in plane `plasma' frequency is $\omega_p= 30 \unit{THz}$ and the Josephson plasma frequency is $\omega_J= 0.5 \unit{THz}$. The gap is $\Delta= 3 \unit{THz}$, $\omega_{Higgs}=4.5 \unit{THz}$, $\omega_{\text{BaSh}}=3.0 \unit{THz}$, $\kappa=0.2$ and $\kappa_{\text{BaSh}}=0.2$. }
	\label{fig:layered_josephson_plasmon}
\end{figure}

For a superconducting slab with thickness $d$ in the continuous limit, the near field reflection coefficient is
\begin{align}
R_{\text{slab}} = \frac{R_p(1-e^{2ik_z d})}{1-e^{2ik_z d} R_p^2}
,\quad
R_{p} = \frac{iq-k_z \epsilon_z(\omega,q,k_z)}{iq+k_z \epsilon_z(\omega,q,k_z)}
\,
\label{eqn:r_p_3D}
\end{align}
where $R_p$ is the reflection coefficient of an infinitely thick sample and $k_z$ is the z component of the EM wave momentum inside the slab. See Appendix~\ref{appendix:rp} for the derivation.
A typical reflection coefficient is shown in Fig.~\ref{fig:layered_josephson_plasmon}, taking into account the nonlocal corrections to the dielectric function due to the Goldstone mode. Note that due to high anisotropy of the EM response, the z direction wavelength $\lambda_z \sim \lambda \omega/\omega_p$ can easily get comparable to the layer spacing where $\lambda$ is the in plane wave length. In that case, the full form \equa{eqn:layered_modes_discrete} should be used as the bulk mode dispersion and the number of hyperbolic plasmon branches is limited by the number of layers $N$. Due to Josephson coupling between the layers, the transfer matrix method does not apply and numerical diagonalization of a set of $N$ coupled linear equations will be needed to calculate the near field reflection coefficient.

\section{Discussion}
\label{sec:discussion}
We studied the non local EM response properties of superconductors which are of great importance to the emerging field of THz near field experiments. With analytical formulas for the non local optical conductivity and plots of reflection coefficients, we have demonstrated that for monolayer or multilayer quasi two dimensional superconductors essentially all of the interesting collective modes (plasmons, hyperbolic interlayer or Josephson plasmons, the Carlson-Goldman mode, the amplitude (Higgs) mode and the Bardasis-Schrieffer mode) couple linearly to the THz EM fields produced by near field probes. As old arguments of Anderson show, the dispersion of the plasmon ($\sqrt{q}$) is essentially unaffected by superconductivity but the gap substantially suppresses the loss at low frequencies. Fig.~\ref{fig:Monolayer_plasmon_cg}(a) shows the plasmon dispersion expected for a monolayer superconductor. In superconducting bilayers, an additional acoustic ($\omega \propto q$) plasmon (phase) mode exists and is also easily observable in near field experiments (Fig.~\ref{fig:double_layer_rp}(b)).
As the temperature becomes close to T$_c$, as shown by Fig.~\ref{fig:Monolayer_plasmon_cg}(b), the Carlson-Goldman mode appears but as a very weak feature across the resonance since it has almost no net charge density fluctuation. Note that this mode is not enhanced in multilayer systems.
The amplitude (Higgs) mode appears in the EM response because it couples to the phase fluctuation with a matrix element that is non-zero if there is no perfect particle hole symmetry (\equa{eqn:higgs_coupling_C0}). The ultimate coupling to THz near field is proportional to the square of the near field momentum $q$ (\equa{eqn:sigma_higgs}), and is strongly enhanced by an anticrossing with the plasmon or phase modes. The Higgs mode is only weakly visible for monolayer materials because the $\sqrt{q}$ plasmon dispersion means that the anticrossing occurs at a very small momentum (Fig.~\ref{fig:Monolayer_plasmon_cg}(a)). The feature is much more easily visible in bilayer systems as an anti crossing with the acoustic plasmon (phase) mode (Fig.~\ref{fig:double_layer_rp}(b)). Note that the Higgs mode does appear in nonlinear far field optics \cite{Krull.2014, Matsunaga.2013,Katsumi2018,Nakamura2019,Cui.2019,Shimano2020}.
The coupling to Bardasis-Schrieffer (subdominant order parameter) mode is very similar to that of the Higgs mode, except that it does not require particle hole symmetry breaking. It is again most easily visible as a large $q$ anti-crossing with the phase (Fig.~\ref{fig:double_layer_rp}(b)) or plasmon mode (Fig.~\ref{fig:Monolayer_plasmon_cg}(a)). Note that an analogy of the BaSh mode in excitonic insulators couples linearly to photons already at zero momentum, developing into BaSh polaritons\cite{sun2020BS}.

In multilayer superconductors, a multiplicity of phase modes exist, coined the hyperbolic Josephson plasmons (Fig.~\ref{fig:layered_josephson_plasmon}). The plasmon dispersion is hyperbolic ($\epsilon <0$ for in plane and $\epsilon >0$ for out of plane), leading to total internal reflection (Fig.~\ref{fig:layered_josephson_plasmon}(a)) and many plasmon branches with Higgs and BaSh modes visible as anti-crossings. The multiplayer nature means there are multiple branches of Higgs modes/BaSh modes, but they are weakly separated and may be difficult to resolve.

On the experimental side, detection of the collective modes offers useful information about both the ground state and the low lying excited states. On the theory side, knowledge of how to excite the collective modes are often the first step towards understanding non equilibrium dynamics \cite{sun2019transient,Muller.2019}. From the technological point of view, the low loss plasmonic modes are promising as information carriers in superconductor wave guides. The multiplayer systems described in Sections~\ref{sec:double_layer} and \ref{sec:bulk_layered} can be viewed as a kind of naturally occuring photonic cavities which enhance light matter coupling. 

The formalism presented here is for $s$-wave superconductors. For $d$-wave superconductors, the qualitative features of the EM response such as the two fluid model \equa{eqn:two_fluid} and all the collective modes \cite{Wu.1995} should be the same. Nevertheless, the CG mode might exist down to much lower temperature because of the large proportion of the normal fluid \cite{Ohashi2000} although it might be heavily damped by normal disorder\cite{Artemenko1997}. Due to the nodes in the $d$-wave gap, the THz plasmons might experience substantial damping even at zero temperature.  The effect of disorder is not explicitly taken into account and would be a useful extension of the present research, e.g., disorder assisted Cherenkov radiation of plasmons by quasiparticles. It is also of interest to study the coupling of photons to the Leggett mode \cite{Leggett.1966,Giorgianni.2019} at nonzero momentum.

\begin{acknowledgements}
We acknowledge support from  the Department of Energy under Grant DE-SC0018218. We thank W. Yang, R. Jing, Y. Shao, G. Ni and Y. He for helpful discussions.
\end{acknowledgements}

\bibliographystyle{apsrev4-1}
\bibliography{Superconductor_collective_modes}

\begin{thebibliography}{78}%
\makeatletter
\providecommand \@ifxundefined [1]{%
 \@ifx{#1\undefined}
}%
\providecommand \@ifnum [1]{%
 \ifnum #1\expandafter \@firstoftwo
 \else \expandafter \@secondoftwo
 \fi
}%
\providecommand \@ifx [1]{%
 \ifx #1\expandafter \@firstoftwo
 \else \expandafter \@secondoftwo
 \fi
}%
\providecommand \natexlab [1]{#1}%
\providecommand \enquote  [1]{``#1''}%
\providecommand \bibnamefont  [1]{#1}%
\providecommand \bibfnamefont [1]{#1}%
\providecommand \citenamefont [1]{#1}%
\providecommand \href@noop [0]{\@secondoftwo}%
\providecommand \href [0]{\begingroup \@sanitize@url \@href}%
\providecommand \@href[1]{\@@startlink{#1}\@@href}%
\providecommand \@@href[1]{\endgroup#1\@@endlink}%
\providecommand \@sanitize@url [0]{\catcode `\\12\catcode `\$12\catcode
  `\&12\catcode `\#12\catcode `\^12\catcode `\_12\catcode `\%12\relax}%
\providecommand \@@startlink[1]{}%
\providecommand \@@endlink[0]{}%
\providecommand \url  [0]{\begingroup\@sanitize@url \@url }%
\providecommand \@url [1]{\endgroup\@href {#1}{\urlprefix }}%
\providecommand \urlprefix  [0]{URL }%
\providecommand \Eprint [0]{\href }%
\providecommand \doibase [0]{http://dx.doi.org/}%
\providecommand \selectlanguage [0]{\@gobble}%
\providecommand \bibinfo  [0]{\@secondoftwo}%
\providecommand \bibfield  [0]{\@secondoftwo}%
\providecommand \translation [1]{[#1]}%
\providecommand \BibitemOpen [0]{}%
\providecommand \bibitemStop [0]{}%
\providecommand \bibitemNoStop [0]{.\EOS\space}%
\providecommand \EOS [0]{\spacefactor3000\relax}%
\providecommand \BibitemShut  [1]{\csname bibitem#1\endcsname}%
\let\auto@bib@innerbib\@empty
\bibitem [{\citenamefont {Mattis}\ and\ \citenamefont
  {Bardeen}(1958)}]{Mattis1958}%
  \BibitemOpen
  \bibfield  {author} {\bibinfo {author} {\bibfnamefont {D.~C.}\ \bibnamefont
  {Mattis}}\ and\ \bibinfo {author} {\bibfnamefont {J.}~\bibnamefont
  {Bardeen}},\ }\href {\doibase 10.1103/PhysRev.111.412} {\bibfield  {journal}
  {\bibinfo  {journal} {Phys. Rev.}\ }\textbf {\bibinfo {volume} {111}},\
  \bibinfo {pages} {412} (\bibinfo {year} {1958})}\BibitemShut {NoStop}%
\bibitem [{\citenamefont {Anderson}(1958{\natexlab{a}})}]{Anderson1958b}%
  \BibitemOpen
  \bibfield  {author} {\bibinfo {author} {\bibfnamefont {P.~W.}\ \bibnamefont
  {Anderson}},\ }\href {\doibase 10.1103/PhysRev.112.1900} {\bibfield
  {journal} {\bibinfo  {journal} {Phys. Rev.}\ }\textbf {\bibinfo {volume}
  {112}},\ \bibinfo {pages} {1900} (\bibinfo {year}
  {1958}{\natexlab{a}})}\BibitemShut {NoStop}%
\bibitem [{\citenamefont {Zimmermann}\ \emph {et~al.}(1991)\citenamefont
  {Zimmermann}, \citenamefont {Brandt}, \citenamefont {Bauer}, \citenamefont
  {Seider},\ and\ \citenamefont {Genzel}}]{Zimmermann1991}%
  \BibitemOpen
  \bibfield  {author} {\bibinfo {author} {\bibfnamefont {W.}~\bibnamefont
  {Zimmermann}}, \bibinfo {author} {\bibfnamefont {E.~H.}\ \bibnamefont
  {Brandt}}, \bibinfo {author} {\bibfnamefont {M.}~\bibnamefont {Bauer}},
  \bibinfo {author} {\bibfnamefont {E.}~\bibnamefont {Seider}}, \ and\ \bibinfo
  {author} {\bibfnamefont {L.}~\bibnamefont {Genzel}},\ }\href {\doibase
  10.1016/0921-4534(91)90771-P} {\bibfield  {journal} {\bibinfo  {journal}
  {Physica C}\ }\textbf {\bibinfo {volume} {183}},\ \bibinfo {pages} {99}
  (\bibinfo {year} {1991})}\BibitemShut {NoStop}%
\bibitem [{\citenamefont {Tinkham}(1974)}]{Tinkham.1974}%
  \BibitemOpen
  \bibfield  {author} {\bibinfo {author} {\bibfnamefont {M.}~\bibnamefont
  {Tinkham}},\ }\href {\doibase 10.1103/RevModPhys.46.587} {\bibfield
  {journal} {\bibinfo  {journal} {Rev. Mod. Phys.}\ }\textbf {\bibinfo {volume}
  {46}},\ \bibinfo {pages} {587} (\bibinfo {year} {1974})}\BibitemShut
  {NoStop}%
\bibitem [{\citenamefont {Basov}\ and\ \citenamefont
  {Timusk}(2005)}]{Basov2005a}%
  \BibitemOpen
  \bibfield  {author} {\bibinfo {author} {\bibfnamefont {D.~N.}\ \bibnamefont
  {Basov}}\ and\ \bibinfo {author} {\bibfnamefont {T.}~\bibnamefont {Timusk}},\
  }\href {\doibase 10.1103/RevModPhys.77.721} {\bibfield  {journal} {\bibinfo
  {journal} {Rev. Mod. Phys.}\ }\textbf {\bibinfo {volume} {77}},\ \bibinfo
  {pages} {721} (\bibinfo {year} {2005})}\BibitemShut {NoStop}%
\bibitem [{\citenamefont {Parks}(1969)}]{Parks1969}%
  \BibitemOpen
  \bibfield  {author} {\bibinfo {author} {\bibfnamefont {R.~D.}\ \bibnamefont
  {Parks}},\ }\href {\doibase https://doi.org/10.1201/9780203737965} {\emph
  {\bibinfo {title} {{Superconductivity}}}}\ (\bibinfo  {publisher} {CRC
  Press},\ \bibinfo {year} {1969})\ Chap.~\bibinfo {chapter} {7}\BibitemShut
  {NoStop}%
\bibitem [{\citenamefont {Damascelli}\ \emph {et~al.}(2003)\citenamefont
  {Damascelli}, \citenamefont {Hussain},\ and\ \citenamefont
  {Shen}}]{Damascelli.2003}%
  \BibitemOpen
  \bibfield  {author} {\bibinfo {author} {\bibfnamefont {A.}~\bibnamefont
  {Damascelli}}, \bibinfo {author} {\bibfnamefont {Z.}~\bibnamefont {Hussain}},
  \ and\ \bibinfo {author} {\bibfnamefont {Z.-X.}\ \bibnamefont {Shen}},\
  }\href {\doibase 10.1103/RevModPhys.75.473} {\bibfield  {journal} {\bibinfo
  {journal} {Rev. Mod. Phys.}\ }\textbf {\bibinfo {volume} {75}},\ \bibinfo
  {pages} {473} (\bibinfo {year} {2003})}\BibitemShut {NoStop}%
\bibitem [{\citenamefont {Stewart}(2011)}]{Stewart.2011}%
  \BibitemOpen
  \bibfield  {author} {\bibinfo {author} {\bibfnamefont {G.~R.}\ \bibnamefont
  {Stewart}},\ }\href {\doibase 10.1103/RevModPhys.83.1589} {\bibfield
  {journal} {\bibinfo  {journal} {Rev. Mod. Phys.}\ }\textbf {\bibinfo {volume}
  {83}},\ \bibinfo {pages} {1589} (\bibinfo {year} {2011})}\BibitemShut
  {NoStop}%
\bibitem [{\citenamefont {Sooryakumar}\ and\ \citenamefont
  {Klein}(1980)}]{Sooryakumar1980}%
  \BibitemOpen
  \bibfield  {author} {\bibinfo {author} {\bibfnamefont {R.}~\bibnamefont
  {Sooryakumar}}\ and\ \bibinfo {author} {\bibfnamefont {M.~V.}\ \bibnamefont
  {Klein}},\ }\href {\doibase 10.1103/PhysRevLett.45.660} {\bibfield  {journal}
  {\bibinfo  {journal} {Phys. Rev. Lett.}\ }\textbf {\bibinfo {volume} {45}},\
  \bibinfo {pages} {660} (\bibinfo {year} {1980})}\BibitemShut {NoStop}%
\bibitem [{\citenamefont {Buzea}\ and\ \citenamefont
  {Yamashita}(2001)}]{Buzea_2001}%
  \BibitemOpen
  \bibfield  {author} {\bibinfo {author} {\bibfnamefont {C.}~\bibnamefont
  {Buzea}}\ and\ \bibinfo {author} {\bibfnamefont {T.}~\bibnamefont
  {Yamashita}},\ }\href {\doibase 10.1088/0953-2048/14/11/201} {\bibfield
  {journal} {\bibinfo  {journal} {Superconductor Science and Technology}\
  }\textbf {\bibinfo {volume} {14}},\ \bibinfo {pages} {R115} (\bibinfo {year}
  {2001})}\BibitemShut {NoStop}%
\bibitem [{\citenamefont {Ni}\ \emph {et~al.}(2018)\citenamefont {Ni},
  \citenamefont {McLeod}, \citenamefont {Sun}, \citenamefont {Wang},
  \citenamefont {Xiong}, \citenamefont {Post}, \citenamefont {Sunku},
  \citenamefont {Jiang}, \citenamefont {Hone}, \citenamefont {Dean},
  \citenamefont {Fogler},\ and\ \citenamefont {Basov}}]{Ni2018}%
  \BibitemOpen
  \bibfield  {author} {\bibinfo {author} {\bibfnamefont {G.~X.}\ \bibnamefont
  {Ni}}, \bibinfo {author} {\bibfnamefont {A.~S.}\ \bibnamefont {McLeod}},
  \bibinfo {author} {\bibfnamefont {Z.}~\bibnamefont {Sun}}, \bibinfo {author}
  {\bibfnamefont {L.}~\bibnamefont {Wang}}, \bibinfo {author} {\bibfnamefont
  {L.}~\bibnamefont {Xiong}}, \bibinfo {author} {\bibfnamefont {K.~W.}\
  \bibnamefont {Post}}, \bibinfo {author} {\bibfnamefont {S.~S.}\ \bibnamefont
  {Sunku}}, \bibinfo {author} {\bibfnamefont {B.-Y.}\ \bibnamefont {Jiang}},
  \bibinfo {author} {\bibfnamefont {J.}~\bibnamefont {Hone}}, \bibinfo {author}
  {\bibfnamefont {C.~R.}\ \bibnamefont {Dean}}, \bibinfo {author}
  {\bibfnamefont {M.~M.}\ \bibnamefont {Fogler}}, \ and\ \bibinfo {author}
  {\bibfnamefont {D.~N.}\ \bibnamefont {Basov}},\ }\href {\doibase
  10.1038/s41586-018-0136-9} {\bibfield  {journal} {\bibinfo  {journal}
  {Nature}\ }\textbf {\bibinfo {volume} {557}},\ \bibinfo {pages} {530}
  (\bibinfo {year} {2018})}\BibitemShut {NoStop}%
\bibitem [{\citenamefont {Basov}\ \emph {et~al.}(2014)\citenamefont {Basov},
  \citenamefont {Fogler}, \citenamefont {Lanzara}, \citenamefont {Wang},\ and\
  \citenamefont {Zhang}}]{Basov2014a}%
  \BibitemOpen
  \bibfield  {author} {\bibinfo {author} {\bibfnamefont {D.~N.}\ \bibnamefont
  {Basov}}, \bibinfo {author} {\bibfnamefont {M.~M.}\ \bibnamefont {Fogler}},
  \bibinfo {author} {\bibfnamefont {A.}~\bibnamefont {Lanzara}}, \bibinfo
  {author} {\bibfnamefont {F.}~\bibnamefont {Wang}}, \ and\ \bibinfo {author}
  {\bibfnamefont {Y.}~\bibnamefont {Zhang}},\ }\href {\doibase
  10.1103/RevModPhys.86.959} {\bibfield  {journal} {\bibinfo  {journal} {Rev.
  Mod. Phys.}\ }\textbf {\bibinfo {volume} {86}},\ \bibinfo {pages} {959}
  (\bibinfo {year} {2014})}\BibitemShut {NoStop}%
\bibitem [{\citenamefont {Lundeberg}\ \emph {et~al.}(2017)\citenamefont
  {Lundeberg}, \citenamefont {Gao}, \citenamefont {Asgari}, \citenamefont
  {Tan}, \citenamefont {{Van Duppen}}, \citenamefont {Autore}, \citenamefont
  {Alonso-Gonz{\'{a}}lez}, \citenamefont {Woessner}, \citenamefont {Watanabe},
  \citenamefont {Taniguchi}, \citenamefont {Hillenbrand}, \citenamefont {Hone},
  \citenamefont {Polini},\ and\ \citenamefont {Koppens}}]{Lundeberg2017}%
  \BibitemOpen
  \bibfield  {author} {\bibinfo {author} {\bibfnamefont {M.~B.}\ \bibnamefont
  {Lundeberg}}, \bibinfo {author} {\bibfnamefont {Y.}~\bibnamefont {Gao}},
  \bibinfo {author} {\bibfnamefont {R.}~\bibnamefont {Asgari}}, \bibinfo
  {author} {\bibfnamefont {C.}~\bibnamefont {Tan}}, \bibinfo {author}
  {\bibfnamefont {B.}~\bibnamefont {{Van Duppen}}}, \bibinfo {author}
  {\bibfnamefont {M.}~\bibnamefont {Autore}}, \bibinfo {author} {\bibfnamefont
  {P.}~\bibnamefont {Alonso-Gonz{\'{a}}lez}}, \bibinfo {author} {\bibfnamefont
  {A.}~\bibnamefont {Woessner}}, \bibinfo {author} {\bibfnamefont
  {K.}~\bibnamefont {Watanabe}}, \bibinfo {author} {\bibfnamefont
  {T.}~\bibnamefont {Taniguchi}}, \bibinfo {author} {\bibfnamefont
  {R.}~\bibnamefont {Hillenbrand}}, \bibinfo {author} {\bibfnamefont
  {J.}~\bibnamefont {Hone}}, \bibinfo {author} {\bibfnamefont {M.}~\bibnamefont
  {Polini}}, \ and\ \bibinfo {author} {\bibfnamefont {F.~H.~L.}\ \bibnamefont
  {Koppens}},\ }\href {\doibase 10.1126/science.aan2735} {\bibfield  {journal}
  {\bibinfo  {journal} {Science}\ }\textbf {\bibinfo {volume} {357}},\ \bibinfo
  {pages} {187} (\bibinfo {year} {2017})}\BibitemShut {NoStop}%
\bibitem [{\citenamefont {Dias}\ \emph {et~al.}(2018)\citenamefont {Dias},
  \citenamefont {Iranzo}, \citenamefont {Gon{\c{c}}alves}, \citenamefont
  {Hajati}, \citenamefont {Bludov}, \citenamefont {Jauho}, \citenamefont
  {Mortensen}, \citenamefont {Koppens},\ and\ \citenamefont
  {Peres}}]{Dias2018}%
  \BibitemOpen
  \bibfield  {author} {\bibinfo {author} {\bibfnamefont {E.~J.~C.}\
  \bibnamefont {Dias}}, \bibinfo {author} {\bibfnamefont {D.~A.}\ \bibnamefont
  {Iranzo}}, \bibinfo {author} {\bibfnamefont {P.~A.~D.}\ \bibnamefont
  {Gon{\c{c}}alves}}, \bibinfo {author} {\bibfnamefont {Y.}~\bibnamefont
  {Hajati}}, \bibinfo {author} {\bibfnamefont {Y.~V.}\ \bibnamefont {Bludov}},
  \bibinfo {author} {\bibfnamefont {A.-P.}\ \bibnamefont {Jauho}}, \bibinfo
  {author} {\bibfnamefont {N.~A.}\ \bibnamefont {Mortensen}}, \bibinfo {author}
  {\bibfnamefont {F.~H.~L.}\ \bibnamefont {Koppens}}, \ and\ \bibinfo {author}
  {\bibfnamefont {N.~M.~R.}\ \bibnamefont {Peres}},\ }\href {\doibase
  10.1103/PhysRevB.97.245405} {\bibfield  {journal} {\bibinfo  {journal} {Phys.
  Rev. B}\ }\textbf {\bibinfo {volume} {97}},\ \bibinfo {pages} {245405}
  (\bibinfo {year} {2018})}\BibitemShut {NoStop}%
\bibitem [{\citenamefont {Wang}\ \emph {et~al.}(2020)\citenamefont {Wang},
  \citenamefont {Zhao}, \citenamefont {Shi}, \citenamefont {Wu}, \citenamefont
  {Zhao}, \citenamefont {Jiang}, \citenamefont {Watanabe}, \citenamefont
  {Taniguchi}, \citenamefont {Zettl}, \citenamefont {Zhou},\ and\ \citenamefont
  {Wang}}]{Wang.2020}%
  \BibitemOpen
  \bibfield  {author} {\bibinfo {author} {\bibfnamefont {S.}~\bibnamefont
  {Wang}}, \bibinfo {author} {\bibfnamefont {S.}~\bibnamefont {Zhao}}, \bibinfo
  {author} {\bibfnamefont {Z.}~\bibnamefont {Shi}}, \bibinfo {author}
  {\bibfnamefont {F.}~\bibnamefont {Wu}}, \bibinfo {author} {\bibfnamefont
  {Z.}~\bibnamefont {Zhao}}, \bibinfo {author} {\bibfnamefont {L.}~\bibnamefont
  {Jiang}}, \bibinfo {author} {\bibfnamefont {K.}~\bibnamefont {Watanabe}},
  \bibinfo {author} {\bibfnamefont {T.}~\bibnamefont {Taniguchi}}, \bibinfo
  {author} {\bibfnamefont {A.}~\bibnamefont {Zettl}}, \bibinfo {author}
  {\bibfnamefont {C.}~\bibnamefont {Zhou}}, \ and\ \bibinfo {author}
  {\bibfnamefont {F.}~\bibnamefont {Wang}},\ }\href
  {https://doi.org/10.1038/s41563-020-0652-5} {\bibfield  {journal} {\bibinfo
  {journal} {Nature Materials}\ ,\ } (\bibinfo {year} {2020})}\BibitemShut
  {NoStop}%
\bibitem [{\citenamefont {Anderson}(1958{\natexlab{b}})}]{Anderson1958a}%
  \BibitemOpen
  \bibfield  {author} {\bibinfo {author} {\bibfnamefont {P.~W.}\ \bibnamefont
  {Anderson}},\ }\href {\doibase 10.1103/PhysRev.110.827} {\bibfield  {journal}
  {\bibinfo  {journal} {Phys. Rev.}\ }\textbf {\bibinfo {volume} {110}},\
  \bibinfo {pages} {827} (\bibinfo {year} {1958}{\natexlab{b}})}\BibitemShut
  {NoStop}%
\bibitem [{\citenamefont {Schrieffer}(1999)}]{Schrieffer1999}%
  \BibitemOpen
  \bibfield  {author} {\bibinfo {author} {\bibfnamefont {J.~R. J.~R.}\
  \bibnamefont {Schrieffer}},\ }\href@noop {} {\emph {\bibinfo {title} {{Theory
  of superconductivity}}}}\ (\bibinfo  {publisher} {Advanced Book Program,
  Perseus Books},\ \bibinfo {year} {1999})\ p.\ \bibinfo {pages}
  {332}\BibitemShut {NoStop}%
\bibitem [{\citenamefont {Arseev}\ \emph {et~al.}(2006)\citenamefont {Arseev},
  \citenamefont {Loiko},\ and\ \citenamefont {Fedorov}}]{Arseev2006}%
  \BibitemOpen
  \bibfield  {author} {\bibinfo {author} {\bibfnamefont {P.~I.}\ \bibnamefont
  {Arseev}}, \bibinfo {author} {\bibfnamefont {S.~O.}\ \bibnamefont {Loiko}}, \
  and\ \bibinfo {author} {\bibfnamefont {N.~K.}\ \bibnamefont {Fedorov}},\
  }\href {\doibase 10.1070/PU2006v049n01ABEH002577} {\bibfield  {journal}
  {\bibinfo  {journal} {Physics-Uspekhi}\ }\textbf {\bibinfo {volume} {49}},\
  \bibinfo {pages} {1} (\bibinfo {year} {2006})}\BibitemShut {NoStop}%
\bibitem [{\citenamefont {Yang}\ and\ \citenamefont {Wu}(2019)}]{Yang2018}%
  \BibitemOpen
  \bibfield  {author} {\bibinfo {author} {\bibfnamefont {F.}~\bibnamefont
  {Yang}}\ and\ \bibinfo {author} {\bibfnamefont {M.~W.}\ \bibnamefont {Wu}},\
  }\href {\doibase 10.1103/PhysRevB.100.104513} {\bibfield  {journal} {\bibinfo
   {journal} {Phys. Rev. B}\ }\textbf {\bibinfo {volume} {100}},\ \bibinfo
  {pages} {104513} (\bibinfo {year} {2019})}\BibitemShut {NoStop}%
\bibitem [{\citenamefont {Bogoljubov}\ \emph {et~al.}(1958)\citenamefont
  {Bogoljubov}, \citenamefont {Tolmachov},\ and\ \citenamefont
  {{\v{S}}irkov}}]{Bogoljubov1958}%
  \BibitemOpen
  \bibfield  {author} {\bibinfo {author} {\bibfnamefont {N.~N.}\ \bibnamefont
  {Bogoljubov}}, \bibinfo {author} {\bibfnamefont {V.~V.}\ \bibnamefont
  {Tolmachov}}, \ and\ \bibinfo {author} {\bibfnamefont {D.~V.}\ \bibnamefont
  {{\v{S}}irkov}},\ }\href {\doibase 10.1002/prop.19580061102} {\bibfield
  {journal} {\bibinfo  {journal} {Fortschritte der Phys.}\ }\textbf {\bibinfo
  {volume} {6}},\ \bibinfo {pages} {605} (\bibinfo {year} {1958})}\BibitemShut
  {NoStop}%
\bibitem [{\citenamefont {Anderson}(1963)}]{Anderson1963}%
  \BibitemOpen
  \bibfield  {author} {\bibinfo {author} {\bibfnamefont {P.~W.}\ \bibnamefont
  {Anderson}},\ }\href {\doibase 10.1103/PhysRev.130.439} {\bibfield  {journal}
  {\bibinfo  {journal} {Phys. Rev.}\ }\textbf {\bibinfo {volume} {130}},\
  \bibinfo {pages} {439} (\bibinfo {year} {1963})}\BibitemShut {NoStop}%
\bibitem [{\citenamefont {Carlson}\ and\ \citenamefont
  {Goldman}(1975)}]{Carlson1975}%
  \BibitemOpen
  \bibfield  {author} {\bibinfo {author} {\bibfnamefont {R.~V.}\ \bibnamefont
  {Carlson}}\ and\ \bibinfo {author} {\bibfnamefont {A.~M.}\ \bibnamefont
  {Goldman}},\ }\href {\doibase 10.1103/PhysRevLett.34.11} {\bibfield
  {journal} {\bibinfo  {journal} {Phys. Rev. Lett.}\ }\textbf {\bibinfo
  {volume} {34}},\ \bibinfo {pages} {11} (\bibinfo {year} {1975})}\BibitemShut
  {NoStop}%
\bibitem [{\citenamefont {{Volkov}}\ and\ \citenamefont
  {{Kogan}}(1974)}]{Volkov.1974}%
  \BibitemOpen
  \bibfield  {author} {\bibinfo {author} {\bibfnamefont {A.~F.}\ \bibnamefont
  {{Volkov}}}\ and\ \bibinfo {author} {\bibfnamefont {S.~M.}\ \bibnamefont
  {{Kogan}}},\ }\href
  {http://www.jetp.ac.ru/cgi-bin/e/index/e/38/5/p1018?a=list} {\bibfield
  {journal} {\bibinfo  {journal} {Soviet Journal of Experimental and
  Theoretical Physics}\ }\textbf {\bibinfo {volume} {38}},\ \bibinfo {pages}
  {1018} (\bibinfo {year} {1974})}\BibitemShut {NoStop}%
\bibitem [{\citenamefont {Littlewood}\ and\ \citenamefont
  {Varma}(1982)}]{Littlewood1982}%
  \BibitemOpen
  \bibfield  {author} {\bibinfo {author} {\bibfnamefont {P.~B.}\ \bibnamefont
  {Littlewood}}\ and\ \bibinfo {author} {\bibfnamefont {C.~M.}\ \bibnamefont
  {Varma}},\ }\href {\doibase 10.1103/PhysRevB.26.4883} {\bibfield  {journal}
  {\bibinfo  {journal} {Phys. Rev. B}\ }\textbf {\bibinfo {volume} {26}},\
  \bibinfo {pages} {4883} (\bibinfo {year} {1982})}\BibitemShut {NoStop}%
\bibitem [{\citenamefont {Cea}\ and\ \citenamefont {Benfatto}(2014)}]{Cea2014}%
  \BibitemOpen
  \bibfield  {author} {\bibinfo {author} {\bibfnamefont {T.}~\bibnamefont
  {Cea}}\ and\ \bibinfo {author} {\bibfnamefont {L.}~\bibnamefont {Benfatto}},\
  }\href {\doibase 10.1103/PhysRevB.90.224515} {\bibfield  {journal} {\bibinfo
  {journal} {Phys. Rev. B}\ }\textbf {\bibinfo {volume} {90}},\ \bibinfo
  {pages} {224515} (\bibinfo {year} {2014})}\BibitemShut {NoStop}%
\bibitem [{\citenamefont {Cea}\ \emph {et~al.}(2015)\citenamefont {Cea},
  \citenamefont {Castellani}, \citenamefont {Seibold},\ and\ \citenamefont
  {Benfatto}}]{Cea2015}%
  \BibitemOpen
  \bibfield  {author} {\bibinfo {author} {\bibfnamefont {T.}~\bibnamefont
  {Cea}}, \bibinfo {author} {\bibfnamefont {C.}~\bibnamefont {Castellani}},
  \bibinfo {author} {\bibfnamefont {G.}~\bibnamefont {Seibold}}, \ and\
  \bibinfo {author} {\bibfnamefont {L.}~\bibnamefont {Benfatto}},\ }\href
  {\doibase 10.1103/PhysRevLett.115.157002} {\bibfield  {journal} {\bibinfo
  {journal} {Phys. Rev. Lett.}\ }\textbf {\bibinfo {volume} {115}},\ \bibinfo
  {pages} {157002} (\bibinfo {year} {2015})}\BibitemShut {NoStop}%
\bibitem [{\citenamefont {Krull}\ \emph {et~al.}(2014)\citenamefont {Krull},
  \citenamefont {Manske}, \citenamefont {Uhrig},\ and\ \citenamefont
  {Schnyder}}]{Krull.2014}%
  \BibitemOpen
  \bibfield  {author} {\bibinfo {author} {\bibfnamefont {H.}~\bibnamefont
  {Krull}}, \bibinfo {author} {\bibfnamefont {D.}~\bibnamefont {Manske}},
  \bibinfo {author} {\bibfnamefont {G.~S.}\ \bibnamefont {Uhrig}}, \ and\
  \bibinfo {author} {\bibfnamefont {A.~P.}\ \bibnamefont {Schnyder}},\ }\href
  {\doibase 10.1103/PhysRevB.90.014515} {\bibfield  {journal} {\bibinfo
  {journal} {Phys. Rev. B}\ }\textbf {\bibinfo {volume} {90}},\ \bibinfo
  {pages} {014515} (\bibinfo {year} {2014})}\BibitemShut {NoStop}%
\bibitem [{\citenamefont {Matsunaga}\ \emph {et~al.}(2013)\citenamefont
  {Matsunaga}, \citenamefont {Hamada}, \citenamefont {Makise}, \citenamefont
  {Uzawa}, \citenamefont {Terai}, \citenamefont {Wang},\ and\ \citenamefont
  {Shimano}}]{Matsunaga.2013}%
  \BibitemOpen
  \bibfield  {author} {\bibinfo {author} {\bibfnamefont {R.}~\bibnamefont
  {Matsunaga}}, \bibinfo {author} {\bibfnamefont {Y.~I.}\ \bibnamefont
  {Hamada}}, \bibinfo {author} {\bibfnamefont {K.}~\bibnamefont {Makise}},
  \bibinfo {author} {\bibfnamefont {Y.}~\bibnamefont {Uzawa}}, \bibinfo
  {author} {\bibfnamefont {H.}~\bibnamefont {Terai}}, \bibinfo {author}
  {\bibfnamefont {Z.}~\bibnamefont {Wang}}, \ and\ \bibinfo {author}
  {\bibfnamefont {R.}~\bibnamefont {Shimano}},\ }\href {\doibase
  10.1103/PhysRevLett.111.057002} {\bibfield  {journal} {\bibinfo  {journal}
  {Phys. Rev. Lett.}\ }\textbf {\bibinfo {volume} {111}},\ \bibinfo {pages}
  {057002} (\bibinfo {year} {2013})}\BibitemShut {NoStop}%
\bibitem [{\citenamefont {Katsumi}\ \emph {et~al.}(2018)\citenamefont
  {Katsumi}, \citenamefont {Tsuji}, \citenamefont {Hamada}, \citenamefont
  {Matsunaga}, \citenamefont {Schneeloch}, \citenamefont {Zhong}, \citenamefont
  {Gu}, \citenamefont {Aoki}, \citenamefont {Gallais},\ and\ \citenamefont
  {Shimano}}]{Katsumi2018}%
  \BibitemOpen
  \bibfield  {author} {\bibinfo {author} {\bibfnamefont {K.}~\bibnamefont
  {Katsumi}}, \bibinfo {author} {\bibfnamefont {N.}~\bibnamefont {Tsuji}},
  \bibinfo {author} {\bibfnamefont {Y.~I.}\ \bibnamefont {Hamada}}, \bibinfo
  {author} {\bibfnamefont {R.}~\bibnamefont {Matsunaga}}, \bibinfo {author}
  {\bibfnamefont {J.}~\bibnamefont {Schneeloch}}, \bibinfo {author}
  {\bibfnamefont {R.~D.}\ \bibnamefont {Zhong}}, \bibinfo {author}
  {\bibfnamefont {G.~D.}\ \bibnamefont {Gu}}, \bibinfo {author} {\bibfnamefont
  {H.}~\bibnamefont {Aoki}}, \bibinfo {author} {\bibfnamefont {Y.}~\bibnamefont
  {Gallais}}, \ and\ \bibinfo {author} {\bibfnamefont {R.}~\bibnamefont
  {Shimano}},\ }\href {\doibase 10.1103/PhysRevLett.120.117001} {\bibfield
  {journal} {\bibinfo  {journal} {Phys. Rev. Lett.}\ }\textbf {\bibinfo
  {volume} {120}},\ \bibinfo {pages} {117001} (\bibinfo {year}
  {2018})}\BibitemShut {NoStop}%
\bibitem [{\citenamefont {Nakamura}\ \emph {et~al.}(2019)\citenamefont
  {Nakamura}, \citenamefont {Iida}, \citenamefont {Murotani}, \citenamefont
  {Matsunaga}, \citenamefont {Terai},\ and\ \citenamefont
  {Shimano}}]{Nakamura2019}%
  \BibitemOpen
  \bibfield  {author} {\bibinfo {author} {\bibfnamefont {S.}~\bibnamefont
  {Nakamura}}, \bibinfo {author} {\bibfnamefont {Y.}~\bibnamefont {Iida}},
  \bibinfo {author} {\bibfnamefont {Y.}~\bibnamefont {Murotani}}, \bibinfo
  {author} {\bibfnamefont {R.}~\bibnamefont {Matsunaga}}, \bibinfo {author}
  {\bibfnamefont {H.}~\bibnamefont {Terai}}, \ and\ \bibinfo {author}
  {\bibfnamefont {R.}~\bibnamefont {Shimano}},\ }\href {\doibase
  10.1103/PhysRevLett.122.257001} {\bibfield  {journal} {\bibinfo  {journal}
  {Phys. Rev. Lett.}\ }\textbf {\bibinfo {volume} {122}},\ \bibinfo {pages}
  {257001} (\bibinfo {year} {2019})}\BibitemShut {NoStop}%
\bibitem [{\citenamefont {Cui}\ \emph {et~al.}(2019)\citenamefont {Cui},
  \citenamefont {Sch\"utt}, \citenamefont {Orth},\ and\ \citenamefont
  {Fernandes}}]{Cui.2019}%
  \BibitemOpen
  \bibfield  {author} {\bibinfo {author} {\bibfnamefont {T.}~\bibnamefont
  {Cui}}, \bibinfo {author} {\bibfnamefont {M.}~\bibnamefont {Sch\"utt}},
  \bibinfo {author} {\bibfnamefont {P.~P.}\ \bibnamefont {Orth}}, \ and\
  \bibinfo {author} {\bibfnamefont {R.~M.}\ \bibnamefont {Fernandes}},\ }\href
  {\doibase 10.1103/PhysRevB.100.144513} {\bibfield  {journal} {\bibinfo
  {journal} {Phys. Rev. B}\ }\textbf {\bibinfo {volume} {100}},\ \bibinfo
  {pages} {144513} (\bibinfo {year} {2019})}\BibitemShut {NoStop}%
\bibitem [{\citenamefont {Shimano}\ and\ \citenamefont
  {Tsuji}(2020)}]{Shimano2020}%
  \BibitemOpen
  \bibfield  {author} {\bibinfo {author} {\bibfnamefont {R.}~\bibnamefont
  {Shimano}}\ and\ \bibinfo {author} {\bibfnamefont {N.}~\bibnamefont
  {Tsuji}},\ }\href {\doibase 10.1146/annurev-conmatphys-031119-050813}
  {\bibfield  {journal} {\bibinfo  {journal} {Annual Review of Condensed Matter
  Physics}\ }\textbf {\bibinfo {volume} {11}},\ \bibinfo {pages} {103}
  (\bibinfo {year} {2020})}\BibitemShut {NoStop}%
\bibitem [{\citenamefont {Cea}\ \emph {et~al.}(2016)\citenamefont {Cea},
  \citenamefont {Castellani},\ and\ \citenamefont {Benfatto}}]{Cea.2016}%
  \BibitemOpen
  \bibfield  {author} {\bibinfo {author} {\bibfnamefont {T.}~\bibnamefont
  {Cea}}, \bibinfo {author} {\bibfnamefont {C.}~\bibnamefont {Castellani}}, \
  and\ \bibinfo {author} {\bibfnamefont {L.}~\bibnamefont {Benfatto}},\ }\href
  {\doibase 10.1103/PhysRevB.93.180507} {\bibfield  {journal} {\bibinfo
  {journal} {Phys. Rev. B}\ }\textbf {\bibinfo {volume} {93}},\ \bibinfo
  {pages} {180507(R)} (\bibinfo {year} {2016})}\BibitemShut {NoStop}%
\bibitem [{\citenamefont {Udina}\ \emph {et~al.}(2019)\citenamefont {Udina},
  \citenamefont {Cea},\ and\ \citenamefont {Benfatto}}]{Udina.2019}%
  \BibitemOpen
  \bibfield  {author} {\bibinfo {author} {\bibfnamefont {M.}~\bibnamefont
  {Udina}}, \bibinfo {author} {\bibfnamefont {T.}~\bibnamefont {Cea}}, \ and\
  \bibinfo {author} {\bibfnamefont {L.}~\bibnamefont {Benfatto}},\ }\href
  {\doibase 10.1103/PhysRevB.100.165131} {\bibfield  {journal} {\bibinfo
  {journal} {Phys. Rev. B}\ }\textbf {\bibinfo {volume} {100}},\ \bibinfo
  {pages} {165131} (\bibinfo {year} {2019})}\BibitemShut {NoStop}%
\bibitem [{\citenamefont {Bardasis}\ and\ \citenamefont
  {Schrieffer}(1961)}]{Bardasis1961}%
  \BibitemOpen
  \bibfield  {author} {\bibinfo {author} {\bibfnamefont {A.}~\bibnamefont
  {Bardasis}}\ and\ \bibinfo {author} {\bibfnamefont {J.~R.}\ \bibnamefont
  {Schrieffer}},\ }\href {\doibase 10.1103/PhysRev.121.1050} {\bibfield
  {journal} {\bibinfo  {journal} {Phys. Rev.}\ }\textbf {\bibinfo {volume}
  {121}},\ \bibinfo {pages} {1050} (\bibinfo {year} {1961})}\BibitemShut
  {NoStop}%
\bibitem [{\citenamefont {Maiti}\ and\ \citenamefont
  {Hirschfeld}(2015)}]{Maiti2015}%
  \BibitemOpen
  \bibfield  {author} {\bibinfo {author} {\bibfnamefont {S.}~\bibnamefont
  {Maiti}}\ and\ \bibinfo {author} {\bibfnamefont {P.~J.}\ \bibnamefont
  {Hirschfeld}},\ }\href {\doibase 10.1103/PhysRevB.92.094506} {\bibfield
  {journal} {\bibinfo  {journal} {Phys. Rev. B}\ }\textbf {\bibinfo {volume}
  {92}},\ \bibinfo {pages} {094506} (\bibinfo {year} {2015})}\BibitemShut
  {NoStop}%
\bibitem [{\citenamefont {Maiti}\ \emph {et~al.}(2016)\citenamefont {Maiti},
  \citenamefont {Maier}, \citenamefont {B\"ohm}, \citenamefont {Hackl},\ and\
  \citenamefont {Hirschfeld}}]{Maiti2016}%
  \BibitemOpen
  \bibfield  {author} {\bibinfo {author} {\bibfnamefont {S.}~\bibnamefont
  {Maiti}}, \bibinfo {author} {\bibfnamefont {T.~A.}\ \bibnamefont {Maier}},
  \bibinfo {author} {\bibfnamefont {T.}~\bibnamefont {B\"ohm}}, \bibinfo
  {author} {\bibfnamefont {R.}~\bibnamefont {Hackl}}, \ and\ \bibinfo {author}
  {\bibfnamefont {P.~J.}\ \bibnamefont {Hirschfeld}},\ }\href {\doibase
  10.1103/PhysRevLett.117.257001} {\bibfield  {journal} {\bibinfo  {journal}
  {Phys. Rev. Lett.}\ }\textbf {\bibinfo {volume} {117}},\ \bibinfo {pages}
  {257001} (\bibinfo {year} {2016})}\BibitemShut {NoStop}%
\bibitem [{\citenamefont {Allocca}\ \emph {et~al.}(2019)\citenamefont
  {Allocca}, \citenamefont {Raines}, \citenamefont {Curtis},\ and\
  \citenamefont {Galitski}}]{Allocca2019}%
  \BibitemOpen
  \bibfield  {author} {\bibinfo {author} {\bibfnamefont {A.~A.}\ \bibnamefont
  {Allocca}}, \bibinfo {author} {\bibfnamefont {Z.~M.}\ \bibnamefont {Raines}},
  \bibinfo {author} {\bibfnamefont {J.~B.}\ \bibnamefont {Curtis}}, \ and\
  \bibinfo {author} {\bibfnamefont {V.~M.}\ \bibnamefont {Galitski}},\ }\href
  {\doibase 10.1103/PhysRevB.99.020504} {\bibfield  {journal} {\bibinfo
  {journal} {Phys. Rev. B}\ }\textbf {\bibinfo {volume} {99}},\ \bibinfo
  {pages} {020504(R)} (\bibinfo {year} {2019})}\BibitemShut {NoStop}%
\bibitem [{\citenamefont {M\"uller}\ \emph {et~al.}(2019)\citenamefont
  {M\"uller}, \citenamefont {Volkov}, \citenamefont {Paul},\ and\ \citenamefont
  {Eremin}}]{Muller.2019}%
  \BibitemOpen
  \bibfield  {author} {\bibinfo {author} {\bibfnamefont {M.~A.}\ \bibnamefont
  {M\"uller}}, \bibinfo {author} {\bibfnamefont {P.~A.}\ \bibnamefont
  {Volkov}}, \bibinfo {author} {\bibfnamefont {I.}~\bibnamefont {Paul}}, \ and\
  \bibinfo {author} {\bibfnamefont {I.~M.}\ \bibnamefont {Eremin}},\ }\href
  {\doibase 10.1103/PhysRevB.100.140501} {\bibfield  {journal} {\bibinfo
  {journal} {Phys. Rev. B}\ }\textbf {\bibinfo {volume} {100}},\ \bibinfo
  {pages} {140501(R)} (\bibinfo {year} {2019})}\BibitemShut {NoStop}%
\bibitem [{\citenamefont {Kretzschmar}\ \emph {et~al.}(2013)\citenamefont
  {Kretzschmar}, \citenamefont {Muschler}, \citenamefont {B\"ohm},
  \citenamefont {Baum}, \citenamefont {Hackl}, \citenamefont {Wen},
  \citenamefont {Tsurkan}, \citenamefont {Deisenhofer},\ and\ \citenamefont
  {Loidl}}]{Kretzschmar.2013}%
  \BibitemOpen
  \bibfield  {author} {\bibinfo {author} {\bibfnamefont {F.}~\bibnamefont
  {Kretzschmar}}, \bibinfo {author} {\bibfnamefont {B.}~\bibnamefont
  {Muschler}}, \bibinfo {author} {\bibfnamefont {T.}~\bibnamefont {B\"ohm}},
  \bibinfo {author} {\bibfnamefont {A.}~\bibnamefont {Baum}}, \bibinfo {author}
  {\bibfnamefont {R.}~\bibnamefont {Hackl}}, \bibinfo {author} {\bibfnamefont
  {H.-H.}\ \bibnamefont {Wen}}, \bibinfo {author} {\bibfnamefont
  {V.}~\bibnamefont {Tsurkan}}, \bibinfo {author} {\bibfnamefont
  {J.}~\bibnamefont {Deisenhofer}}, \ and\ \bibinfo {author} {\bibfnamefont
  {A.}~\bibnamefont {Loidl}},\ }\href {\doibase 10.1103/PhysRevLett.110.187002}
  {\bibfield  {journal} {\bibinfo  {journal} {Phys. Rev. Lett.}\ }\textbf
  {\bibinfo {volume} {110}},\ \bibinfo {pages} {187002} (\bibinfo {year}
  {2013})}\BibitemShut {NoStop}%
\bibitem [{\citenamefont {B{\"{o}}hm}\ \emph {et~al.}(2014)\citenamefont
  {B{\"{o}}hm}, \citenamefont {Kemper}, \citenamefont {Moritz}, \citenamefont
  {Kretzschmar}, \citenamefont {Muschler}, \citenamefont {Eiter}, \citenamefont
  {Hackl}, \citenamefont {Devereaux}, \citenamefont {Scalapino},\ and\
  \citenamefont {Wen}}]{Bohm2014}%
  \BibitemOpen
  \bibfield  {author} {\bibinfo {author} {\bibfnamefont {T.}~\bibnamefont
  {B{\"{o}}hm}}, \bibinfo {author} {\bibfnamefont {A.~F.}\ \bibnamefont
  {Kemper}}, \bibinfo {author} {\bibfnamefont {B.}~\bibnamefont {Moritz}},
  \bibinfo {author} {\bibfnamefont {F.}~\bibnamefont {Kretzschmar}}, \bibinfo
  {author} {\bibfnamefont {B.}~\bibnamefont {Muschler}}, \bibinfo {author}
  {\bibfnamefont {H.-M.}\ \bibnamefont {Eiter}}, \bibinfo {author}
  {\bibfnamefont {R.}~\bibnamefont {Hackl}}, \bibinfo {author} {\bibfnamefont
  {T.~P.}\ \bibnamefont {Devereaux}}, \bibinfo {author} {\bibfnamefont {D.~J.}\
  \bibnamefont {Scalapino}}, \ and\ \bibinfo {author} {\bibfnamefont {H.-H.}\
  \bibnamefont {Wen}},\ }\href {\doibase 10.1103/PhysRevX.4.041046} {\bibfield
  {journal} {\bibinfo  {journal} {Phys. Rev. X}\ }\textbf {\bibinfo {volume}
  {4}},\ \bibinfo {pages} {041046} (\bibinfo {year} {2014})}\BibitemShut
  {NoStop}%
\bibitem [{\citenamefont {Jost}\ \emph {et~al.}(2018)\citenamefont {Jost},
  \citenamefont {Scholz}, \citenamefont {Zweck}, \citenamefont {Meier},
  \citenamefont {B\"ohmer}, \citenamefont {Canfield}, \citenamefont
  {Lazarevi\ifmmode~\acute{c}\else \'{c}\fi{}},\ and\ \citenamefont
  {Hackl}}]{Jost.2018}%
  \BibitemOpen
  \bibfield  {author} {\bibinfo {author} {\bibfnamefont {D.}~\bibnamefont
  {Jost}}, \bibinfo {author} {\bibfnamefont {J.-R.}\ \bibnamefont {Scholz}},
  \bibinfo {author} {\bibfnamefont {U.}~\bibnamefont {Zweck}}, \bibinfo
  {author} {\bibfnamefont {W.~R.}\ \bibnamefont {Meier}}, \bibinfo {author}
  {\bibfnamefont {A.~E.}\ \bibnamefont {B\"ohmer}}, \bibinfo {author}
  {\bibfnamefont {P.~C.}\ \bibnamefont {Canfield}}, \bibinfo {author}
  {\bibfnamefont {N.}~\bibnamefont {Lazarevi\ifmmode~\acute{c}\else
  \'{c}\fi{}}}, \ and\ \bibinfo {author} {\bibfnamefont {R.}~\bibnamefont
  {Hackl}},\ }\href {\doibase 10.1103/PhysRevB.98.020504} {\bibfield  {journal}
  {\bibinfo  {journal} {Phys. Rev. B}\ }\textbf {\bibinfo {volume} {98}},\
  \bibinfo {pages} {020504(R)} (\bibinfo {year} {2018})}\BibitemShut {NoStop}%
\bibitem [{\citenamefont {Altland}\ and\ \citenamefont
  {Simons}(2010)}]{Altland.2010}%
  \BibitemOpen
  \bibfield  {author} {\bibinfo {author} {\bibfnamefont {A.}~\bibnamefont
  {Altland}}\ and\ \bibinfo {author} {\bibfnamefont {B.~D.}\ \bibnamefont
  {Simons}},\ }\href@noop {} {\emph {\bibinfo {title} {{Condensed Matter Field
  Theory}}}}\ (\bibinfo  {publisher} {Cambridge University Press},\ \bibinfo
  {address} {Cambridge, England},\ \bibinfo {year} {2010})\BibitemShut
  {NoStop}%
\bibitem [{\citenamefont {Kosterlitz}\ and\ \citenamefont
  {Thouless}(1973)}]{Kosterlitz_1973}%
  \BibitemOpen
  \bibfield  {author} {\bibinfo {author} {\bibfnamefont {J.~M.}\ \bibnamefont
  {Kosterlitz}}\ and\ \bibinfo {author} {\bibfnamefont {D.~J.}\ \bibnamefont
  {Thouless}},\ }\href {\doibase 10.1088/0022-3719/6/7/010} {\bibfield
  {journal} {\bibinfo  {journal} {Journal of Physics C: Solid State Physics}\
  }\textbf {\bibinfo {volume} {6}},\ \bibinfo {pages} {1181} (\bibinfo {year}
  {1973})}\BibitemShut {NoStop}%
\bibitem [{\citenamefont {Grasset}\ \emph {et~al.}(2018)\citenamefont
  {Grasset}, \citenamefont {Cea}, \citenamefont {Gallais}, \citenamefont
  {Cazayous}, \citenamefont {Sacuto}, \citenamefont {Cario}, \citenamefont
  {Benfatto},\ and\ \citenamefont {M\'easson}}]{Grasset.2018}%
  \BibitemOpen
  \bibfield  {author} {\bibinfo {author} {\bibfnamefont {R.}~\bibnamefont
  {Grasset}}, \bibinfo {author} {\bibfnamefont {T.}~\bibnamefont {Cea}},
  \bibinfo {author} {\bibfnamefont {Y.}~\bibnamefont {Gallais}}, \bibinfo
  {author} {\bibfnamefont {M.}~\bibnamefont {Cazayous}}, \bibinfo {author}
  {\bibfnamefont {A.}~\bibnamefont {Sacuto}}, \bibinfo {author} {\bibfnamefont
  {L.}~\bibnamefont {Cario}}, \bibinfo {author} {\bibfnamefont
  {L.}~\bibnamefont {Benfatto}}, \ and\ \bibinfo {author} {\bibfnamefont
  {M.-A.}\ \bibnamefont {M\'easson}},\ }\href {\doibase
  10.1103/PhysRevB.97.094502} {\bibfield  {journal} {\bibinfo  {journal} {Phys.
  Rev. B}\ }\textbf {\bibinfo {volume} {97}},\ \bibinfo {pages} {094502}
  (\bibinfo {year} {2018})}\BibitemShut {NoStop}%
\bibitem [{\citenamefont {Paramekanti}\ \emph {et~al.}(2000)\citenamefont
  {Paramekanti}, \citenamefont {Randeria}, \citenamefont {Ramakrishnan},\ and\
  \citenamefont {Mandal}}]{Paramekanti.2000}%
  \BibitemOpen
  \bibfield  {author} {\bibinfo {author} {\bibfnamefont {A.}~\bibnamefont
  {Paramekanti}}, \bibinfo {author} {\bibfnamefont {M.}~\bibnamefont
  {Randeria}}, \bibinfo {author} {\bibfnamefont {T.~V.}\ \bibnamefont
  {Ramakrishnan}}, \ and\ \bibinfo {author} {\bibfnamefont {S.~S.}\
  \bibnamefont {Mandal}},\ }\href {\doibase 10.1103/PhysRevB.62.6786}
  {\bibfield  {journal} {\bibinfo  {journal} {Phys. Rev. B}\ }\textbf {\bibinfo
  {volume} {62}},\ \bibinfo {pages} {6786} (\bibinfo {year}
  {2000})}\BibitemShut {NoStop}%
\bibitem [{\citenamefont {Benfatto}\ \emph {et~al.}(2004)\citenamefont
  {Benfatto}, \citenamefont {Toschi},\ and\ \citenamefont
  {Caprara}}]{Benfatto.2004}%
  \BibitemOpen
  \bibfield  {author} {\bibinfo {author} {\bibfnamefont {L.}~\bibnamefont
  {Benfatto}}, \bibinfo {author} {\bibfnamefont {A.}~\bibnamefont {Toschi}}, \
  and\ \bibinfo {author} {\bibfnamefont {S.}~\bibnamefont {Caprara}},\ }\href
  {\doibase 10.1103/PhysRevB.69.184510} {\bibfield  {journal} {\bibinfo
  {journal} {Phys. Rev. B}\ }\textbf {\bibinfo {volume} {69}},\ \bibinfo
  {pages} {184510} (\bibinfo {year} {2004})}\BibitemShut {NoStop}%
\bibitem [{\citenamefont {Mahan}(1990)}]{Mahan1990}%
  \BibitemOpen
  \bibfield  {author} {\bibinfo {author} {\bibfnamefont {G.~D.}\ \bibnamefont
  {Mahan}},\ }\href
  {https://books.google.com/books/about/Many_Particle_Physics.html?id=v8du6cp0vUAC}
  {\emph {\bibinfo {title} {{Many-particle physics}}}}\ (\bibinfo  {publisher}
  {Plenum Press},\ \bibinfo {year} {1990})\ p.\ \bibinfo {pages}
  {1032}\BibitemShut {NoStop}%
\bibitem [{\citenamefont {Sun}\ \emph {et~al.}(2018{\natexlab{a}})\citenamefont
  {Sun}, \citenamefont {Basov},\ and\ \citenamefont {Fogler}}]{Sun.2018}%
  \BibitemOpen
  \bibfield  {author} {\bibinfo {author} {\bibfnamefont {Z.}~\bibnamefont
  {Sun}}, \bibinfo {author} {\bibfnamefont {D.~N.}\ \bibnamefont {Basov}}, \
  and\ \bibinfo {author} {\bibfnamefont {M.~M.}\ \bibnamefont {Fogler}},\
  }\href {\doibase 10.1073/pnas.1717010115} {\bibfield  {journal} {\bibinfo
  {journal} {Proc. Natl. Acad. Sci.}\ }\textbf {\bibinfo {volume} {115}},\
  \bibinfo {pages} {3285} (\bibinfo {year} {2018}{\natexlab{a}})}\BibitemShut
  {NoStop}%
\bibitem [{\citenamefont {Torre}\ \emph {et~al.}(2019)\citenamefont {Torre},
  \citenamefont {{Vieira de Castro}}, \citenamefont {{Van Duppen}},
  \citenamefont {{Barcons Ruiz}}, \citenamefont {Peeters}, \citenamefont
  {Koppens},\ and\ \citenamefont {Polini}}]{Torre2019}%
  \BibitemOpen
  \bibfield  {author} {\bibinfo {author} {\bibfnamefont {I.}~\bibnamefont
  {Torre}}, \bibinfo {author} {\bibfnamefont {L.}~\bibnamefont {{Vieira de
  Castro}}}, \bibinfo {author} {\bibfnamefont {B.}~\bibnamefont {{Van
  Duppen}}}, \bibinfo {author} {\bibfnamefont {D.}~\bibnamefont {{Barcons
  Ruiz}}}, \bibinfo {author} {\bibfnamefont {F.~M.}\ \bibnamefont {Peeters}},
  \bibinfo {author} {\bibfnamefont {F.~H.~L.}\ \bibnamefont {Koppens}}, \ and\
  \bibinfo {author} {\bibfnamefont {M.}~\bibnamefont {Polini}},\ }\href
  {\doibase 10.1103/PhysRevB.99.144307} {\bibfield  {journal} {\bibinfo
  {journal} {Phys. Rev. B}\ }\textbf {\bibinfo {volume} {99}},\ \bibinfo
  {pages} {144307} (\bibinfo {year} {2019})}\BibitemShut {NoStop}%
\bibitem [{\citenamefont {Sun}\ \emph {et~al.}(2018{\natexlab{b}})\citenamefont
  {Sun}, \citenamefont {Basov},\ and\ \citenamefont {Fogler}}]{Sun2018b}%
  \BibitemOpen
  \bibfield  {author} {\bibinfo {author} {\bibfnamefont {Z.}~\bibnamefont
  {Sun}}, \bibinfo {author} {\bibfnamefont {D.~N.}\ \bibnamefont {Basov}}, \
  and\ \bibinfo {author} {\bibfnamefont {M.~M.}\ \bibnamefont {Fogler}},\
  }\href {\doibase 10.1103/PhysRevB.97.075432} {\bibfield  {journal} {\bibinfo
  {journal} {Phys. Rev. B}\ }\textbf {\bibinfo {volume} {97}},\ \bibinfo
  {pages} {075432} (\bibinfo {year} {2018}{\natexlab{b}})}\BibitemShut
  {NoStop}%
\bibitem [{\citenamefont {Warren}\ and\ \citenamefont
  {Ferrell}(1960)}]{Warren1960}%
  \BibitemOpen
  \bibfield  {author} {\bibinfo {author} {\bibfnamefont {J.~L.}\ \bibnamefont
  {Warren}}\ and\ \bibinfo {author} {\bibfnamefont {R.~A.}\ \bibnamefont
  {Ferrell}},\ }\href {\doibase 10.1103/PhysRev.117.1252} {\bibfield  {journal}
  {\bibinfo  {journal} {Phys. Rev.}\ }\textbf {\bibinfo {volume} {117}},\
  \bibinfo {pages} {1252} (\bibinfo {year} {1960})}\BibitemShut {NoStop}%
\bibitem [{\citenamefont {Abrikosov}\ \emph {et~al.}(1975)\citenamefont
  {Abrikosov}, \citenamefont {Dzyaloshinskii}, \citenamefont {Gorkov},\ and\
  \citenamefont {Silverman}}]{Abrikosov:1975}%
  \BibitemOpen
  \bibfield  {author} {\bibinfo {author} {\bibfnamefont {A.~A.}\ \bibnamefont
  {Abrikosov}}, \bibinfo {author} {\bibfnamefont {I.}~\bibnamefont
  {Dzyaloshinskii}}, \bibinfo {author} {\bibfnamefont {L.~P.}\ \bibnamefont
  {Gorkov}}, \ and\ \bibinfo {author} {\bibfnamefont {R.~A.}\ \bibnamefont
  {Silverman}},\ }\href {https://cds.cern.ch/record/107441} {\emph {\bibinfo
  {title} {{Methods of quantum field theory in statistical physics}}}}\
  (\bibinfo  {publisher} {Dover},\ \bibinfo {address} {New York, NY},\ \bibinfo
  {year} {1975})\BibitemShut {NoStop}%
\bibitem [{\citenamefont {Orenstein}\ \emph {et~al.}(1990)\citenamefont
  {Orenstein}, \citenamefont {Thomas}, \citenamefont {Millis}, \citenamefont
  {Cooper}, \citenamefont {Rapkine}, \citenamefont {Timusk}, \citenamefont
  {Schneemeyer},\ and\ \citenamefont {Waszczak}}]{Orenstein1990}%
  \BibitemOpen
  \bibfield  {author} {\bibinfo {author} {\bibfnamefont {J.}~\bibnamefont
  {Orenstein}}, \bibinfo {author} {\bibfnamefont {G.~A.}\ \bibnamefont
  {Thomas}}, \bibinfo {author} {\bibfnamefont {A.~J.}\ \bibnamefont {Millis}},
  \bibinfo {author} {\bibfnamefont {S.~L.}\ \bibnamefont {Cooper}}, \bibinfo
  {author} {\bibfnamefont {D.~H.}\ \bibnamefont {Rapkine}}, \bibinfo {author}
  {\bibfnamefont {T.}~\bibnamefont {Timusk}}, \bibinfo {author} {\bibfnamefont
  {L.~F.}\ \bibnamefont {Schneemeyer}}, \ and\ \bibinfo {author} {\bibfnamefont
  {J.~V.}\ \bibnamefont {Waszczak}},\ }\href {\doibase
  10.1103/PhysRevB.42.6342} {\bibfield  {journal} {\bibinfo  {journal} {Phys.
  Rev. B}\ }\textbf {\bibinfo {volume} {42}},\ \bibinfo {pages} {6342}
  (\bibinfo {year} {1990})}\BibitemShut {NoStop}%
\bibitem [{\citenamefont {Schmid}\ and\ \citenamefont
  {Sch{\"{o}}n}(1975)}]{schmid.1975}%
  \BibitemOpen
  \bibfield  {author} {\bibinfo {author} {\bibfnamefont {A.}~\bibnamefont
  {Schmid}}\ and\ \bibinfo {author} {\bibfnamefont {G.}~\bibnamefont
  {Sch{\"{o}}n}},\ }\href {\doibase 10.1103/PhysRevLett.34.941} {\bibfield
  {journal} {\bibinfo  {journal} {Phys. Rev. Lett.}\ }\textbf {\bibinfo
  {volume} {34}},\ \bibinfo {pages} {941} (\bibinfo {year} {1975})}\BibitemShut
  {NoStop}%
\bibitem [{\citenamefont {{Artemenko}}\ and\ \citenamefont
  {{Volkov}}(1976)}]{Artemenko.1976}%
  \BibitemOpen
  \bibfield  {author} {\bibinfo {author} {\bibfnamefont {S.~N.}\ \bibnamefont
  {{Artemenko}}}\ and\ \bibinfo {author} {\bibfnamefont {A.~F.}\ \bibnamefont
  {{Volkov}}},\ }\href
  {http://www.jetp.ac.ru/cgi-bin/e/index/e/42/5/p896?a=list} {\bibfield
  {journal} {\bibinfo  {journal} {Soviet Journal of Experimental and
  Theoretical Physics}\ }\textbf {\bibinfo {volume} {42}},\ \bibinfo {pages}
  {896} (\bibinfo {year} {1976})}\BibitemShut {NoStop}%
\bibitem [{\citenamefont {Pethick}\ and\ \citenamefont
  {Smith}(1979)}]{Pethick1979}%
  \BibitemOpen
  \bibfield  {author} {\bibinfo {author} {\bibfnamefont {C.}~\bibnamefont
  {Pethick}}\ and\ \bibinfo {author} {\bibfnamefont {H.}~\bibnamefont
  {Smith}},\ }\href {\doibase 10.1016/0003-4916(79)90253-7} {\bibfield
  {journal} {\bibinfo  {journal} {Annals of Physics}\ }\textbf {\bibinfo
  {volume} {119}},\ \bibinfo {pages} {133} (\bibinfo {year}
  {1979})}\BibitemShut {NoStop}%
\bibitem [{\citenamefont {Artemenko}\ and\ \citenamefont
  {Volkov}(1979)}]{Artemenko1979}%
  \BibitemOpen
  \bibfield  {author} {\bibinfo {author} {\bibfnamefont {S.~N.}\ \bibnamefont
  {Artemenko}}\ and\ \bibinfo {author} {\bibfnamefont {A.~F.}\ \bibnamefont
  {Volkov}},\ }\href {\doibase 10.1070/PU1979v022n05ABEH005495} {\bibfield
  {journal} {\bibinfo  {journal} {Sov. Phys. Uspekhi}\ }\textbf {\bibinfo
  {volume} {22}},\ \bibinfo {pages} {295} (\bibinfo {year} {1979})}\BibitemShut
  {NoStop}%
\bibitem [{\citenamefont {{Kulik}}\ \emph {et~al.}(1981)\citenamefont
  {{Kulik}}, \citenamefont {{Entin-Wohlman}},\ and\ \citenamefont
  {{Orbach}}}]{Kulik.1981}%
  \BibitemOpen
  \bibfield  {author} {\bibinfo {author} {\bibfnamefont {I.~O.}\ \bibnamefont
  {{Kulik}}}, \bibinfo {author} {\bibfnamefont {O.}~\bibnamefont
  {{Entin-Wohlman}}}, \ and\ \bibinfo {author} {\bibfnamefont {R.}~\bibnamefont
  {{Orbach}}},\ }\href {\doibase 10.1007/BF00115617} {\bibfield  {journal}
  {\bibinfo  {journal} {Journal of Low Temperature Physics}\ }\textbf {\bibinfo
  {volume} {43}},\ \bibinfo {pages} {591} (\bibinfo {year} {1981})}\BibitemShut
  {NoStop}%
\bibitem [{\citenamefont {Goldman}(2007)}]{Goldman2007}%
  \BibitemOpen
  \bibfield  {author} {\bibinfo {author} {\bibfnamefont {A.~M.}\ \bibnamefont
  {Goldman}},\ }\href {\doibase 10.1007/s10948-006-0166-7} {\bibfield
  {journal} {\bibinfo  {journal} {J. Supercond. Nov. Magn.}\ }\textbf {\bibinfo
  {volume} {19}},\ \bibinfo {pages} {317} (\bibinfo {year} {2007})}\BibitemShut
  {NoStop}%
\bibitem [{\citenamefont {Sun}\ \emph {et~al.}(2016)\citenamefont {Sun},
  \citenamefont {Basov},\ and\ \citenamefont {Fogler}}]{Sun2016a}%
  \BibitemOpen
  \bibfield  {author} {\bibinfo {author} {\bibfnamefont {Z.}~\bibnamefont
  {Sun}}, \bibinfo {author} {\bibfnamefont {D.~N.}\ \bibnamefont {Basov}}, \
  and\ \bibinfo {author} {\bibfnamefont {M.~M.}\ \bibnamefont {Fogler}},\
  }\href {\doibase 10.1103/PhysRevLett.117.076805} {\bibfield  {journal}
  {\bibinfo  {journal} {Phys. Rev. Lett.}\ }\textbf {\bibinfo {volume} {117}},\
  \bibinfo {pages} {076805} (\bibinfo {year} {2016})}\BibitemShut {NoStop}%
\bibitem [{\citenamefont {Tinkham}(2004)}]{Tinkham}%
  \BibitemOpen
  \bibfield  {author} {\bibinfo {author} {\bibfnamefont {M.}~\bibnamefont
  {Tinkham}},\ }\href@noop {} {\emph {\bibinfo {title} {{Introduction to
  Superconductivity}}}}\ (\bibinfo  {publisher} {Dover Publications},\ \bibinfo
  {address} {Mineola, New York},\ \bibinfo {year} {2004})\BibitemShut {NoStop}%
\bibitem [{\citenamefont {Artemenko}\ and\ \citenamefont
  {Kobelkov}(1997)}]{Artemenko1997}%
  \BibitemOpen
  \bibfield  {author} {\bibinfo {author} {\bibfnamefont {S.~N.}\ \bibnamefont
  {Artemenko}}\ and\ \bibinfo {author} {\bibfnamefont {A.~G.}\ \bibnamefont
  {Kobelkov}},\ }\href {\doibase 10.1016/S0921-4534(97)01082-4} {\bibfield
  {journal} {\bibinfo  {journal} {Physica C}\ }\textbf {\bibinfo {volume}
  {282-287}},\ \bibinfo {pages} {1845} (\bibinfo {year} {1997})}\BibitemShut
  {NoStop}%
\bibitem [{\citenamefont {Orenstein}(2003)}]{Orenstein2003}%
  \BibitemOpen
  \bibfield  {author} {\bibinfo {author} {\bibfnamefont {J.}~\bibnamefont
  {Orenstein}},\ }\href {\doibase 10.1016/S0921-4534(03)00705-6} {\bibfield
  {journal} {\bibinfo  {journal} {Physica C}\ }\textbf {\bibinfo {volume}
  {390}},\ \bibinfo {pages} {243} (\bibinfo {year} {2003})}\BibitemShut
  {NoStop}%
\bibitem [{\citenamefont {Barabash}\ and\ \citenamefont
  {Stroud}(2003)}]{Barabash2003}%
  \BibitemOpen
  \bibfield  {author} {\bibinfo {author} {\bibfnamefont {S.~V.}\ \bibnamefont
  {Barabash}}\ and\ \bibinfo {author} {\bibfnamefont {D.}~\bibnamefont
  {Stroud}},\ }\href {\doibase 10.1016/j.physb.2003.08.028} {\bibfield
  {journal} {\bibinfo  {journal} {Physica B}\ }\textbf {\bibinfo {volume}
  {338}},\ \bibinfo {pages} {224} (\bibinfo {year} {2003})}\BibitemShut
  {NoStop}%
\bibitem [{\citenamefont {Chiao}\ \emph {et~al.}(2000)\citenamefont {Chiao},
  \citenamefont {Hill}, \citenamefont {Lupien}, \citenamefont {Taillefer},
  \citenamefont {Lambert}, \citenamefont {Gagnon},\ and\ \citenamefont
  {Fournier}}]{Chiao2000}%
  \BibitemOpen
  \bibfield  {author} {\bibinfo {author} {\bibfnamefont {M.}~\bibnamefont
  {Chiao}}, \bibinfo {author} {\bibfnamefont {R.~W.}\ \bibnamefont {Hill}},
  \bibinfo {author} {\bibfnamefont {C.}~\bibnamefont {Lupien}}, \bibinfo
  {author} {\bibfnamefont {L.}~\bibnamefont {Taillefer}}, \bibinfo {author}
  {\bibfnamefont {P.}~\bibnamefont {Lambert}}, \bibinfo {author} {\bibfnamefont
  {R.}~\bibnamefont {Gagnon}}, \ and\ \bibinfo {author} {\bibfnamefont
  {P.}~\bibnamefont {Fournier}},\ }\href {\doibase 10.1103/PhysRevB.62.3554}
  {\bibfield  {journal} {\bibinfo  {journal} {Phys. Rev. B}\ }\textbf {\bibinfo
  {volume} {62}},\ \bibinfo {pages} {3554} (\bibinfo {year}
  {2000})}\BibitemShut {NoStop}%
\bibitem [{\citenamefont {Pokrovsky}(1996)}]{Pokrovsky1996}%
  \BibitemOpen
  \bibfield  {author} {\bibinfo {author} {\bibfnamefont {V.~L.}\ \bibnamefont
  {Pokrovsky}},\ }in\ \href {\doibase 10.1117/12.241743} {\emph {\bibinfo
  {booktitle} {Spectroscopic Studies of Superconductors}}},\ Vol.\ \bibinfo
  {volume} {2696},\ \bibinfo {editor} {edited by\ \bibinfo {editor}
  {\bibfnamefont {I.}~\bibnamefont {Bozovic}}\ and\ \bibinfo {editor}
  {\bibfnamefont {D.}~\bibnamefont {van~der Marel}}},\ \bibinfo {organization}
  {International Society for Optics and Photonics}\ (\bibinfo  {publisher}
  {SPIE},\ \bibinfo {year} {1996})\ pp.\ \bibinfo {pages} {137 --
  159}\BibitemShut {NoStop}%
\bibitem [{\citenamefont {Stinson}\ \emph {et~al.}(2014)\citenamefont
  {Stinson}, \citenamefont {Wu}, \citenamefont {Jiang}, \citenamefont {Fei},
  \citenamefont {Rodin}, \citenamefont {Chapler}, \citenamefont {McLeod},
  \citenamefont {{Castro Neto}}, \citenamefont {Lee}, \citenamefont {Fogler},\
  and\ \citenamefont {Basov}}]{Stinson2014a}%
  \BibitemOpen
  \bibfield  {author} {\bibinfo {author} {\bibfnamefont {H.~T.}\ \bibnamefont
  {Stinson}}, \bibinfo {author} {\bibfnamefont {J.~S.}\ \bibnamefont {Wu}},
  \bibinfo {author} {\bibfnamefont {B.~Y.}\ \bibnamefont {Jiang}}, \bibinfo
  {author} {\bibfnamefont {Z.}~\bibnamefont {Fei}}, \bibinfo {author}
  {\bibfnamefont {A.~S.}\ \bibnamefont {Rodin}}, \bibinfo {author}
  {\bibfnamefont {B.~C.}\ \bibnamefont {Chapler}}, \bibinfo {author}
  {\bibfnamefont {A.~S.}\ \bibnamefont {McLeod}}, \bibinfo {author}
  {\bibfnamefont {A.}~\bibnamefont {{Castro Neto}}}, \bibinfo {author}
  {\bibfnamefont {Y.~S.}\ \bibnamefont {Lee}}, \bibinfo {author} {\bibfnamefont
  {M.~M.}\ \bibnamefont {Fogler}}, \ and\ \bibinfo {author} {\bibfnamefont
  {D.~N.}\ \bibnamefont {Basov}},\ }\href {\doibase 10.1103/PhysRevB.90.014502}
  {\bibfield  {journal} {\bibinfo  {journal} {Phys. Rev. B}\ }\textbf {\bibinfo
  {volume} {90}},\ \bibinfo {pages} {014502} (\bibinfo {year}
  {2014})}\BibitemShut {NoStop}%
\bibitem [{\citenamefont {Sun}\ \emph {et~al.}(2014)\citenamefont {Sun},
  \citenamefont {Litchinitser},\ and\ \citenamefont {Zhou}}]{Zhou2014}%
  \BibitemOpen
  \bibfield  {author} {\bibinfo {author} {\bibfnamefont {J.}~\bibnamefont
  {Sun}}, \bibinfo {author} {\bibfnamefont {N.~M.}\ \bibnamefont
  {Litchinitser}}, \ and\ \bibinfo {author} {\bibfnamefont {J.}~\bibnamefont
  {Zhou}},\ }\href {\doibase 10.1021/ph4000983} {\bibfield  {journal} {\bibinfo
   {journal} {ACS Photonics}\ }\textbf {\bibinfo {volume} {1}},\ \bibinfo
  {pages} {293} (\bibinfo {year} {2014})}\BibitemShut {NoStop}%
\bibitem [{\citenamefont {Sun}\ \emph {et~al.}(2015)\citenamefont {Sun},
  \citenamefont {Guti{\'{e}}rrez-Rubio}, \citenamefont {Basov},\ and\
  \citenamefont {Fogler}}]{Sun.2015}%
  \BibitemOpen
  \bibfield  {author} {\bibinfo {author} {\bibfnamefont {Z.}~\bibnamefont
  {Sun}}, \bibinfo {author} {\bibfnamefont {{\'{A}}.}~\bibnamefont
  {Guti{\'{e}}rrez-Rubio}}, \bibinfo {author} {\bibfnamefont {D.~N.}\
  \bibnamefont {Basov}}, \ and\ \bibinfo {author} {\bibfnamefont {M.~M.}\
  \bibnamefont {Fogler}},\ }\href {\doibase 10.1021/acs.nanolett.5b00814}
  {\bibfield  {journal} {\bibinfo  {journal} {Nano Lett.}\ }\textbf {\bibinfo
  {volume} {15}},\ \bibinfo {pages} {4455} (\bibinfo {year}
  {2015})}\BibitemShut {NoStop}%
\bibitem [{\citenamefont {Basov}\ \emph {et~al.}(2016)\citenamefont {Basov},
  \citenamefont {Fogler},\ and\ \citenamefont {{Garc{\'{i}}a de
  Abajo}}}]{Basov2016}%
  \BibitemOpen
  \bibfield  {author} {\bibinfo {author} {\bibfnamefont {D.~N.}\ \bibnamefont
  {Basov}}, \bibinfo {author} {\bibfnamefont {M.~M.}\ \bibnamefont {Fogler}}, \
  and\ \bibinfo {author} {\bibfnamefont {F.~J.}\ \bibnamefont {{Garc{\'{i}}a de
  Abajo}}},\ }\href
  {http://science.sciencemag.org/content/354/6309/aag1992.abstract} {\bibfield
  {journal} {\bibinfo  {journal} {Science}\ }\textbf {\bibinfo {volume}
  {354}},\ \bibinfo {pages} {195} (\bibinfo {year} {2016})}\BibitemShut
  {NoStop}%
\bibitem [{\citenamefont {Sun}\ and\ \citenamefont {Millis}(2020)}]{sun2020BS}%
  \BibitemOpen
  \bibfield  {author} {\bibinfo {author} {\bibfnamefont {Z.}~\bibnamefont
  {Sun}}\ and\ \bibinfo {author} {\bibfnamefont {A.~J.}\ \bibnamefont
  {Millis}},\ }\href@noop {} {\enquote {\bibinfo {title} {Bardasis-schrieffer
  polaritons in excitonic insulators},}\ } (\bibinfo {year} {2020}),\ \Eprint
  {http://arxiv.org/abs/2003.02997} {arXiv:2003.02997 [cond-mat.str-el]}
  \BibitemShut {NoStop}%
\bibitem [{\citenamefont {Sun}\ and\ \citenamefont
  {Millis}(2019)}]{sun2019transient}%
  \BibitemOpen
  \bibfield  {author} {\bibinfo {author} {\bibfnamefont {Z.}~\bibnamefont
  {Sun}}\ and\ \bibinfo {author} {\bibfnamefont {A.~J.}\ \bibnamefont
  {Millis}},\ }\href@noop {} {\enquote {\bibinfo {title} {Transient trapping
  into metastable states in systems with competing orders},}\ } (\bibinfo
  {year} {2019}),\ \Eprint {http://arxiv.org/abs/1905.05341} {arXiv:1905.05341
  [cond-mat.str-el]} \BibitemShut {NoStop}%
\bibitem [{\citenamefont {Wu}\ and\ \citenamefont {Griffin}(1995)}]{Wu.1995}%
  \BibitemOpen
  \bibfield  {author} {\bibinfo {author} {\bibfnamefont {W.-C.}\ \bibnamefont
  {Wu}}\ and\ \bibinfo {author} {\bibfnamefont {A.}~\bibnamefont {Griffin}},\
  }\href {\doibase 10.1103/PhysRevB.51.1190} {\bibfield  {journal} {\bibinfo
  {journal} {Phys. Rev. B}\ }\textbf {\bibinfo {volume} {51}},\ \bibinfo
  {pages} {1190} (\bibinfo {year} {1995})}\BibitemShut {NoStop}%
\bibitem [{\citenamefont {Ohashi}\ and\ \citenamefont
  {Takada}(2000)}]{Ohashi2000}%
  \BibitemOpen
  \bibfield  {author} {\bibinfo {author} {\bibfnamefont {Y.}~\bibnamefont
  {Ohashi}}\ and\ \bibinfo {author} {\bibfnamefont {S.}~\bibnamefont
  {Takada}},\ }\href {\doibase 10.1103/PhysRevB.62.5971} {\bibfield  {journal}
  {\bibinfo  {journal} {Phys. Rev. B}\ }\textbf {\bibinfo {volume} {62}},\
  \bibinfo {pages} {5971} (\bibinfo {year} {2000})}\BibitemShut {NoStop}%
\bibitem [{\citenamefont {Leggett}(1966)}]{Leggett.1966}%
  \BibitemOpen
  \bibfield  {author} {\bibinfo {author} {\bibfnamefont {A.~J.}\ \bibnamefont
  {Leggett}},\ }\href {\doibase 10.1143/PTP.36.901} {\bibfield  {journal}
  {\bibinfo  {journal} {Progress of Theoretical Physics}\ }\textbf {\bibinfo
  {volume} {36}},\ \bibinfo {pages} {901} (\bibinfo {year} {1966})}\BibitemShut
  {NoStop}%
\bibitem [{\citenamefont {Giorgianni}\ \emph {et~al.}(2019)\citenamefont
  {Giorgianni}, \citenamefont {Cea}, \citenamefont {Vicario}, \citenamefont
  {Hauri}, \citenamefont {Withanage}, \citenamefont {Xi},\ and\ \citenamefont
  {Benfatto}}]{Giorgianni.2019}%
  \BibitemOpen
  \bibfield  {author} {\bibinfo {author} {\bibfnamefont {F.}~\bibnamefont
  {Giorgianni}}, \bibinfo {author} {\bibfnamefont {T.}~\bibnamefont {Cea}},
  \bibinfo {author} {\bibfnamefont {C.}~\bibnamefont {Vicario}}, \bibinfo
  {author} {\bibfnamefont {C.~P.}\ \bibnamefont {Hauri}}, \bibinfo {author}
  {\bibfnamefont {W.~K.}\ \bibnamefont {Withanage}}, \bibinfo {author}
  {\bibfnamefont {X.}~\bibnamefont {Xi}}, \ and\ \bibinfo {author}
  {\bibfnamefont {L.}~\bibnamefont {Benfatto}},\ }\href
  {https://doi.org/10.1038/s41567-018-0385-4} {\bibfield  {journal} {\bibinfo
  {journal} {Nature Physics}\ }\textbf {\bibinfo {volume} {15}},\ \bibinfo
  {pages} {341} (\bibinfo {year} {2019})}\BibitemShut {NoStop}%
\bibitem [{\citenamefont {Lemonik}\ and\ \citenamefont
  {Mitra}(2017)}]{Lemonik.2017}%
  \BibitemOpen
  \bibfield  {author} {\bibinfo {author} {\bibfnamefont {Y.}~\bibnamefont
  {Lemonik}}\ and\ \bibinfo {author} {\bibfnamefont {A.}~\bibnamefont
  {Mitra}},\ }\href {\doibase 10.1103/PhysRevB.96.104506} {\bibfield  {journal}
  {\bibinfo  {journal} {Phys. Rev. B}\ }\textbf {\bibinfo {volume} {96}},\
  \bibinfo {pages} {104506} (\bibinfo {year} {2017})}\BibitemShut {NoStop}%
\end{thebibliography}%

\appendix
\begin{widetext}
\section{Correlation functions}
\label{app:correlation_function}
The correlation function $\chi_{\sigma_i \sigma_j}$ shown in Fig.~\ref{fig:bubble} is defined as
\begin{align}
\chi_{\sigma_i \sigma_j}(q) = 
\left\langle \hat{T} 
\left(\psi^\dagger \sigma_i \psi \right)_{(r,t)}
\left(\psi^\dagger \sigma_j \psi \right)_0  
\right\rangle \bigg|_q 
=
\sum_{\omega_n, k}
Tr\left[ G(k,i\omega_n) \sigma_i G(k+q,i(\omega_n+\Omega))  \sigma_j \right]
\,.
\label{eqn:chi_defi}
\end{align}
where $\hat{T}$ is the time order symbol, $x=(\mathbf{r},t)$, $q=(\mathbf{q},i\Omega)$ and
\begin{align}
G(k,i\omega_n) &= G_\Delta(k,i\omega_n) = 
\left\langle \hat{T} 
\psi (x)
\psi^\dagger(0) 
\right\rangle \bigg|_{k,i\omega_n}
=\frac{1}{i\omega_n - \xi_k \sigma_3 -\Delta\sigma_1 }
\,
\end{align}
is the electron Green's function. Rotation from imaginary to real time makes the time ordered correlation functions into retarded ones.
In the correlation functions involving the currents, one should change the $\sigma$ vertex to the current vertex. For example, 
\begin{align}
\chi_{j_l \sigma_m}(q) &= 
\left\langle \hat{T} 
\left(\psi^\dagger v_l\sigma_0 \psi \right)_x
\left(\psi^\dagger \sigma_m \psi \right)_0  
\right\rangle \bigg|_q 
=
\sum_{\omega_n, k} 
\frac{1}{2} \left( v(k)+v(k+q) \right)
Tr\left[ G(k,i\omega_n) \sigma_0 G(k+q,i(\omega_n+\Omega))  \sigma_l \right]
\,.
\end{align}
Evaluating the correlation function \equa{eqn:chi_defi} renders
\begin{align}
\chi_{\sigma_i \sigma_j}(q) &= 
\sum_{\omega_n, k}
Tr\left[ 
\frac{
\left(i\omega_n + \xi \sigma_3 + \Delta \sigma_1 \right) \sigma_i
\left( i(\omega_n +\Omega) + \xi^\prime \sigma_3 + \Delta \sigma_1 \right) \sigma_j
}{\left( (i\omega_n)^2 - E^2) \right) 
\left(
(i(\omega_n+\Omega))^2 - E^{\prime 2}) \right)
}
\right]
\notag\\
&= \frac{1}{4}
\sum_{k}
\Bigg\{ 
Tr\left[
\sigma_i \sigma_j 
\right]
\left(
\frac{f(E^\prime)-f(E)}{i\Omega- (E-E^\prime)}+
\frac{1-f(E^\prime)-f(E)}{i\Omega- (E+E^\prime)} +
\frac{f(E^\prime)+f(E)-1}{i\Omega + (E+E^\prime)} +
\frac{f(E)-f(E^\prime)}{i\Omega- (E^\prime-E)}
\right)
\notag\\
&+
Tr\left[
\frac{\sigma_i (\xi^\prime \sigma_3 + \Delta \sigma_1) \sigma_j}{E^\prime}
\right]
\left(
-\frac{f(E^\prime)-f(E)}{i\Omega- (E-E^\prime)}+
\frac{1-f(E^\prime)-f(E)}{i\Omega- (E+E^\prime)} -
\frac{f(E^\prime)+f(E)-1}{i\Omega + (E+E^\prime)} +
\frac{f(E)-f(E^\prime)}{i\Omega- (E^\prime-E)}
\right)
\notag\\
&+
Tr\left[
\frac{(\xi \sigma_3 + \Delta \sigma_1) \sigma_i \sigma_j}{E}
\right]
\left(
-\frac{f(E^\prime)-f(E)}{i\Omega- (E-E^\prime)}
-\frac{1-f(E^\prime)-f(E)}{i\Omega- (E+E^\prime)} +
\frac{f(E^\prime)+f(E)-1}{i\Omega + (E+E^\prime)} +
\frac{f(E)-f(E^\prime)}{i\Omega- (E^\prime-E)}
\right)
\notag\\
&+
Tr\left[
\frac{(\xi \sigma_3 + \Delta \sigma_1) \sigma_i (\xi^\prime \sigma_3 + \Delta \sigma_1) \sigma_j}{E E^\prime}
\right]
\left(
\frac{f(E^\prime)-f(E)}{i\Omega- (E-E^\prime)}
-\frac{1-f(E^\prime)-f(E)}{i\Omega- (E+E^\prime)} 
-\frac{f(E^\prime)+f(E)-1}{i\Omega + (E+E^\prime)} +
\frac{f(E)-f(E^\prime)}{i\Omega- (E^\prime-E)}
\right)
\Bigg\}
\label{eqn:chi_final}
\end{align}
where $\xi$/$E$ means $\xi(k)$/$E(k)$, $\xi^\prime$/$E^\prime$ means $\xi(k+q)$/$E(k+q)$ and $f(E)$ is the fermion occupation number at energy $E$. At zero temperature, rotating $i\Omega$ to $\omega$, \equa{eqn:chi_final} simplifies to 
\begin{align}
\chi_{\sigma_i \sigma_j}(\omega, q) 
= \frac{1}{4}
\sum_{k}
\Bigg\{& 
Tr\left[
\sigma_i \sigma_j 
-
\frac{(\xi \sigma_3 + \Delta \sigma_1) \sigma_i (\xi^\prime \sigma_3 + \Delta \sigma_1) \sigma_j}{E E^\prime}
\right]
\frac{2(E+E^\prime)}{\omega^2 - (E+E^\prime)^2}
\notag\\
&+
Tr\left[
\frac{\sigma_i (\xi^\prime \sigma_3 + \Delta \sigma_1) \sigma_j}{E^\prime}
-
\frac{(\xi \sigma_3 + \Delta \sigma_1) \sigma_i \sigma_j}{E}
\right]
\frac{2\omega}{\omega^2 - (E+E^\prime)^2}
\Bigg\}
\label{eqn:chi_zero_t}
\end{align}

\begin{figure}
	\includegraphics[width= 3 in]{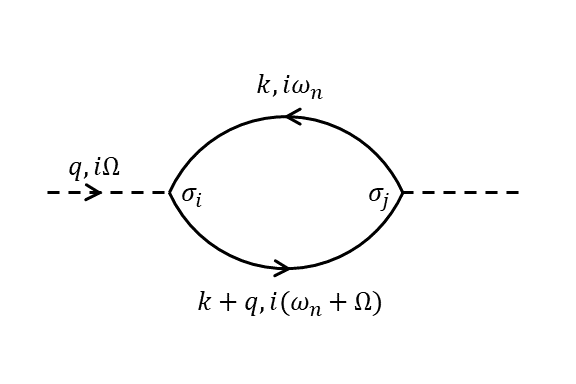}
	\caption{The bubble diagram for correlation function $\chi_{\sigma_i \sigma_j}$.}
	\label{fig:bubble}
\end{figure}

\subsection{The Higgs propagator}
\label{appendix:Higgs}
The Higgs propagator involves the correlation in $\sigma_1$ channel: 
\begin{align}
\chi_{\sigma_1 \sigma_1}(\omega, q) 
&=
\sum_{k}
\Bigg\{
\left(
1- \frac{\Delta^2-\xi\xi^\prime}{EE^\prime}
\right)
\frac{E+E^\prime}{\omega^2 - (E+E^\prime)^2}
\Bigg\}
\label{eqn:chi_zero_t}
\end{align}
At zero momentum, it becomes 
\begin{align}
\chi_{\sigma_1 \sigma_1}(\omega, 0) 
&= 
\sum_{k}
\frac{\xi^2}{E}
\frac{4}{\omega^2 - 4E^2}
\label{eqn:chi_sigma1}
\end{align}
With the knowledge of the gap equation, the Higgs propagator is thus \cite{Cea2015}
\begin{align}
G_a^{-1}(\omega)
&= \frac{1}{g} +
\chi_{\sigma_1 \sigma_1}(\omega, 0)
= (\omega^2 - 4\Delta^2)
\sum_{k}
\frac{1}{E(\omega^2 - 4E^2)}
=-(\omega^2 - 4\Delta^2) F(\omega)
\,.
\label{eqn:chi_11}
\end{align}
where 
\begin{align}
F(\omega)
&=
\sum_{k}
\frac{1}{E(-\omega^2 + 4E^2)}
\approx \frac{1}{2}\nu \int d\xi \frac{1}{E(-\omega^2+4E^2)}
=\frac{\nu}{4\Delta^2} \frac{2\Delta}{\omega} \frac{\mathrm{sin}^{-1}\left(\frac{\omega}{2\Delta}\right)}{\sqrt{1-\left(\frac{\omega}{2\Delta}\right)^2}}
\notag\\
&=
\frac{\nu}{2\omega \Delta }
\left\{
\begin{array}{lc}
\frac{\sin^{-1}(\frac{\omega}{2\Delta})}{\sqrt{1-(\frac{\omega}{2\Delta})^2}} & \omega \le 2\Delta
\\
\frac{-\sinh^{-1}\left(\sqrt{-1+(\frac{\omega}{2\Delta})^2}\right)}
{\sqrt{-1+(\frac{\omega}{2\Delta})^2}} 
+ i\frac{\pi}{2\sqrt{-1+(\frac{\omega}{2\Delta})^2}}
& \omega > 2\Delta
\end{array}
\right.
\,
\label{eqn:chi_11}
\end{align}
is shown in Fig.~\ref{fig:f_omega}.

\begin{figure}
	\includegraphics[width=0.4 \linewidth]{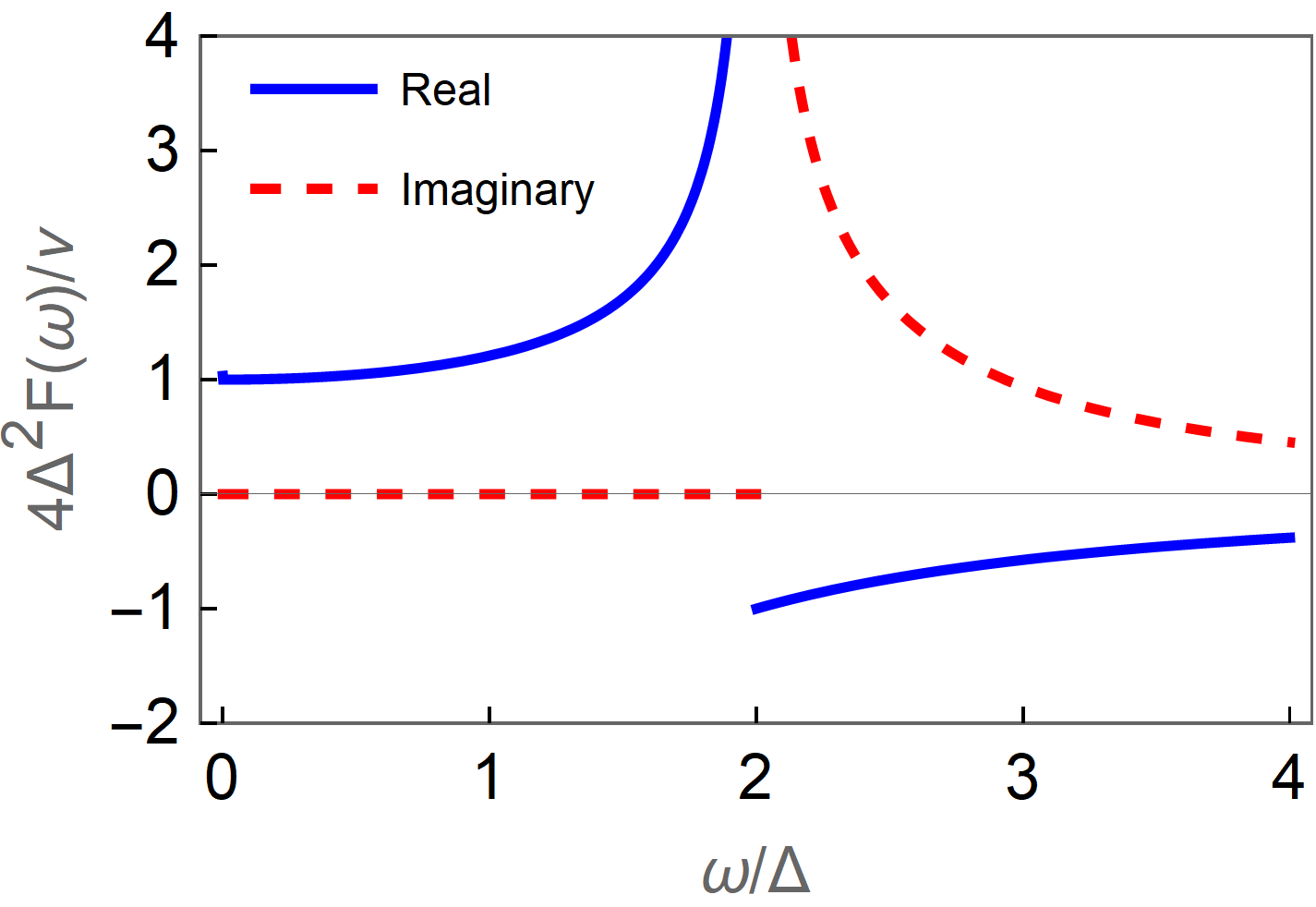}
	\caption{Real and imaginary parts of $F(\omega)$ as functions of $\omega$.}
	\label{fig:f_omega}
\end{figure}

At non-zero momentum, the propagator is \cite{Littlewood1982}
\begin{align}
G_a^{-1}(\omega, q) 
&=
\frac{1}{2}\sum_{k}
\frac{E+E^\prime}{EE^\prime}
\frac{\omega^2 - (\xi-\xi^\prime)^2-4\Delta^2}{\omega^2 - (E+E^\prime)^2}
\label{eqn:chi_finite_q}
\end{align}
whose $O(q^2)$ expansion gives 
\begin{align}
G_a^{-1}(\omega, q) 
\approx 
\left(-\omega^2 + 4\Delta^2 + \frac{1}{d}v_F^2q^2 \right)
F(\omega)
\,.
\label{eqn:chi_q_expand}
\end{align}

\subsection{The Bardasis-Schrieffer propagator}
\label{appendix:BS}
The total order parameter can be written as $\Delta_k=\Delta+ \sum_l\Delta_l(r,t) f_l(k)$ where we have chosen the mean field gap $\Delta$ to be real.
The subdominant pairing order parameter fluctuations $\Delta_l$ can have two possible directions: 1, orthogonal to $\Delta$ on the complex plane or in the `imaginary' direction; 2, parallel to $\Delta$ or in the `real' direction.  The `imaginary' fluctuations are the BaSh modes while the `real' ones don't have poles and are not collective modes.

We first consider the BaSh mode correlator  
\begin{align}
\chi_{f_l(k)\sigma_2, f_l(k)\sigma_2}(i\Omega,q) =
\sum_{\omega_n, k} Tr\left[ G(k,i\omega_n) f_l(k)\sigma_2 G(k+q,i(\omega_n+\Omega))   f_l(k)\sigma_2 \right]
\,.
\end{align}
In two dimension with rotational symmetry, the $d_{x^2-y^2}$ BaSh mode correlator is in the $\cos(2\theta_k)\sigma_2$ channel:
\begin{align}
\chi_{\cos(2\theta_k)\sigma_2,\cos(2\theta_k)\sigma_2}(\omega, 0) 
&= \sum_{k}
\frac{4 \cos^2(2\theta_k) E_k }{\omega^2 - 4E^2}
=-\frac{1}{2}\left(\frac{1}{g} +\omega^2 F(\omega) \right)
\,.
\label{eqn:bs_correlation}
\end{align}
The BaSh mode inverse propagator is 
\begin{align}
G_{\text{BaSh}}^{-1}(\omega) = \frac{1}{g_d} +
\chi_{\cos(2\theta_k)\sigma_2,\cos(2\theta_k)\sigma_2}(\omega, 0) 
= \frac{1}{g_d}  -\frac{1}{2g} -  \frac{1}{2}\omega^2 F(\omega)
\,
\label{eqn:bs_propagator}
\end{align}
which crosses zero at $\omega_{\text{BaSh}}$ below the gap, as shown by Fig.~\ref{fig:GBS_omega}(a). For momentum along $x$, extending the correlator to $O(q^2)$ gives 
\begin{align}
G_{\text{BaSh}}^{-1}(\omega) \approx \frac{1}{g_d}  -\frac{1}{2g} -  \frac{1}{2}\omega^2 F(\omega) 
+\frac{1}{16} \frac{\nu}{\Delta^2} v_F^2 q^2
\,
\label{eqn:bs_propagator_q}
\end{align}
in two dimension. In the case of $\omega_{\text{BaSh}} \ll 2\Delta$, the propagator is simplified to 
\begin{align}
G_{\text{BaSh}}^{-1}(\omega) \approx \frac{\nu}{8\Delta^2}
\left( \omega_{\text{BaSh}}^2 + \frac{1}{2} v_F^2 q^2 -\omega^2 \right) 
\,
\label{eqn:bs_propagator_q_simple}
\end{align}
where $\omega_{\text{BaSh}}^2=8\Delta^2 (\frac{1}{\nu g_d} - \frac{1}{2 \nu g})$. Thus the BaSh mode frequency disperses as $\omega_{\text{BaSh}}(q)^2= \omega_{\text{BaSh}}^2 + \frac{1}{2} v_F^2 q^2$.

We now consider the correlator of the `real' fluctuations:
\begin{align}
\chi_{\cos(2\theta_k)\sigma_1,\cos(2\theta_k)\sigma_1}(\omega, 0) 
&= \sum_{k}
\frac{\xi^2}{E}
\frac{4 \cos^2(2\theta_k)}{\omega^2 - 4E^2}
=\frac{1}{2} \left[ -\frac{1}{g}-(\omega^2 - 4\Delta^2) F(\omega) \right]
\,
\label{eqn:bs_real_correlation}
\end{align}
which is different from the Higgs correlator \equa{eqn:chi_sigma1} only by the $\cos^2(2\theta_k)$ factor. The resulting propagator is 
\begin{align}
G^{-1}_{\text{real}}(\omega, 0) 
= \frac{1}{g_d} + \chi_{\cos(2\theta_k)\sigma_1,\cos(2\theta_k)\sigma_1}(\omega, 0) 
= \frac{1}{g_d} -\frac{1}{2g} -\frac{1}{2}(\omega^2 - 4\Delta^2) F(\omega)
\,
\label{eqn:bs_real_correlation}
\end{align}
which never crosses zero as shown by Fig.~\ref{fig:GBS_omega}(b).

\begin{figure}
	\includegraphics[width=0.8 \linewidth]{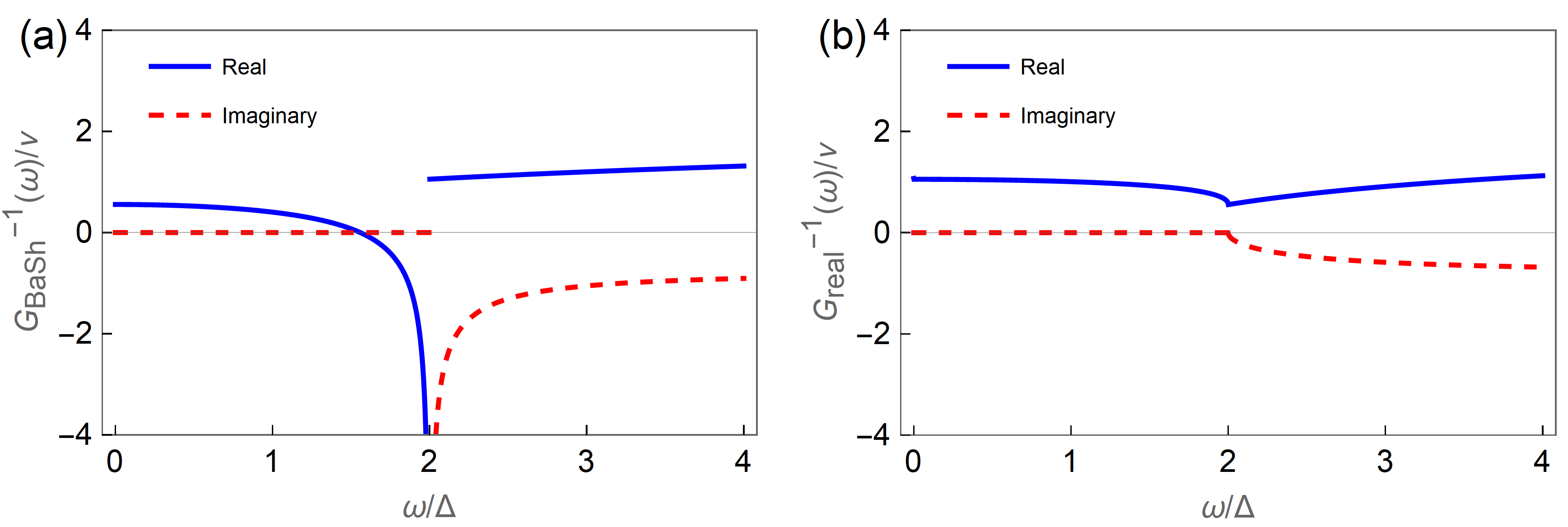}
	\caption{(a) Real and imaginary parts of the BaSh mode propagator $G_{\text{BaSh}}^{-1}(\omega)$ as a function of $\omega$. (b) Those of the real fluctuation $G_{\text{real}}^{-1}(\omega)$.}
	\label{fig:GBS_omega}
\end{figure}

\subsection{The linear coupling between phase and Bardasis-Schrieffer/Higgs modes}
\label{appendix:coupling_constants}
The coupling between phase fluctuation and the Higgs mode requires particle hole symmetry breaking which we model using an energy dependent DOS $g(\xi) = \nu (1+\lambda \xi/E_F)$. The coupling constants are derived from the correlation functions in \equa{eqn:phase_higgs_C}. From the general formula \equa{eqn:chi_zero_t} at zero temperature, the temporal part is
\begin{align}
\chi_{\sigma_3 \sigma_1}(\omega, q) 
&=
\sum_{k}
\Bigg\{ \Delta
\frac{\xi+\xi^\prime}{E E^\prime}
\frac{(E+E^\prime)}{(E+E^\prime)^2-\omega^2 }
\Bigg\}
\xrightarrow{q=0}
4\Delta\sum_{k}
\frac{\xi}{E \left(4E^2-\omega^2 \right) }
=4\Delta \int d\xi  \frac{g(\xi) \xi}{E \left(4E^2-\omega^2 \right) }
\notag \\
&= \lambda \frac{\Delta}{2E_F} \nu 
\left[ 
-\sqrt{\left(\frac{2\Delta}{\omega} \right)^2 -1} 
\tan^{-1} \left(  \frac{1}{\sqrt{\left(\frac{2\Delta}{\omega} \right)^2 -1} \sqrt{\left(\frac{\Delta}{\omega_D} \right)^2 +1} }  \right)
+\sinh^{-1} \left( \frac{\omega_D}{\Delta} \right)
\right]
\notag\\
& \approx  \lambda \nu \frac{\Delta}{2E_F}\sinh^{-1} \left( \frac{\omega_D}{\Delta} \right) 
\,
\label{eqn:phase_Higgs_C0}
\end{align}
which gives $C_0$ in \equa{eqn:higgs_coupling_C0}.
The spatial part is 
\begin{align}
\chi_{v_i\sigma_0,\, \sigma_1}(\omega, q) 
&=
\frac{\Delta}{2}
\sum_{k}
\Bigg\{
(v_i+v_i^\prime)
\frac{E-E^\prime}{E E^\prime}
\frac{\omega}{(E+E^\prime)^2-\omega^2 }
\Bigg\}
\approx  
\Delta \omega q_j 
\sum_{k}
\frac{ v_i v_j \xi}{\left(4 E^2-\omega^2\right) E^3 }
= \frac{1}{12d} \lambda \nu  \frac{\Delta}{E_F} \left(\frac{v_F}{\Delta}  \right)^2 \omega q_i
\,
\label{eqn:phase_Higgs_Ci}
\end{align}
which gives $C_i$ in \equa{eqn:higgs_coupling_Ci}.

The coupling of phase to the `real' $d$-wave order parameter fluctuations is similar to \equa{eqn:phase_Higgs_C0} and \eqref{eqn:phase_Higgs_Ci} except that another $f_d(k)$ term should be added to the momentum summation. The coupling constants are also suppressed by the small particle-hole breaking factor $\lambda \Delta/E_F$. We don't calculate them here since the `real' fluctuations are not collective modes.

We now calculate the coupling of phase to the $d$-wave BaSh fluctuations which are in the $\sigma_2 f_d(k)$ channel. The temporal part is 
\begin{align}
\chi_{\sigma_3,\, \sigma_2 f_d(k)}(\omega, q) 
&=
i\Delta \omega
\sum_{k}
\Bigg\{ f_d(k)
\frac{E+E^\prime}{E E^\prime}
\frac{-1}{(E+E^\prime)^2-\omega^2 }
\Bigg\}
\approx \frac{i}{4}\Delta \omega  
\sum_{k}
f_d(k)
\frac{1}{E^6}
\left( -\frac{5}{4} \frac{\xi^2}{E}
+ \frac{3}{4} E 
\right) (\mathbf{v} \mathbf{q})^2 
\,.
\label{eqn:phase_BS_B0}
\end{align}
The expansion to $O(q^2)$ is necessary because of the $d$-wave symmetry of $f_d(k)$.
It proves the temporal term in \equa{eqn:BS_phase_coupling} but we don't calculate it since this term affects the EM response at higher orders in $q$. The spatial part is 
\begin{align}
\chi_{v_i \sigma_0,\, \sigma_2 f_d(k)}(\omega, q) 
&=
i\Delta \sum_{k}
\Bigg\{ f_d(k) v_i
\frac{\xi-\xi^\prime}{E E^\prime}
\frac{E+E^\prime}{(E+E^\prime)^2-\omega^2 }
\Bigg\}
\approx 
i2\Delta q_j \sum_{k}
f_d(k) v_i v_j
\frac{1}{\left(4E^2-\omega^2 \right) E }
\,.
\label{eqn:phase_BS_B1}
\end{align}
There are two $d$-wave BaSh modes in two dimension: the $d_{x^2-y^2}$ and $d_{xy}$ modes which correspond to $f_{d1}=\cos 2\theta_k$ and $f_{d2}=\sin 2\theta_k$ respectively. Since they are different only by a $\pi/4$ rotation, we focus on the $d_{x^2-y^2}$ mode only. Replacing $f_d$ by $\cos 2\theta_k$ in \equa{eqn:phase_BS_B1} renders
\begin{align}
\chi_{v_i \sigma_0,\, \sigma_2 f_d(k)}(\omega, q) 
= i\pi \Delta v_F^2 F(\omega) M_{ij} q_j
\,
\label{eqn:phase_BS_B1_final}
\end{align}
where $\hat{M}=\sigma_3$. 

\subsection{The density density correlation}
The density density correlation is in the $\sigma_3$ channel:
\begin{align}
\chi^{(0)}_{\rho \rho} =
\chi_{\sigma_3 \sigma_3}(\omega, q) 
&= 
\frac{1}{2}\sum_{k}
\Bigg\{
\left(
1- \frac{\xi\xi^\prime-\Delta^2}{EE^\prime}
\right)
\frac{2(E+E^\prime)}{\omega^2 - (E+E^\prime)^2}
\Bigg\}
\label{eqn:chi_rho_rho}
\end{align}
At zero momentum it becomes 
\begin{align}
\chi_{\sigma_3 \sigma_3}(\omega, 0) 
&= 
\sum_{k}
\frac{\Delta^2}{E}
\frac{4}{\omega^2 - 4E^2}
= -4\Delta^2 F(\omega)
\label{eqn:chi_pho_pho_zero_q}
\end{align}
In the limit of $\omega \ll \Delta,\, q\ll \xi^{-1}$, we have $\chi_{\sigma_3 \sigma_3} =-\nu$.

\section{Near field reflection coefficients}
\label{appendix:rp}
\begin{figure}
	\includegraphics[width= \linewidth]{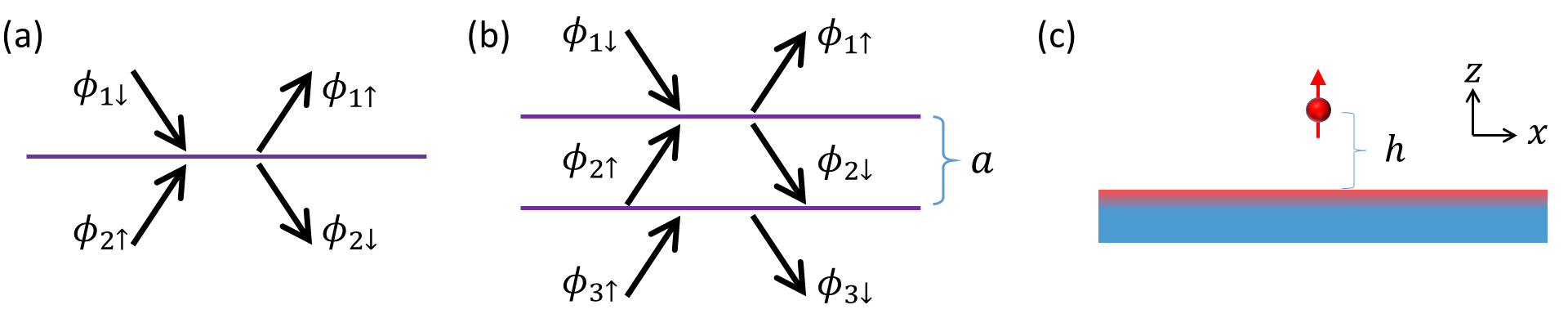}
	\caption{Schematics of the near field reflection problem in (a) monolayer, (b) double layer and (c) slab systems. The tip is shown as a dipole moment polarized along $\hat{z}$ direction.}
	\label{fig:nano_terahertz_setup}
\end{figure}
\subsection{Monolayer}
\label{appendix:rp_monolayer}
In the near field limit, there is only longitudinal electric field and no magnetic field. The incident and reflected fields can be described simply using electric potentials $\phi(r,t)$, as shown in Fig.~\ref{fig:nano_terahertz_setup}(a). We write the electrical potential as
\begin{align}
\phi_i(r,t)=e^{-i\omega t} \left(\phi_{i\uparrow}e^{iqx - q z}+\phi_{i\downarrow}e^{iqx + q z} \right)
\,,
\end{align}
where $\phi_{i\uparrow}$/$\phi_{i\downarrow}$ are the amplitude of up going/down going fields in the ith vacuum medium. We have explicitly noted that the z direction momentum is $\pm iq$ due to the Laplace equation satisfied by $\phi$ in vacuum, i.e., the electric potentials are evanescent waves. The reflection problem is described by the boundary conditions of $E_\parallel$ being continuous across the 2D layer and $E_\perp$ satisfying Gauss's law, or equivalently
\begin{align}
\phi_{1\uparrow} + \phi_{1\downarrow} = \phi_{2\uparrow} + \phi_{2\downarrow}
\,,\quad
(q\phi_{1\uparrow} - q\phi_{1\downarrow}) - (q\phi_{2\uparrow} - q \phi_{2\downarrow}) = 4\pi \rho_{2D} = 4\pi \frac{q}{\omega} j_{2D}= 4\pi \frac{q}{\omega} \sigma(\omega,q) (-iq) (\phi_{1\uparrow} + \phi_{1\downarrow})
\,.
\label{eqn:bc}
\end{align}
Written in matrix form, \equa{eqn:bc} becomes
\begin{align}
\begin{pmatrix}
1  & 1 \\
q + \frac{i4\pi q^2}{\omega} \sigma & -q + \frac{i4\pi q^2}{\omega} \sigma
\end{pmatrix}
\begin{pmatrix}
\phi_{1\uparrow} \\
\phi_{1\downarrow}
\end{pmatrix}
=
\begin{pmatrix}
1  & 1 \\
q  & -q
\end{pmatrix}
\begin{pmatrix}
\phi_{2\uparrow} \\
\phi_{2\downarrow}
\end{pmatrix}
\end{align}
whose solution gives the linear relation between the fields each side of the 2D layer 
\begin{align}
\begin{pmatrix}
\phi_{1\uparrow} \\
\phi_{1\downarrow}
\end{pmatrix}
=
\frac{1}{-2q}
\begin{pmatrix}
-q + \frac{i4\pi q^2}{\omega} \sigma  & -1 \\
-q - \frac{i4\pi q^2}{\omega} \sigma &  1
\end{pmatrix}
\begin{pmatrix}
1  & 1 \\
q  & -q
\end{pmatrix}
\begin{pmatrix}
\phi_{2\uparrow} \\
\phi_{2\downarrow}
\end{pmatrix}
=
\begin{pmatrix}
1 - \frac{i2\pi q}{\omega} \sigma  & -\frac{i2\pi q}{\omega} \sigma  \\
\frac{i2\pi q}{\omega} \sigma &  1 + \frac{i2\pi q}{\omega} \sigma
\end{pmatrix}
\begin{pmatrix}
\phi_{2\uparrow} \\
\phi_{2\downarrow}
\end{pmatrix}
\equiv
\hat{M}
\begin{pmatrix}
\phi_{2\uparrow} \\
\phi_{2\downarrow}
\end{pmatrix}
\end{align}
where $\hat{M}$ is the transfer matrix. Setting $\phi_{2\uparrow}=0$, one obtains the near field reflection coefficient for a 2D layer
\begin{align}
R_p \equiv \frac{\phi_{1\uparrow}}{\phi_{1\downarrow}} = \frac{-\frac{i2\pi q}{\omega} \sigma}{ 1 + \frac{i2\pi q}{\omega} \sigma} = 1- \frac{1}{\epsilon_{2D}}
\end{align} 
where $\epsilon_{2D}= 1 + \frac{i2\pi q}{\omega} \sigma$ is the dielectric function in 2D.

\subsection{Double layer}
As shown in Fig.~\ref{fig:nano_terahertz_setup}(b), applying the reflection problem twice, one obtains
\begin{align}
\begin{pmatrix}
\phi_{1\uparrow} \\
\phi_{1\downarrow}
\end{pmatrix}
=
\hat{M}
\begin{pmatrix}
\phi_{2\uparrow} \\
\phi_{2\downarrow}
\end{pmatrix}
=
\hat{M}
\begin{pmatrix}
e^{-qa}  & 0 \\
0 & e^{qa}
\end{pmatrix}
\hat{M}
\begin{pmatrix}
\phi_{3\uparrow} \\
\phi_{3\downarrow}
\end{pmatrix}
\,.
\end{align}
Setting $\phi_{3\uparrow}=0$ yields the reflection coefficient \equa{eqn:r_p_double_layer} for the double layer system. A characteristic plot of the reflection coefficient of the double layer system is Fig.~\ref{fig:doublelayer_rp_q} where resonances due to the symmetric and anti symmetric plasmons show up. The plasmonic field excited by a spatially local source can be obtained by summing up the reflection coefficients at all momentums, as shown in Fig.~\ref{fig:real_space_fringes}.

\begin{figure}
	\includegraphics[width=3.4 in]{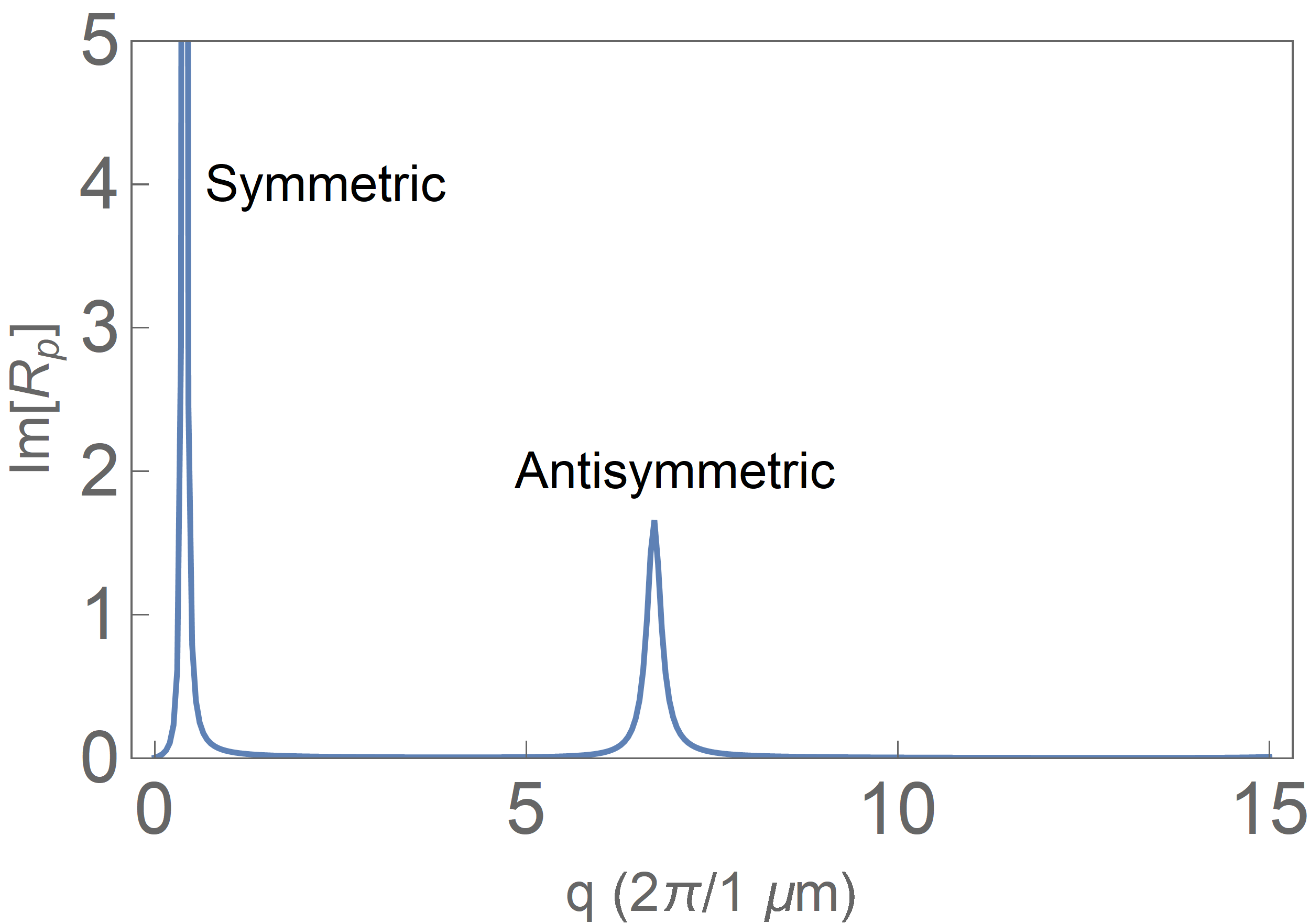}
	\caption{Horizontal cut of the color plot in Fig.~\ref{fig:double_layer_rp}(b) at $\omega= 4 \unit{THz}$, i.e., $\mathrm{Im}[R_p(4 \unit{THz},q)]$ as a function of $q$.}
	\label{fig:doublelayer_rp_q}
\end{figure}
\begin{figure}
	\includegraphics[width=0.7 \linewidth]{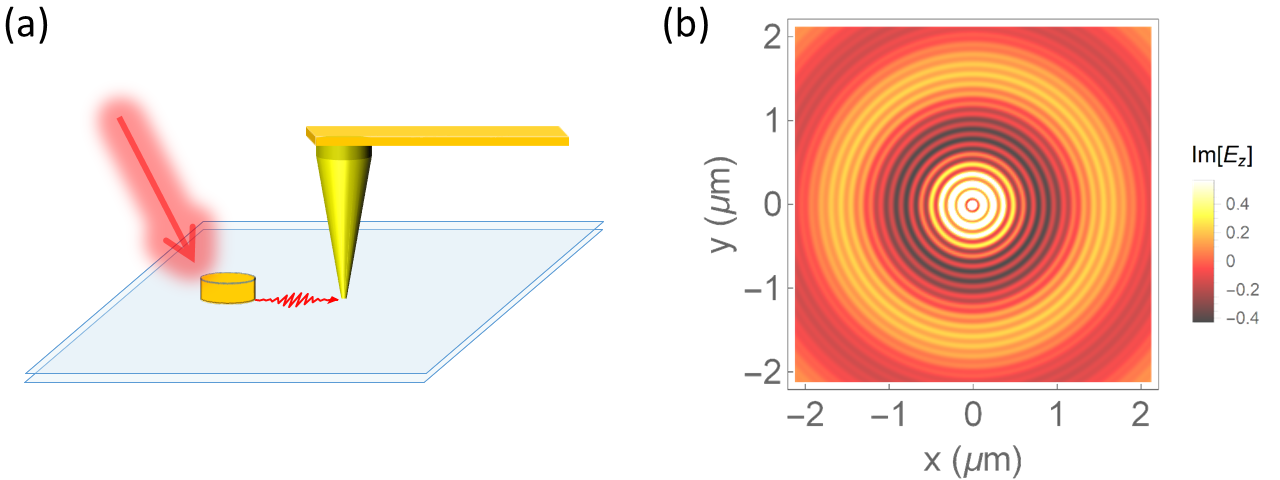}
	\caption{(a) Schematic of the near field experiment. (b) Distribution of z direction electric field $E_z(x,y,z=30 \unit{nm})$ excited by a dipole oscillating at the frequency $\omega=5 \unit{THz}$ placed on top of a superconductor double layer system shown in Fig.~\ref{fig:double_layer_rp}(a). The large/small period is due to the symmetric/antisymmetric mode. The dipole is polarized in z direction and is placed at $(x,y,z)=(0,0,30 \unit{nm})$ above the top layer. The parameters are $k_F = 2\pi/(3 \unit{nm})$, $v_F = 2.5\times 10^5 \unit{m/s} $,  $\gamma= 30 \unit{THz}$, $a= 3 \unit{nm}$,  $\Delta= 3.0 \unit{THz}$, $\kappa=0.2$ and $\kappa_{\text{BaSh}}=0.2$. Higgs/BaSh frequencies are assumed to be $4.5 \unit{THz}$/$3.0 \unit{THz}$ at zero momentum. }
	\label{fig:real_space_fringes}
\end{figure}

\subsection{A slab with nonlocal optical response}
If the polarization function is nonlocal, more unfortunately, if it depends also on the $z$ direction momentum $k_z$ such as that of the layered superconductor \equa{eqn:polarization_layered_sc}, the near field reflection coefficient of the vacuum-infinite superconductor interface should be modified to 
\begin{align}
R_p(\omega,q) = \frac{iq-k_z \epsilon_z(\omega,q,k_z)}{iq+k_z \epsilon_z(\omega,q,k_z)}
\,
\label{eqn:r_p_bscco}
\end{align}
where $q$ is the in-plane momentum which is a conserved quantity and $k_z$ is that in the nonlocal medium determined by the condition $\epsilon(\omega,q,k_z)=0$. \equa{eqn:r_p_bscco} can be derived in a similar fashion to Appendix~\ref{appendix:rp_monolayer}.
For a slab with finite thickness $a$, as shown in Fig.~\ref{fig:nano_terahertz_setup}(c), the transfer matrix method for solving the reflection problem renders
\begin{align}
R_{slab}(\omega,q) = R_p\frac{1-e^{2ik_z d}}{1-e^{2ik_z d} R_p^2}
\,
\label{eqn:r_p_bscco_slab}
\end{align}
where $R_p$ is from \equa{eqn:r_p_bscco}.

\end{widetext}

\section{Derivation of the two fluid model}
\label{appendix:two_fluid}
In principle, the two fluid formula \equa{eqn:two_fluid} can be obtained from the general derivation \equa{eqn:EM_response_tensor} with electron-impurity or electron-phonon scattering taken into account.
Here we sketch the derivation of \equa{eqn:two_fluid} by calculating the polarization function from \equa{eqn:EM_response_tensor}:
\begin{align}
\chi_{\rho\rho} = \chi_{\rho\rho}^{(0)}  - \frac{\left(\omega \chi_{\rho\rho}^{(0)} + \mathbf{q} \chi_{\rho\mathbf{j}}^{(0)} \right)^2}
{\omega^2 \chi_{\rho\rho}^{(0)} + q_i q_j \chi_{j_ij_j}^{(0)} + \frac{n}{m}q^2 + 2\omega \mathbf{q}\chi_{\rho\mathbf{j}}^{(0)}  }
\,.
\label{eqn:chi}
\end{align}
Close to $T_c$, we have 
\begin{align}
\chi_{\rho\rho}^{(0)} = \chi^{(0)} + \chi_s =   \chi^{(0)} - \frac{\pi}{4} \frac{\Delta}{T_c} \nu
\,
\label{eqn:chi_rho}
\end{align}
where $\chi^{(0)}$ come from the `intra band' process among thermally excited quasiparticles while $\chi_s$ is the `interband' contribution from exciting quasi particle pairs which has the interpretation of superfluid susceptibility (compressibility). Close to $T_c$, the $\chi^{(0)}$ should resemble the polarization function of a normal fermi liquid, i.e., the Lindhard function \cite{Mahan1990, Warren1960} with non-zero scattering rate. Similarly \cite{Altland.2010,Abrikosov:1975}, 
\begin{align}
\chi_{j_xj_x}^{(0)} + \frac{n}{m} = -i\omega \sigma_n + \frac{n_s}{m}
\,
\label{eqn:chi_jj}
\end{align}
where $\sigma_n $ is the `intraband' part and
\begin{align}
n_s &= n \int d\xi \left( \partial_E f(E) - \partial_\xi f(\xi) \right) 
\notag\\
&= n\frac{\Delta^2}{2}  \int d\xi \frac{1}{\xi} \partial^2_\xi f(\xi) + O\left(\frac{\Delta^4}{T^4}\right)
\notag\\
&\approx \frac{7\zeta(3)}{4\pi^2} \frac{\Delta^2}{T_c^2} n =2(1-T/T_c) n
\,
\label{eqn:chi_jj}
\end{align}
is the superfluid density of a clean superconductor. Moreover, close to $T_c$, the intraband contributions to the correlation functions should approximately satisfy the continuity equations
$\omega \chi^{(0)} +  \mathbf{q} \chi_{\rho\mathbf{j}}^{(0)} =0$ and $q_j \sigma_{nij} +  \omega \chi_{\rho j_i}^{(0)} =0$
since they are identical to those of the normal state at $\Delta=0$.
With these simplifications \equa{eqn:chi} becomes
\begin{align}
\chi_{\rho\rho} &= \chi^{(0)} + \chi_s  - \frac{\left(\omega \chi_s  \right)^2}
{\omega^2 \chi_s  +\frac{n_s}{m}q^2  }
\notag\\
&=\chi^{(0)}  + \frac{q^2 }
{\omega^2  - v_g^2 q^2  } D_s/\pi
\,.
\label{eqn:chi_two_fluid}
\end{align}
The corresponding conductivity is just \equa{eqn:two_fluid} with $\sigma_n = \chi^{(0)} i\omega/q^2$ and $v_g = \sqrt{\frac{n_s}{m}/\chi_s}$. 

\section{Longitudinal optical conductivity of the normal fermi liquid}
In the low frequency hydrodynamic regime ($\omega \ll \Gamma_{ee}$, $q \ll l_{ee}^{-1}$) of a fermi liquid, the longitudinal optical conductivity reads \cite{Sun.2018}
\begin{align}
\sigma(\omega,q) = i \frac{n e^2 / m}{\omega +i\Gamma_d -v_d^2 q^2/\omega} \,.
\label{eqn:hydrodynamic_conductivity}
\end{align}
In the above formula, $n$ is the electron density, $m$ is the electron effective mass, $\Gamma_d$ is the momentum relaxation rate and $v_d=\sqrt{\frac{1}{m}\left(\frac{\partial P}{\partial n} \right)_{ise}}$ is the first sound velocity of a neutral fermi liquid. 
Neglecting the effect of the Landau parameter $F_{0s}$, $v_d=v_F/\sqrt{D}$
where $D$ is the space dimension. In the limit of $\omega \gg v_d q ,\, D_f q^2$ where $D_f = v_d^2/\Gamma_d$ is the diffusion constant, \equa{eqn:hydrodynamic_conductivity} becomes the Drude formula. In the opposite limit, $\omega \ll v_d q,\, D_f q^2$, it crossovers to the Thomas-Fermi case.

\section{Ginzburg Landau action around T$_c$}
In this section, we derive the Ginzburg Landau action around $T_c$ of an $s$-wave BCS superconductor, where the collectives modes (except for the CG mode) are all overdamped. Without EM field, the action reads \cite{Altland.2010}
\begin{align}
S(\Delta) = \mathrm{Tr} \ln G_{\Delta} +   \int d\tau dr  \frac{1}{2g}|\Delta|^2 = \int d\tau dr \mathcal{L}
\label{eqn:GL_action_tc}
\end{align}
where $\mathcal{L}$ is the `Lagrangian'. Expansion of $S$ up to $|\Delta|^2$ gives 
\begin{align}
S = \sum_{\omega, q} \left( \frac{1}{2g} + \chi(\omega,q) \right) \Delta(-\omega,-q) \Delta(\omega,q)
\label{eqn:S_tc}
\end{align}
where 
\begin{align}
\chi(\omega, q)& = \frac{1}{V} \sum_k \frac{1-f(\xi_k)-f{\xi(-k+q)}}{i\omega_n -\xi_k -\xi_{-k+q}}
\notag\\
&=
\nu \left( -\frac{1}{2} \ln \frac{c_0 \omega_D}{T}
- \frac{c_1 i\omega}{T} -  \frac{ c_2 \omega^2}{T^2} 
+  \frac{c_s n}{mT^2} q^2
+ O(\omega^3, q^4)
\right)
\label{eqn:S_tc}
\end{align}
is the susceptibility for superconducting fluctuations and $c_0$, $c_1$, $c_2$, $c_s$ are $O(1)$ positive constants. We note that $c_0=2e^{\gamma_E}/\pi$ where $\gamma_E\approx 0.577$ is Euler's constant. The $|\Delta|^4$ contribution from a uniform static order parameter has the coefficient 
\begin{align}
\chi_4 & = \frac{2}{V} \sum_{k,i\omega_n} \frac{1}{\left( (i\omega_n)^2 -\xi_k^2 \right)^2}= \nu \frac{c_\beta}{T^2} 
\label{eqn:S_4}
\end{align}
where $c_\beta$ is an $O(1)$ positive constant. Thus the effective Lagrangian reads
\begin{align}
\mathcal{L} = \nu \Bigg(& \frac{c_1}{T}\Delta^\ast \partial_t \Delta 
 -
\frac{c_2}{T^2}|\partial_t \Delta|^2
+  \frac{c_s n}{m T^2} |\nabla \Delta|^2 
\notag \\
&+ \frac{1}{2}\left( \frac{1}{g\nu}- \ln\frac{\Lambda}{T} \right) |\Delta|^2 + \frac{c_\beta}{T^2} |\Delta|^4 
\Bigg)
\,.
\label{eqn:L_tc}
\end{align}
The EM field enters through the gauge invariant form $\partial_\mu \rightarrow \partial_\mu  + ieA_\mu $. 

The $\Delta^\ast \partial_t \Delta$ term should not exist in a well defined Lagrangian but should be understood as describing a dissipative term.  It is apparent that close to $T_c$, the amplitude dynamics is over damped \cite{Lemonik.2017} with the damping rate $\sim T$. As temperature lowers to $T\ll T_c$, this Lagrangian predicts that the amplitude dynamics crossovers to an under damped one. However, the power expansion in $\Delta$ is no longer valid there and the formalism in the main text should be employed.

\end{document}